\pdfoutput=1
\documentclass[cernpreprint, atlasdraft=true, texlive=2016, UKenglish, texmf]{atlasdoc}
 
\usepackage{atlaspackage}
\usepackage{pdflscape}
\usepackage{comment}
\usepackage{siunitx}
\usepackage{atlasbiblatex}
 
\usepackage{atlasphysics}

\graphicspath{{logos/}{figures/}}
 
\usepackage{ANA-EXOT-2019-23-PAPER-defs}
 
\newcommand{\numR}[1]{\num[round-mode=figures,round-precision=2]{#1}}
 
\AtlasTitle{Search for neutral long-lived particles in $pp$ collisions at $\sqrt{s}=13$ TeV that decay into displaced hadronic jets in the ATLAS calorimeter}
 
\AtlasAbstract{
A search for decays of pair-produced neutral long-lived particles (LLPs) is presented using 139 fb$^{-1}$ of proton--proton collision data collected by the ATLAS detector at the LHC in 2015--2018 at a centre-of-mass energy of 13 TeV.
Dedicated techniques were developed for the reconstruction of displaced jets produced by LLPs decaying hadronically in the ATLAS hadronic calorimeter.
Two search regions are defined for different LLP kinematic regimes.
The observed numbers of events are consistent with the expected background, and limits for several benchmark signals are determined.
For a SM Higgs boson with a mass of 125 GeV, branching ratios above 10\% are excluded at 95\% confidence level for values of $c$ times LLP mean proper lifetime in the range between 20 mm and 10 m depending on the model.
Upper limits are also set on the cross-section times branching ratio for scalars with a mass of 60 GeV and for masses between 200 GeV and 1 TeV.
}
 
\PreprintIdNumber{CERN-EP-2022-002}

\author{The ATLAS Collaboration}
 
\AtlasRefCode{EXOT-2019-23}
 
\AtlasJournal{JHEP}
\AtlasJournalRef{JHEP 06 (2022) 005}
\AtlasDOI{10.1007/JHEP06(2022)005}

\hypersetup{pdftitle={ATLAS document},pdfauthor={The ATLAS Collaboration}}
 
\begin{document}
 
\maketitle

\section{Introduction}
 
Many of the scenarios proposed to address some of the open questions of the Standard Model (SM) predict the existence of new non-SM particles whose lifetimes can be long enough for their decays to be significantly displaced from the interaction point but still within the ATLAS detector.
Examples of theories involving such suitably long-lived particles (LLPs) are various supersymmetric (SUSY) models~\cite{Arvanitaki:2012ps,ArkaniHamed:2012gw,Giudice:1998bp,Barbier:2004ez,Csaki:2015fea,Fan:2011yu,Fan:2012jf}; neutral naturalness models~\cite{Chacko:2015fbc,Burdman:2006tz,Cai:2008au,Chacko:2005pe} which feature a hidden sector (HS)~\cite{Strassler:2006im,Strassler:951513,Chan:2011aa} that addresses the hierarchy problem; models that seek to incorporate dark matter~\cite{Baumgart:2009tn,Kaplan:2009ag,Dienes:2011ja,Dienes:2012yz}, or explain the matter--antimatter asymmetry of the universe~\cite{Cui:2014twa}; and models that lead to massive neutrinos~\cite{Helo:2013esa,Batell:2016zod} which provide an explanation for the origin of light neutrino masses and mixings.
 
The decay of an LLP created in proton--proton collisions in the ATLAS detector could produce one of a variety of highly unconventional signatures, depending on the detector subsystem in which the LLP decays.
Searches for promptly decaying particles often have very low sensitivity to such signatures and hence the study of LLPs requires dedicated analyses.
This paper presents a search sensitive to neutral LLPs decaying chiefly in the calorimeters of the ATLAS detector.
This allows the analysis to probe values of $c\tau$ (where $c$ is the speed of light and $\tau$ is the mean proper lifetime of the LLP) ranging between a few centimetres and a few tens of metres.

\begin{figure}[ht]
\centering
\includegraphics[width=0.35\textwidth]{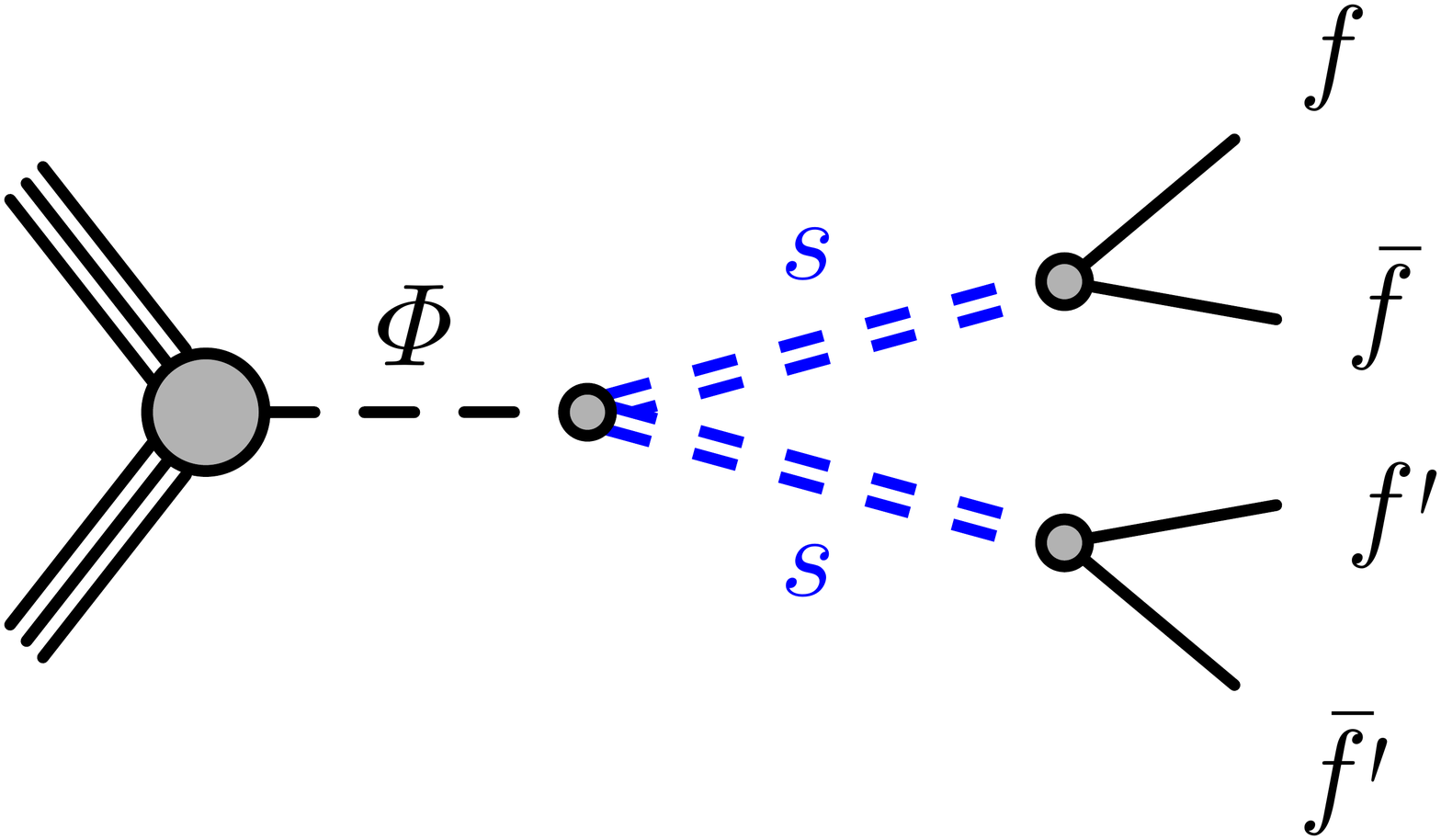}
\caption{A diagram showing the \higgsd decay used as the benchmark model.
The $s$ inherits Yukawa couplings to SM fermions from the $\Phi$, and therefore decays primarily to heavy quarks.}
\label{fig:HSfd}
\end{figure}

A HS benchmark model~\cite{Chung2006,Chang2008,Strassler:2006im,Strassler:951513,Chan:2011aa} is studied, in which the SM and HS are connected via a heavy neutral boson ($\Phi$), which decays into two long-lived neutral scalar bosons ($s$).
While $\Phi$ could be the SM Higgs boson, this analysis considers mediators with masses ($\mPhi$) ranging from $\mPhiLo \GeV$ to $\mPhiHi \GeV$, and scalars with masses between $\mSLo \GeV$ and $\mSHi \GeV$.
The decay \higgsd is considered (see Figure~\ref{fig:HSfd}), where $f$ refers to fermions and $\bar{f}$ refers to anti-fermions.
The branching ratio (BR, with value $B$) of each decay mode depends on the mass of the scalar, with the dominant decay being to the heaviest quark that is kinematically accessible.
In the hypothesised physics process with the largest scalar mass $\mS$ (475~\GeV), decays to top quarks dominate ($B>99\%$).
Conversely, in the hypothesis with the lightest $\mS$ (5~\GeV), decays to charm quarks dominate ($B\sim 75\%$), followed by decays to $\tau$-leptons ($B\sim 25\%$).
In the other cases, the relation among the BRs is approximately constant and typically 85:8:5 for $b\bar{b}$, $c\bar{c}$, and $\tau^{+}\tau^{-}$, respectively.
A single analysis strategy, described in Section~\ref{sec:selection}, is used for all mass hypotheses regardless of the scalar LLP's decay mode.
The $c\tau$ of LLPs in HS models are usually not constrained by theory, apart from a rough upper limit of $c\tau \lesssim 10^8$~m given by the cosmological constraint of Big Bang Nucleosynthesis~\cite{Jedamzik:2006xz}, and could be short enough for the LLPs to decay inside the ATLAS detector volume.
 
The SM fermions from the LLP decay result in jets whose origins may be far from the interaction point (IP) of the colliding protons, leading to so-called displaced vertices or displaced jets.
If the LLP decay occurs in the calorimeters, the decay products are collimated enough to be reconstructed as a single jet which is narrow, trackless and with an unusually high proportion of its energy in the hadronic calorimeter.
Since pair-produced LLPs are considered, this analysis requires two such non-standard jets, with two selections targeting different LLP kinematic regimes. One is optimised for models with $\mPhi\leq 200\gev$ (referred to as low-\ET models), and the other for models with $\mPhi> 200\gev$ (high-\ET models), where \ET denotes transverse energy.
The dominant background process that mimics this signal is SM multijet production, in cases where the jets are composed mainly of neutral hadrons or where some of the tracks are misreconstructed. 
Despite the low probability for a prompt jet to produce a signal-like jet, the SM multijet cross-section is high enough for this to be the dominant background in this search.
Other contributions come from jets reconstructed from non-collision background consisting of cosmic rays and beam-induced background (BIB)~\cite{CERN-EP-2016-029}.
The latter is composed of LHC beam--gas interactions and beam-halo interactions with the collimators upstream of the ATLAS detector, resulting in muons travelling parallel to the beam-pipe.
This analysis makes use of a new per-jet neural network to discriminate signal-like jets from non-displaced jets or BIB-like jets, and a boosted decision tree to separate signal from background events. The per-jet neural network replaces a boosted decision tree used in the previous version of the analysis, using an adversarial training scheme to minimise the effect of mis-modelling in the input variables. This is the first time such a technique has been used in an ATLAS analysis.
The background estimation is performed using a data-driven method.

Previous searches for similar signatures (pair-produced neutral LLPs decaying hadronically) at hadron colliders have been performed at the Tevatron~\cite{HVD0,HVCDF} and at the LHC.
The search for displaced jets at LHCb in Ref.~\cite{LHCb:2017xxn} is sensitive to $c\tau$ values from ${\sim}1$~mm to ${\sim}0.1$~m.
The most recent CMS searches at 13~\TeV~\cite{CMS-EXO-17-018,CMS-EXO-19-001,CMS-EXO-19-005,CMS-EXO-19-011} involve jets with displaced vertices in the tracking system, and are sensitive to $c\tau$ values from ${\sim}1$~mm to ${\sim}1$~m.
Previous ATLAS searches at 13~\TeV looked for displaced vertices in the tracking system~\cite{HIGG-2017-05,EXOT-2018-57}, 
pairs of reconstructed vertices in the muon spectrometer~\cite{EXOT-2017-05}, or the combination of one displaced vertex in the muon spectrometer and one in the inner tracking detector~\cite{EXOT-2018-61}.
These ATLAS searches are complementary, and together provide coverage of $c\tau$ values extending from effectively prompt to ${\sim}200$~m.
This analysis is an update of a search for pairs of displaced hadronic jets in the ATLAS calorimeters~\cite{EXOT-2017-25}, with significant improvements to the displaced-jet identification, and using the full LHC Run~2 dataset with 139~fb$^{-1}$ of 13~\TeV data instead of only data from 2016. It is sensitive to long-lived particle $c\tau$ values between approximately 20~mm and 10~m, depending on the model.

The paper is structured as follows. The ATLAS detector is described in Section~\ref{sec:detector}. The collection of the data and generation of samples of simulated events are then discussed in Section~\ref{sec:datasamples}. The trigger and event selection are detailed in Section~\ref{sec:selection}, followed by a discussion of the estimation of the background yield in the search regions in Section~\ref{sec:background}. The systematic uncertainties are summarised in Section~\ref{sec:systematics}. The statistical interpretation of the data is described in Section~\ref{sec:interpretation}, and the conclusions are given in Section~\ref{sec:conclusion}.

\section{ATLAS detector}
\label{sec:detector}
 
\newcommand{\AtlasCoordFootnote}{
ATLAS uses a right-handed coordinate system with its origin at the nominal interaction point
in the centre of the detector and the $z$-axis along the beam pipe.
The $x$-axis points from the IP to the centre of the LHC ring,
and the $y$-axis points upwards.
Cylindrical coordinates $(r,\phi)$ are used in the transverse plane,
$\phi$ being the azimuthal angle around the $z$-axis.
The pseudorapidity is defined in terms of the polar angle $\theta$ as $\eta = -\ln \tan(\theta/2)$.
Angular distance is measured in units of $\Delta R \equiv \sqrt{(\Delta\eta)^{2} + (\Delta\phi)^{2}}$.}

The ATLAS detector~\cite{PERF-2007-01} at the LHC covers nearly the entire solid angle around the collision point.\footnote{\AtlasCoordFootnote}
It is a multipurpose detector consisting of an inner tracking detector surrounded by a thin superconducting solenoid, electromagnetic and hadronic calorimeters,
and a muon spectrometer incorporating three large superconducting toroidal magnets.
The inner-detector system is immersed in a \SI{2}{\tesla} axial magnetic field
and provides charged-particle tracking in the range $|\eta| < 2.5$.
 
The high-granularity silicon pixel detector covers the vertex region and typically provides four measurements per track.
The layer closest to the interaction point is known as the insertable B-Layer~\cite{ATLAS-TDR-2010-19,PIX-2018-001}. It was added in 2014 and provides high-resolution hits at small radius to improve the tracking performance.
The pixel detector is surrounded by the silicon microstrip tracker, which usually provides four three-dimensional measurement points per track.
These silicon detectors are complemented by the transition radiation tracker, with coverage up to $|\eta| = 2.0$. 
 
The calorimeter system covers the pseudorapidity range $|\eta| < 4.9$.
Within the region $|\eta|< 3.2$, electromagnetic calorimetry is provided by barrel and endcap high-granularity lead/liquid-argon (LAr) electromagnetic calorimeters (together referred to as the ECal), with an additional thin LAr presampler covering $|\eta| < 1.8$ to correct for energy loss in material upstream of the calorimeters.
The ECal extends from $1.5$~m to $2.0$~m in radial distance $r$ in the barrel and from $3.6$~m to $4.25$~m in $|z|$ in the endcaps.
Hadronic calorimetry is provided by a steel/scintillator-tile calorimeter (HCal), segmented into three barrel structures within $|\eta| < 1.7$, and two copper/LAr hadronic endcap calorimeters covering $|\eta| > 1.5$.
The HCal covers the region from $2.25$~m to $4.25$~m in $r$ in the barrel (although the HCal active material extends only up to $3.9$~m) and from $4.3$~m to $6.05$~m in $|z|$ in the endcaps.
The solid angle coverage is completed with forward copper/LAr and tungsten/LAr calorimeter modules optimised for electromagnetic and hadronic measurements, respectively.
 
The calorimeters have a highly granular lateral and longitudinal segmentation. Including the presamplers, there are seven sampling layers in the combined central calorimeters (the LAr presampler, three in the ECal barrel and three in the HCal barrel), and the endcap regions provide up to eight sampling layers (the presampler, three in ECal endcaps and four in HCal endcaps). The forward calorimeter modules provide three sampling layers in the forward region.
The total amount of material in the ECal corresponds to 24--35 radiation lengths in the barrel and 35--40 radiation lengths in the endcaps. The combined depth of the calorimeters for hadronic energy measurements is more than 9~hadronic interaction lengths nearly everywhere across the full detector acceptance.
 
The muon spectrometer comprises separate trigger and
high-precision tracking chambers measuring the deflection of muons in the magnetic field generated by the superconducting air-core toroids.
The field integral of the toroids ranges between \num{2.0} and \SI{6.0}{\tesla\metre}
across most of the detector.
 
The ATLAS detector selects events using a tiered trigger system~\cite{TRIG-2019-04}. The level-1 (L1) trigger is implemented in custom electronics and reduces the event rate from the LHC crossing frequency of 40~MHz to a design value of 100~kHz. The second level, known as the high-level trigger (HLT), is implemented in software running on a commodity PC farm that processes the events and reduces the rate of recorded events to 1~kHz. An extensive software suite~\cite{ATL-SOFT-PUB-2021-001} is used in the reconstruction and analysis of real and simulated data, in detector operations, and in the trigger and data acquisition systems of the experiment.
 
\section{Data and simulation samples}
\label{sec:datasamples}
Data collected by the ATLAS detector during the period 2015--2018 from proton--proton ($pp$) collisions at $\sqrt{s}=13\TeV$ are used in this search.
Only data collected during stable beam conditions in which all detector subsystems were operational are considered~\cite{DAPR-2018-01}.
These data were collected with a set of dedicated LLP signature-driven triggers and separated into four datasets, defined according to the triggers used to collect them.
The search is performed on the main dataset, composed of all data events passing at least one of the two types of CalRatio triggers~\cite{TRIG-2012-02} running on bunch crossings where protons were present in both beams.
The name ``CalRatio'' refers to the ratio of energies deposited in the hadronic and electromagnetic calorimeters.
As described in detail in Section~\ref{sec:selection}, these include the low-\et CalRatio triggers and high-\et CalRatio triggers. 
Different versions of these triggers ran during the full data-taking period.
The amount of data collected in each case is summarised in Table~\ref{tab:data-datasets-by-year}, with at least one of these triggers running at any given time during the entire data-taking.
Two additional datasets were collected for the study of non-collision backgrounds.
The BIB dataset, used for the study of BIB events faking signal events,
was collected from events failing the CalRatio trigger BIB-removal algorithm.
The cosmics dataset, used for the estimation of cosmic-ray events passing the analysis selection, was collected from events recorded during empty bunch crossings, as described in Section~\ref{sec:selection}.
Finally, a dijet dataset is selected using a single-jet-based trigger and vetoing on the CalRatio triggers to make it orthogonal to the main dataset. This dataset is used in the neural network training described in Section~\ref{sec:NN} and in the calculation of some of the systematic uncertainties involved in the analysis.

\begin{table}[h]
\begin{center}
\begin{tabular}{|c|c|c|c|c|c|}
\hline
Trigger & \multicolumn{5}{c|}{Collected integrated luminosity [fb$^{-1}$]} \\
& 2015 & 2016 & 2017 & 2018 & Total\\
\hline
High-\et CalRatio trigger with $\et>60\gev$ & 3 & 33 & 41 & 40 &  117 \\
High-\et CalRatio trigger with $\et>100\gev$ & -- & -- & 44 & 59 & 103\\
Low-\et CalRatio trigger (2016 version) & -- & 11 & 43 & -- & ~~54\\
Low-\et CalRatio trigger (2018 version) & -- & -- & -- & 59 & ~~59\\
\hline
\end{tabular}
\vspace{0.25cm}
\caption{CalRatio triggers which were available during the LHC Run~2 data-taking, and corresponding integrated luminosity collected in each period. The high-\et CalRatio trigger with $\et>60\gev$ was disabled in 2017 for instantaneous luminosities higher than $1.4 \times 10^{34}~\textrm{cm}^{-2} \textrm{s}^{-1}$. Two versions of the low-\et CalRatio trigger were used, with slight differences in their algorithms. The details are reported in Section~\ref{sec:selection}.
}
\label{tab:data-datasets-by-year}
\end{center}
\end{table}

The HS $\Phi \rightarrow s s$ signal samples were generated using the \MGNLO v2.6.2~\cite{MadGraphNLO}  generator at leading order (LO) with the \NNPDF[2.3lo] parton distribution function (PDF) set~\cite{Ball:2012cx}.
In these samples, $\Phi$ is produced via gluon--gluon fusion.
The $\Phi$ transverse momentum distribution for the samples is reweighted to match that obtained for corresponding next-to-leading-order (NLO) predictions using the same event generator.
A production cross-section of $48.6$~pb, taken from a next-to-next-to leading-order calculation~\cite{Cepeda:2019klc}, is assumed when normalising results for the case where the mediator is the SM Higgs boson (the mass of which was set to 125~\gev).
Parton showering and hadronisation was modelled using \Pythia[8.230]~\cite{Sjostrand:2014zea} with the \textsc{A14} set of tuned parameters (tune)~\cite{ATL-PHYS-PUB-2014-021}.
Several sets of samples were generated, with different assumptions for the masses of the mediator ($\mPhi \in [60,1000]\gev$) and LLPs ($\mS \in [5, 475]\gev$).
These mass ranges ensure an extensive LLP boost spectrum, allowing a wide variety of topologies to be studied.
The LLP mean proper lifetime at which each of the signal samples was generated ($\tau_{gen}$) was determined so as to maximise the fraction of decays in the ATLAS hadronic calorimeter and muon system, and is quoted in relevant tables and figures throughout this document.
Some of the samples were generated for two assumptions about the LLP generated lifetime: one sample is used to study the signal throughout the analysis, while the other sample (with the alternative lifetime assumption but exactly the same mass choices) is used to validate the procedure for extrapolating limits to different mean proper lifetimes of the long-lived scalar.

The dominant SM background in this analysis is SM multijet production. Although a data-driven method is used to perform the background estimation, Monte Carlo (MC) simulated multijet events are needed to train the per-jet neural network to discriminate between signal jets and the multijet background, and to evaluate some of the systematic uncertainties.
The samples were generated at LO with \Pythia[8.186]~\cite{Sjostrand:2007gs} using the \textsc{A14} tune for parton showering and hadronisation. The \NNPDF[2.3lo] PDF set~\cite{Ball:2012cx} was used.

The effect of multiple interactions in the same and neighbouring bunch
crossings (\pileup) was modelled by overlaying the simulated hard-scattering event with
inelastic $pp$ collision events generated with \PYTHIA[8.186]~\cite{Sjostrand:2007gs}
using the \NNPDF[2.3lo] PDF set and the
A3 tune~\cite{ATL-PHYS-PUB-2016-017}.
 
The detector geometry and response were simulated with \GEANT~\cite{SOFT-2010-01, Geant4}.
The standard ATLAS reconstruction software is used for both simulation and collision data.

\section{Trigger and event selection}
\label{sec:selection}
 
This section describes how the data were collected, processed and selected for the analysis. The main dataset was collected with dedicated CalRatio triggers, and auxiliary datasets are also defined so as to be enriched in background events such as BIB, cosmic rays, or SM dijets. Using a combination of data and simulated samples, per-jet neural networks are trained to distinguish three classes of objects: signal-like jets, BIB-like jets, and non-displaced jets. The neural network training architecture involves an adversarial network, which is used to ensure that differences between data and simulation in the training samples are not exploited. Details are provided in Section~\ref{sec:NN}.
Per-event boosted decision trees (one each for low-\ET and high-\ET signals) make use of the per-jet neural network scores as well as several other event-level variables to construct discriminants which separate signal from all sources of background. Some additional requirements are applied to maximise the sensitivity of the final selection. Finally, the data-driven ABCD method, explained in detail in Section~\ref{sec:background}, is used to estimate the remaining background in the final selection. In this method the boosted decision tree score is used as one of two axes which define the ABCD plane.
 
\subsection{Triggers and event preselection}
 
Events are selected by the CalRatio triggers~\cite{TRIG-2012-02}, which are designed to identify jets that result from neutral LLPs decaying near the outer radius of the ECal or within the HCal.
The combined L1 and HLT selections make use of the main characteristics of jets resulting from the decay of a neutral LLP in the calorimeters: due to the late development of these jets, they appear narrower than standard jets, they have a high fraction of their energy deposited in the HCal, and they are isolated from tracks.
Two types of such triggers are used, differing only in the L1 trigger selection.
 
\begin{figure}[t]
\centering
\subfloat[]{\includegraphics[width=0.48\linewidth]{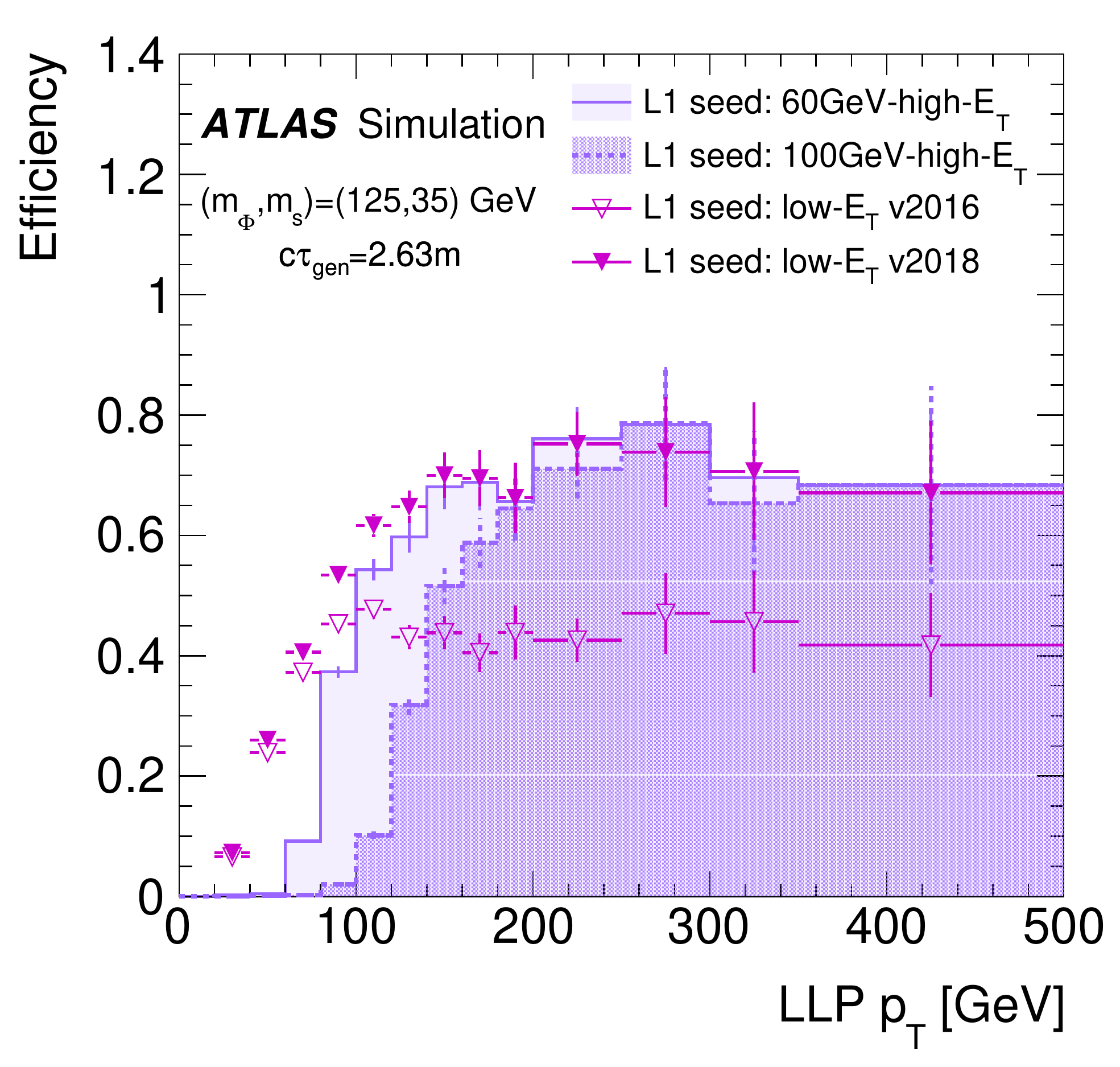}}
\subfloat[]{\includegraphics[width=0.48\linewidth]{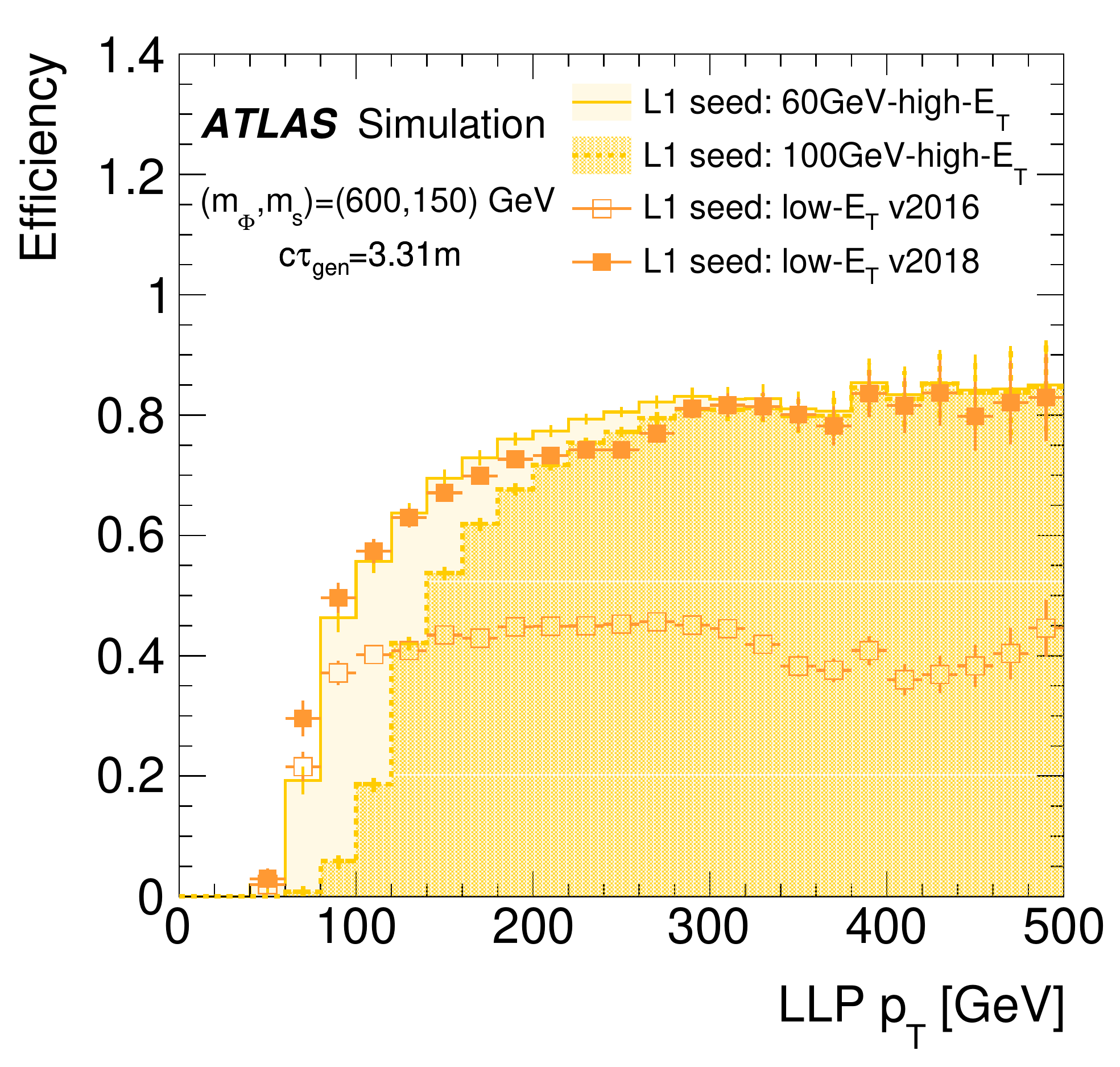}}
\caption{Trigger efficiency for simulated signal events as a function of the LLP \pt\ for (a) one of the low-\ET signal samples and (b) one of the high-\ET signal samples for HLT CalRatio triggers seeded by the high-\et L1 triggers with \ET thresholds of 60~\GeV and 100~\GeV and by the two versions of the low-\et L1 triggers. Only statistical uncertainties are shown.
}
\label{fig:trigEffpt}
\end{figure}

\begin{figure}[ht]
\centering
\subfloat[]{\includegraphics[width=0.48\linewidth]{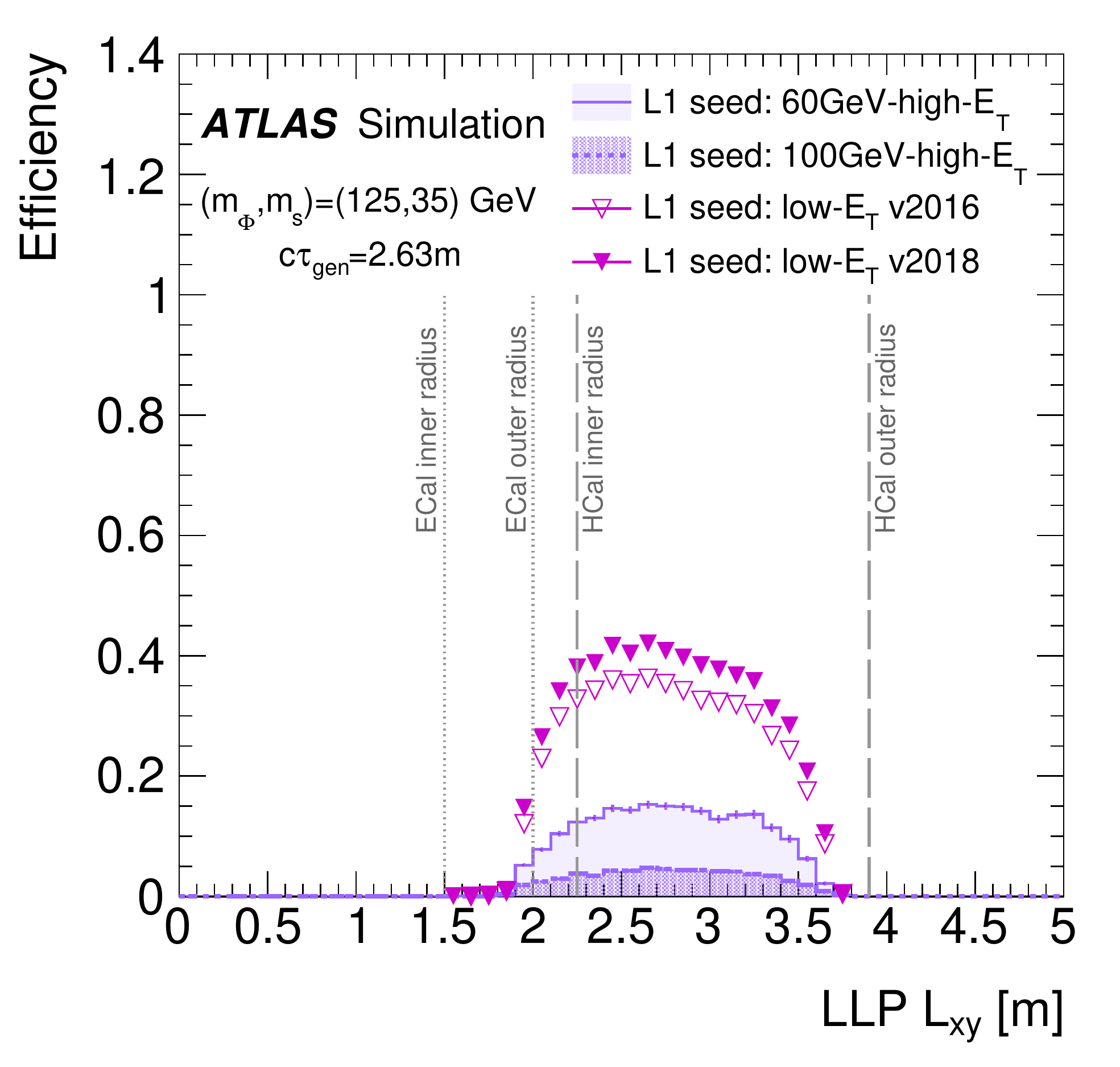}}
\subfloat[]{\includegraphics[width=0.48\linewidth]{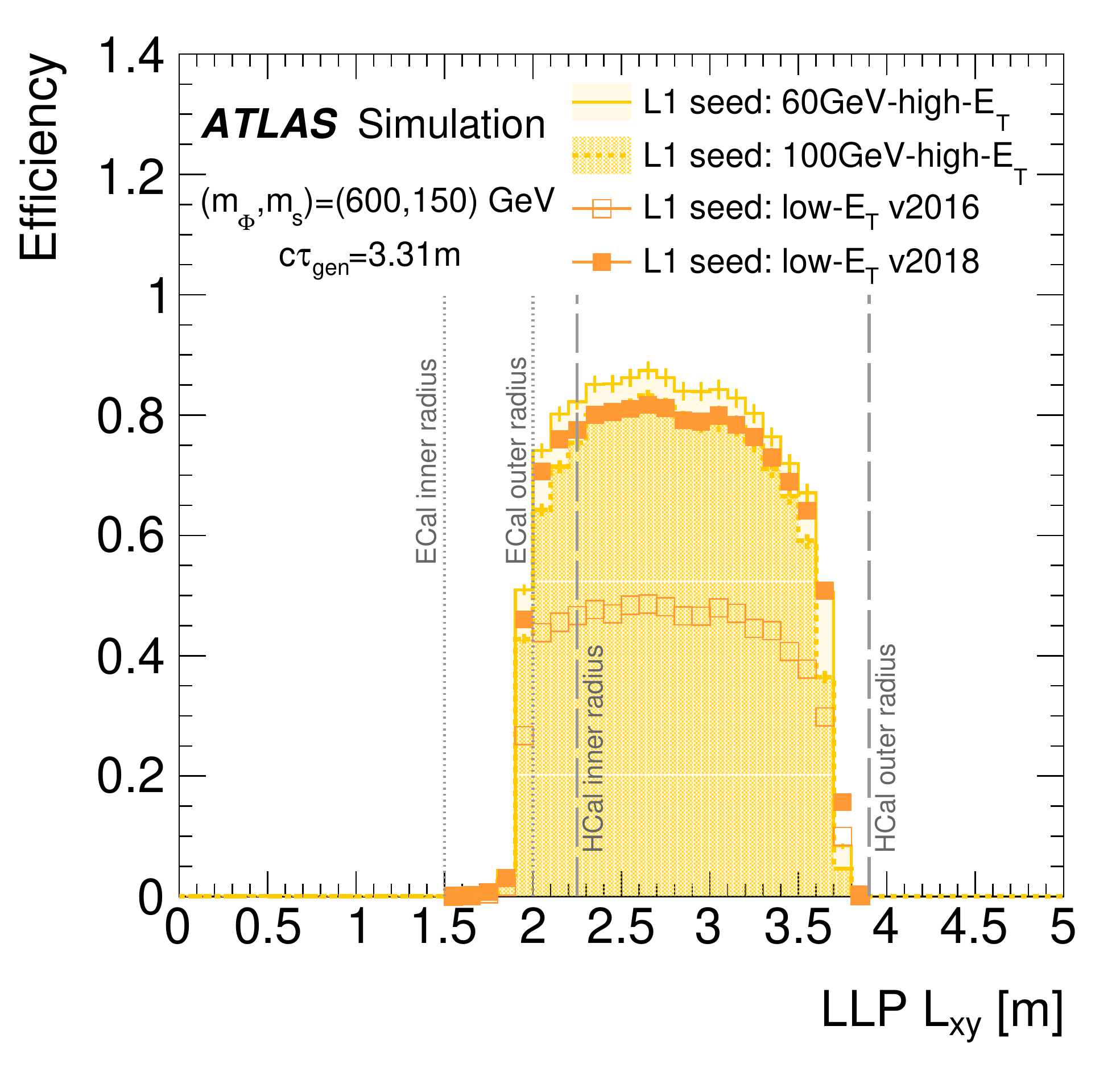}}
\caption{Trigger efficiency for simulated signal events as a function of the LLP decay position in the $x$--$y$ plane ($L_{xy}$) for LLPs decaying in the barrel ($|\eta|<1.4$) for (a) one of the low-\ET signal samples and (b) one of the high-\ET signal samples for HLT CalRatio triggers seeded by the high-\et L1 triggers with \ET thresholds of 60~\GeV and 100~\GeV and by the two versions of the low-\et L1 triggers. Only statistical uncertainties are shown.
}
\label{fig:trigEffLxy}
\end{figure}

At L1, two different triggers are used. The high-\et L1 triggers select narrow jets with energy deposits above certain thresholds (either $\et>60\gev$ or $\et>100\gev$) in a $0.2 \times 0.2~ (\Delta \eta \times \Delta \phi)$ region of the ECal and HCal combined~\cite{ATLAS-CONF-2017-061}.
The low-\et L1 trigger makes use of the L1 topological trigger system~\cite{TRIG-2019-02} by accepting events
where the largest energy deposit, with total-\et threshold at 30~\GeV, does not geometrically overlap in the $\eta$–$\phi$ plane with any energy deposits with $\et>3\GeV$ in the ECal.
The veto on ECal deposits ensures a high value of the ratio at L1 of energy deposited in the HCal to energy deposited in the ECal, $\calrat$ > 9, rejecting a large portion of the background with typical values of $\calrat$ <1, and allowing the \et threshold to be kept low.
This looser \et requirement increases the efficiency for lower-\pt LLPs.
In its 2016 version, the trigger requires that if there is more than one HCal deposit, the second-most energetic one is also isolated from energy deposits in the ECal.
This condition was removed in 2018, which helped to increase the trigger efficiency as seen in Figures~\ref{fig:trigEffpt} and~\ref{fig:trigEffLxy}.
 
At the HLT, a two-step selection algorithm is applied to CalRatio triggers, regardless of the L1 selection.
In the first step, calorimeter deposits are clustered into jets using the anti-$k_{t}$ algorithm~\cite{Cacciari:2008gp}~\cite{Fastjet} with radius parameter $R=0.4$.
The standard jet-cleaning requirements~\cite{ATLAS-CONF-2015-029} applied offline in most ATLAS analyses reject jets with high values of $\calrat$, one of the key characteristics of the displaced hadronic jets, and are therefore not included in these triggers.
A dedicated cleaning algorithm (referred to as CalRatio jet cleaning and applied starting in 2016) is applied instead. This algorithm is based on the standard jet cleaning described in Ref.~\cite{ATLAS-CONF-2015-029} but with the requirements on the jet $\calrat$ substituted by a condition to reject any jet with more than 85\% of its energy associated with a single calorimeter sampling layer and with an absolute value of its negative energy higher than 4~\GeV, calculated as the sum of the energy in all cells with negative energy, caused by electronic and pile-up noise.
At least one of the HLT jets passing the CalRatio jet cleaning is required to satisfy $E_{\mathrm{T}} > 30$~\GeV, $|\eta| < 2.5$ and $\lograt > 1.2$.
Jets satisfying these requirements are used to determine $0.8 \times 0.8$ regions in $\Delta \eta \times \Delta\phi$ centred on the jet axis in which to perform charged-particle reconstruction (tracking).
Triggering jets are required to have no tracks with $p_{\mathrm{T}} > 2$~\GeV\ within $\Delta R = 0.2$ of the jet axis.
 
In the second step, events passing the CalRatio triggers are required to pass a BIB-removal algorithm. Events which pass the first step but fail this algorithm are collected to form the BIB dataset.
Muons from BIB enter the HCal horizontally and may radiate a photon via bremsstrahlung, generating an energy deposit that may be reconstructed as a signal-like jet.
BIB deposits are more likely to be aligned in $\phi$ since they travel parallel to the LHC beam pipe. Furthermore, muons from BIB leave these energy deposits in the calorimeter earlier than particles resulting from a collision in the same bunch crossing.
The BIB-removal algorithm applied at the HLT therefore identifies events as containing BIB if the triggering jet has at least four HCal-barrel cells close together in $\phi$ and in the same calorimeter layer but not belonging to the same jet and requiring that every cell's timing be consistent with that of a BIB deposit~\cite{DAPR-2012-01}.
This algorithm has been estimated to have an efficiency of approximately $70\%$ in BIB identification.
 
All events in the main dataset are required to pass at least one of the two CalRatio triggers described in this section.
The same HLT selection is also active in empty bunch crossings. These are crossings where protons are absent in both beams and isolated from filled bunches by at least five unfilled bunches on either side. Events in empty bunch crossings that have at least one $0.2 \times 0.2~ (\Delta \eta \times \Delta \phi)$ calorimeter energy deposit with $\et>30\GeV$ at L1, and which pass the two steps of the HLT selection algorithm, are stored in the cosmics dataset.
 
The dijet dataset is collected using the single-jet L1 trigger with the lowest energy threshold in any given data-taking period, and at the HLT, events are required to have at least one jet with $\pt>400$~\GeV or $\pt>420$~\GeV depending on the data-taking period. Events passing any of the CalRatio triggers described above are vetoed from this selection.
 
The CalRatio trigger efficiency for simulated signal LLPs is defined as the fraction of jets spatially matched ($\Delta R < 0.2$) to one of the generated LLPs that fire the trigger.
The trigger efficiency as a function of triggering LLP particle-level \pt is shown in Figure~\ref{fig:trigEffpt} for two signal samples.
Only generated LLPs decaying in the calorimeters are considered in this plot.
The 60-\GeV-high-\et CalRatio trigger, which is seeded by the high-\et L1 trigger with an \ET threshold of 60~\GeV, provides the highest efficiency (about 80\%) to select LLPs with $\pt > 100\GeV$.
The 2018 version of the low-\et CalRatio trigger (seeded by the low-\et L1 trigger) has a similar efficiency for high-\pt LLPs, although it is slightly lower in some \pt ranges.
For LLPs with $\pt<100\GeV$, both versions of the low-\et CalRatio trigger are able to recover part of the inefficiency that is present in the high-\et CalRatio trigger.
This is especially relevant in the low-\et samples, which are mostly populated by low-\pt LLPs.

The trigger efficiency also depends strongly on the LLP decay position, as shown in Figure~\ref{fig:trigEffLxy} for two samples of simulated signal events.
The efficiency as a function of LLP decay length in the $x$--$y$ plane is shown for LLPs decaying in the calorimeter's barrel ($|\eta|<1.4$).
The selection is most efficient in the HCal for both triggers.
The fact that the low-\ET samples contain a large fraction of LLPs with low \pt makes these samples have a lower trigger efficiency.
The total per-event trigger efficiency in signal is obtained by applying a logical OR of the CalRatio triggers, according to the conditions and luminosities listed in Table~\ref{tab:data-datasets-by-year} on a per-event basis, to the complete signal sample integrated over all LLP transverse momenta and for any decay position.
Given that the CalRatio triggers accept events where at least one of the jets fulfils the two-step HLT conditions, this efficiency accounts for events where one or both LLPs decay in the calorimeters.
In the HS signal, the total trigger efficiency rises from $1.6\%$ in the sample with $\mPhi = 125 \gev$ and $\mS = 55 \gev$ to $28.8\%$ for the $\mPhi = 1000 \gev$ and $\mS = 50 \gev$ sample.

For every event, primary vertices (PV) with at least two tracks with $\pt > 500$~\MeV are reconstructed.
If more than one PV candidate is found, the one with the largest sum of the squares of the transverse momenta of all its associated tracks is chosen as the PV.
Clean jets are used at several steps of the event selection.
They are required to have $\pt > 40\GeV$, $|\eta| < 2.5$ and to pass the CalRatio jet cleaning.
To select events with trackless jets, an additional event-level variable, $\sumMinDR{\mathrm{(jet, tracks)}}$, is used.
The quantity $\Delta{R}_\mathrm{min}{\mathrm{(jet, tracks)}}$ is defined as the angular distance between the jet axis and the closest track with $\pt>2\GeV$ that is associated with the PV,
and $\sumMinDR{\mathrm{(jet, tracks)}}$ is calculated by summing this distance over all the clean jets with $\pt>50\GeV$.
Events with no displaced decays have a very small value of this variable.
Every displaced jet contributing to the sum causes a considerable increase in its value, making this variable a good discriminator between signal and multijet background.
 
For an event to pass the analysis preselection, it is required to have passed at least one of the CalRatio triggers, to contain a PV and at least two clean jets, and to have $\sumMinDR{\mathrm{(jet, tracks)}}>0.5$.
After preselection, $\sumMinDR{\mathrm{(jet, tracks)}}$ still has good discrimination power and it is used in the data-driven background estimation described in Section~\ref{sec:background}.

\subsection{Displaced-jet-tagging neural network}
\label{sec:NN}
 
A neural network (NN) is trained on a per-jet basis to discriminate displaced jets in signal from non-displaced jets and from jets generated by BIB.
The architecture chosen for the network is a set of convolutional layers fed into a long short-term memory (LSTM) layer using Keras~\cite{keras} (version 2.2.4) with TensorFlow~\cite{tensorflow} (version 1.12.0) as the backend.
The displaced-jet-tagging NN is trained on SM multijet MC samples, the BIB dataset and a combination of the simulated HS signal samples.
Each signal sample is split into two orthogonal subsets, one used for training and the other for application in the rest of the analysis chain.
In the signal samples, only jets matched to MC generated LLPs decaying after passing through the ATLAS inner detector are used in the training.
In the SM multijet samples, the two highest-\pt clean jets in the event are used in the training.
In the BIB dataset, only the triggering jet is used. About 1~million jets of each class are obtained, with an 80:10:10 training, testing, validation split.
Signal jets with small boosts are in general harder to discriminate from BIB and SM multijets than those with large boosts.
In order to obtain the optimal training for all boost regimes, two separate NNs are trained.
The low-\et NN is trained using the HS signal samples with $\mPhi=60, 125, 200$~\GeV, which typically contain low-\pt signal jets with small boosts. The high-\et NN is trained on the HS samples with $\mPhi=400, 600, 1000$~\GeV, dominated by high-\pt jets.
 
The input variables are low-level features associated with each jet, using information from the entirety of the ATLAS detector: positions, momenta, impact parameter and quality variables of inner-detector tracks which are within $\Delta R = 0.2$ of jets; momenta, timing information and positions of calorimeter topoclusters (collections of calorimeter cells used in jet reconstruction~\cite{PERF-2014-07}) associated with each jet, as well as the fraction of the jet energy deposited in each layer of the ECal and HCal; and spatial and timing information for muon track segments within $\Delta \phi = 0.2$ of a jet. In addition, the momenta and positions of the jets themselves are also used. Overall, the input to the neural network consists of 10 track variables for up to 20 tracks per jet, 12 topocluster variables for up to 30 topoclusters per jet, 6 muon-segment variables for up to 30 muon segments per jet, and 3 jet variables.
To check the modelling of variables prior to training, simulated distributions of the per-jet NN inputs are compared with data and fair agreement is generally observed, although some variables display some mismodelling; this is addressed during training by using the adversarial technique described below.

The displaced-jet-tagger uses a combination of two specialised network architectures.
The first one is a one-dimensional (1D) convolutional neural network (CNN) which acts as a feature extractor.
A convolutional layer performs a series of convolution operations on a 1D array of inputs, corresponding to the variables per subdetector listed above, separately for the tracker, calorimeter and muon-segment variables. A kernel size of 1 is used for the 1D convolutions, with a series of filters being applied which up-sample the features per object in the first filter, then down-sample as the filter goes down in size, working as a feature extractor.
In this way the network can exploit the correlations between the input variables.
The second step is the LSTM~\cite{lstm} network, where the memory uses relations between subsequent inputs (for example, between subsequent topoclusters) which allows the network to exploit the correlations between the different topoclusters, tracks and muon segments separately. The topocluster and track inputs to the LSTM are ordered in descending \pT, which was shown to enhance the performance of the LSTM. No ordering of the muon segments is applied, but up to 30 muon segments per jet are included which is enough to contain all muon segments for the vast majority of input jets.

The outputs of the individual networks learning from the different subdetectors are concatenated with the jet variables as input to one network which has a number of fully connected layers, named `dense layers'. The output of this final, concatenated network is three predictions: signal NN score, SM multijet NN score and BIB NN score. Each prediction is a score between 0 and 1 and it is closer to 1 the more confident the NN is that a jet belongs to this particular class; the three scores add up to 1.
Parameters such as learning rate, regularisation and learning nodes per layer are optimised by scanning a multidimensional grid.

\begin{figure}[ht]
\centering
\includegraphics[width=\textwidth]{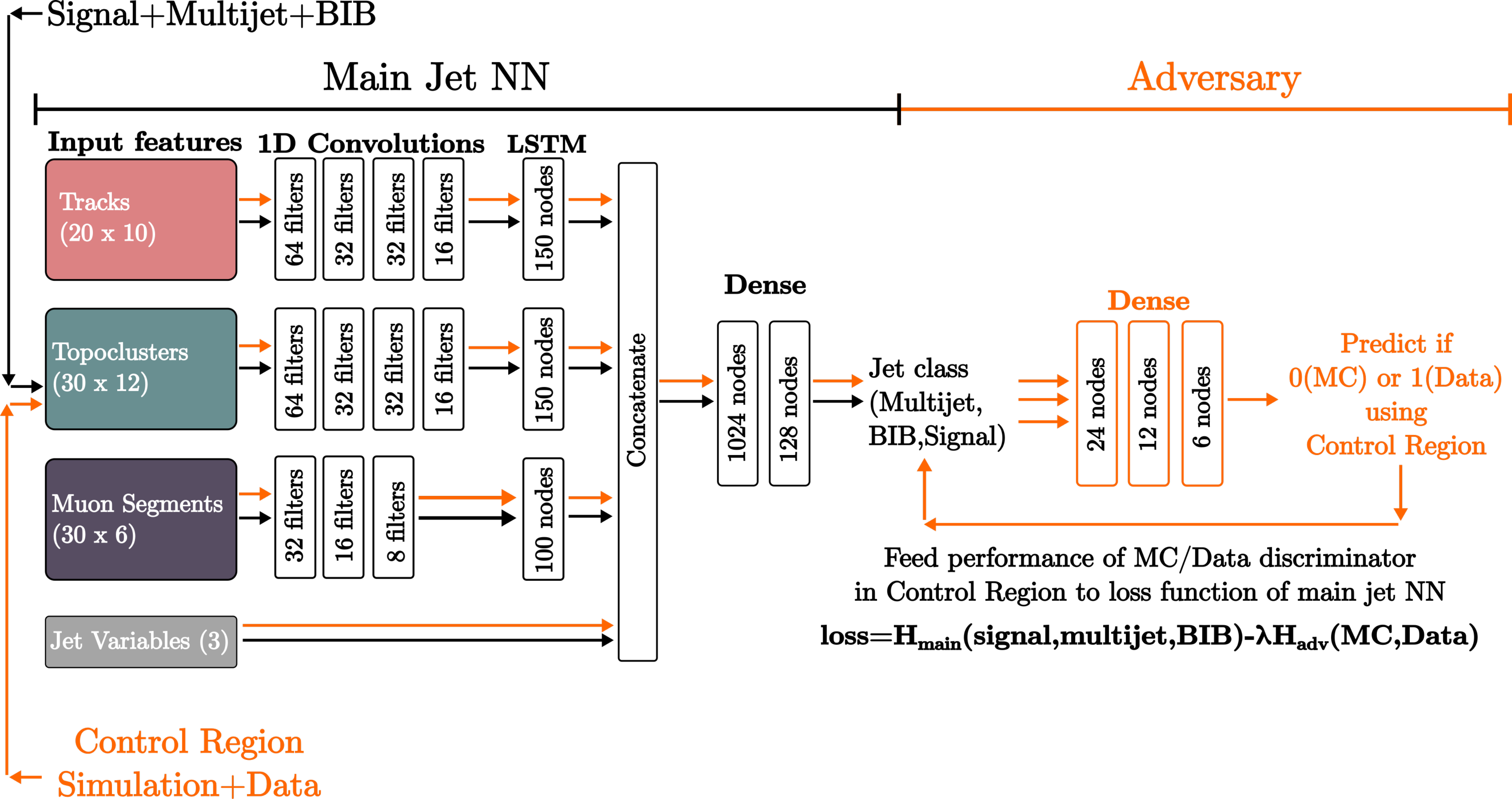}
\caption{A simplified diagram of how the adversarial network is structured. The principal network which discriminates between signal, SM multijets and BIB is shown with the inputs moving through the network as black arrows. Dense layers refer to fully-connected layers, while LSTM refers to Long-Short Term Memory layers described in the text. The adversary which classifies jets as being MC simulation or data is shown with the inputs moving through the network as orange arrows. The input variables are labelled $n\times m$ where $n$ is the maximum of the number of tracks, topoclusters or muon segments per jet and $m$ is the number of distinct variables per input object.}
\label{fig:adv_diagram_main}
\end{figure}

Finally, given that the NN is trained on simulation samples as well as on data samples, an adversarial network is added to the output of the jet-tagging NN with the objective of mitigating the dependence of the displaced-jet-tagger outputs on potential mismodelling of any of the input variables in simulation. The adversarial network concept is based on Ref.~\cite{adversary_ref} but built specifically for the input variable mismodelling circumstances of this search.
An illustration of the architecture is shown in Figure~\ref{fig:adv_diagram_main}.
The adversarial network is trained on a selection of SM multijet MC and collision data events in a dijet control region separate from the signal selection used to train the principal NN.
This dijet control region is constructed from events in the dijet dataset which pass the following requirements:
the leading and subleading jets must pass standard quality selections and have $\pT > 400$~\GeV and $\pT > 60$~\GeV respectively;
they must have high angular separation ($|\Delta\phi| > 1.6$);
and the magnitude of the vectorial sum of transverse momenta of these jets ($\mht$) is required to be below 120~\GeV.
 
\begin{figure}[ht]
\centering
\subfloat[]{\includegraphics[width=0.48\textwidth]{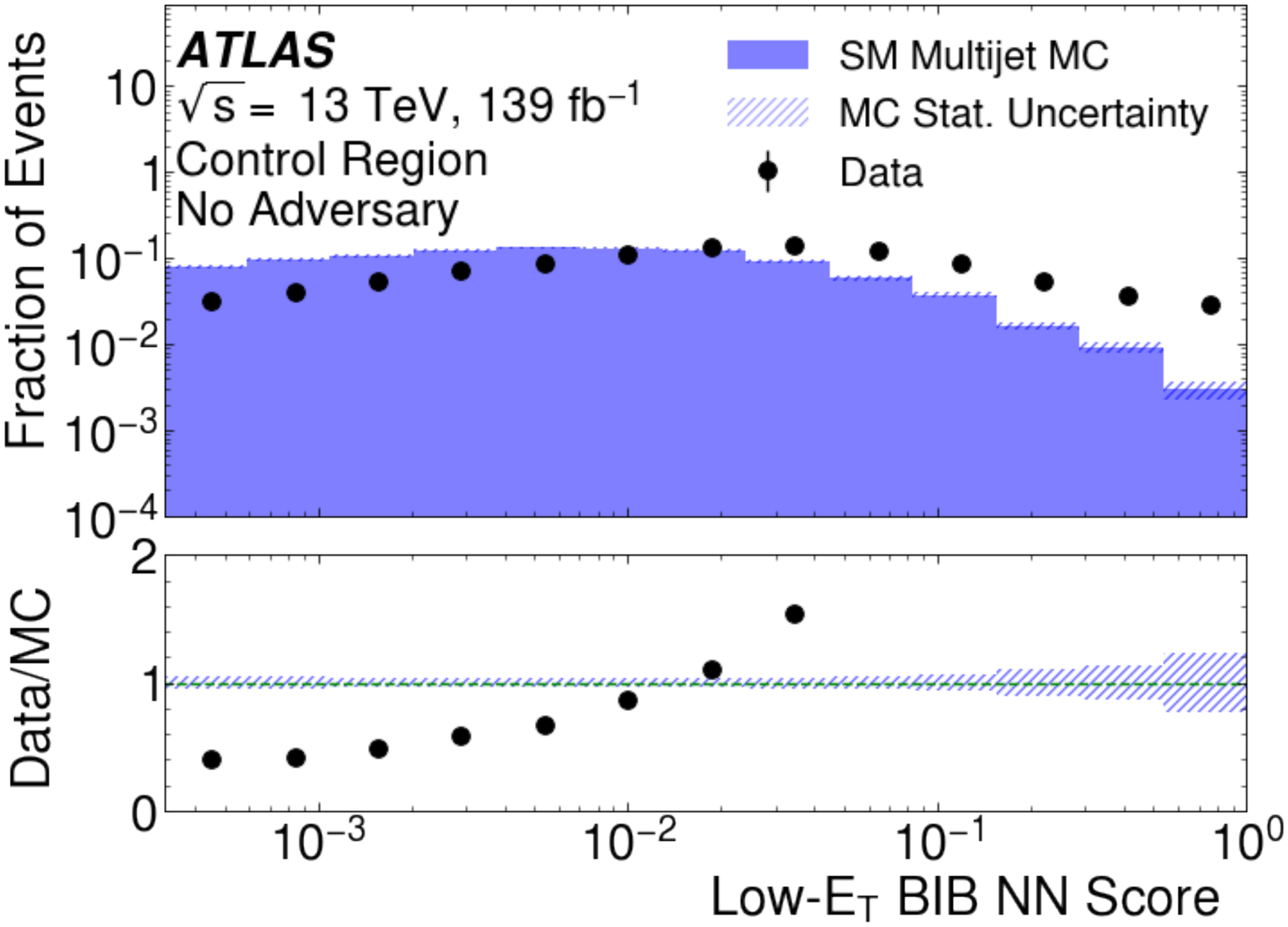}}
\subfloat[]{\includegraphics[width=0.48\textwidth]{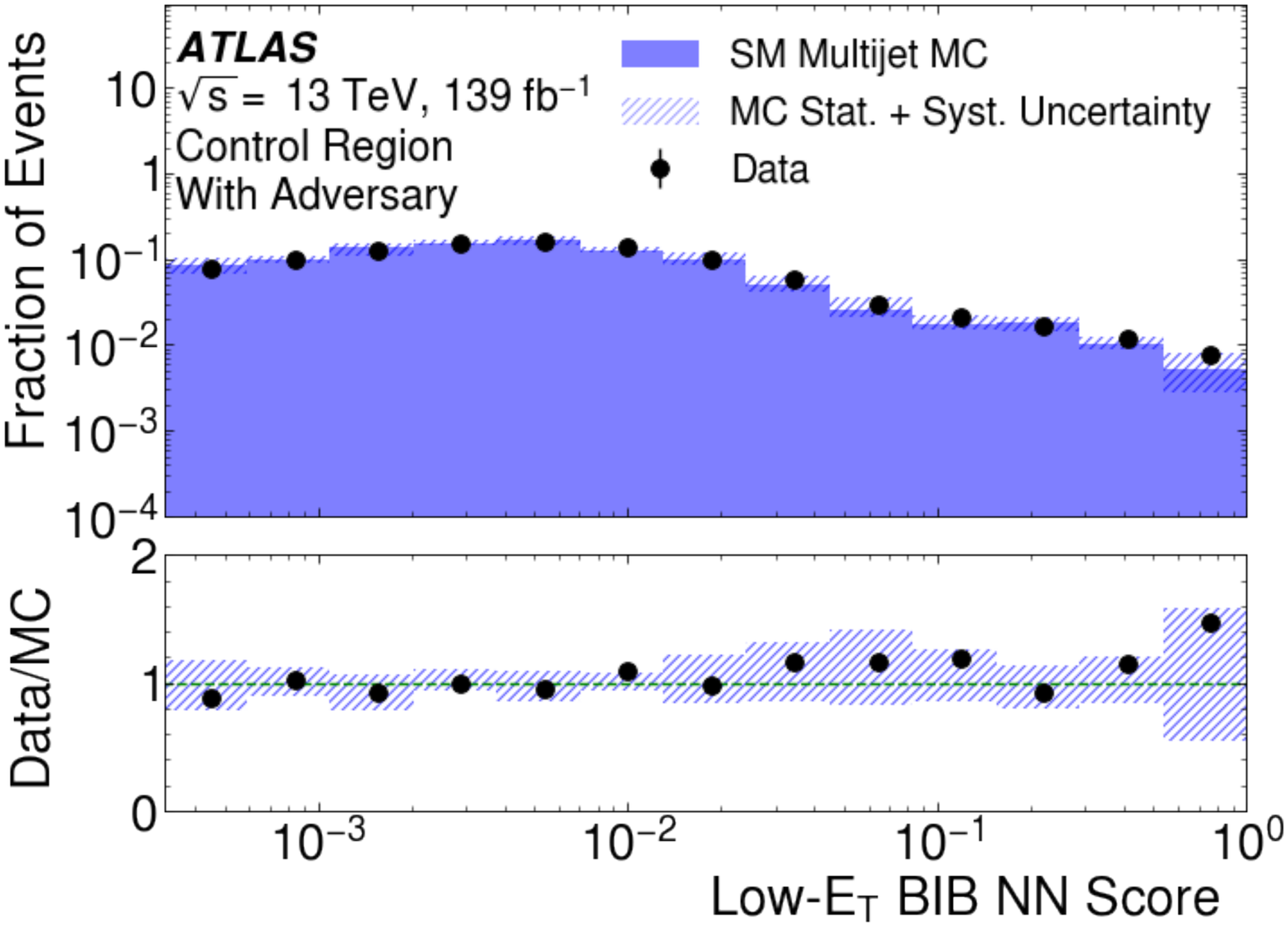}}\\
\subfloat[]{\includegraphics[width=0.48\textwidth]{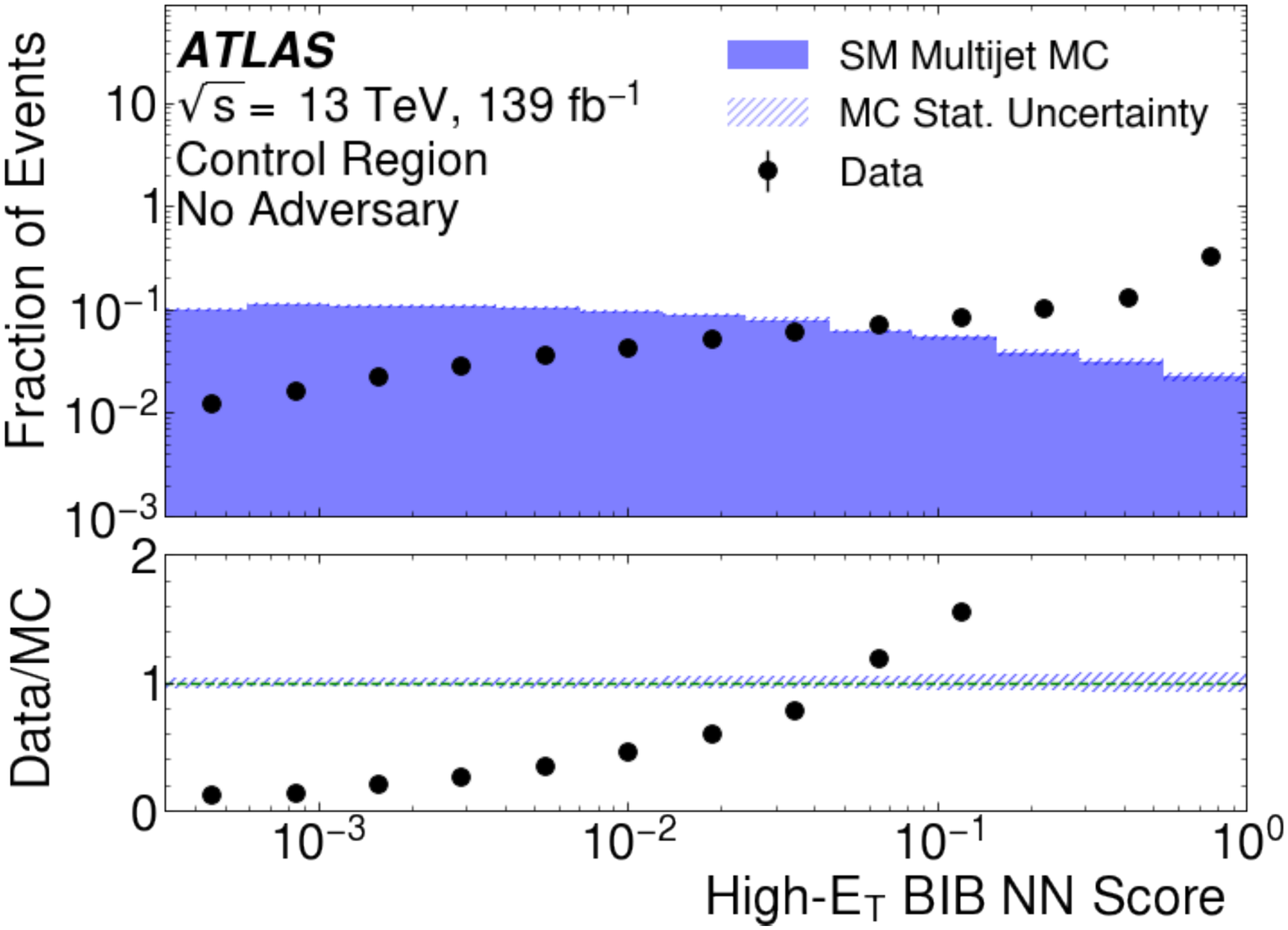}}
\subfloat[]{\includegraphics[width=0.48\textwidth]{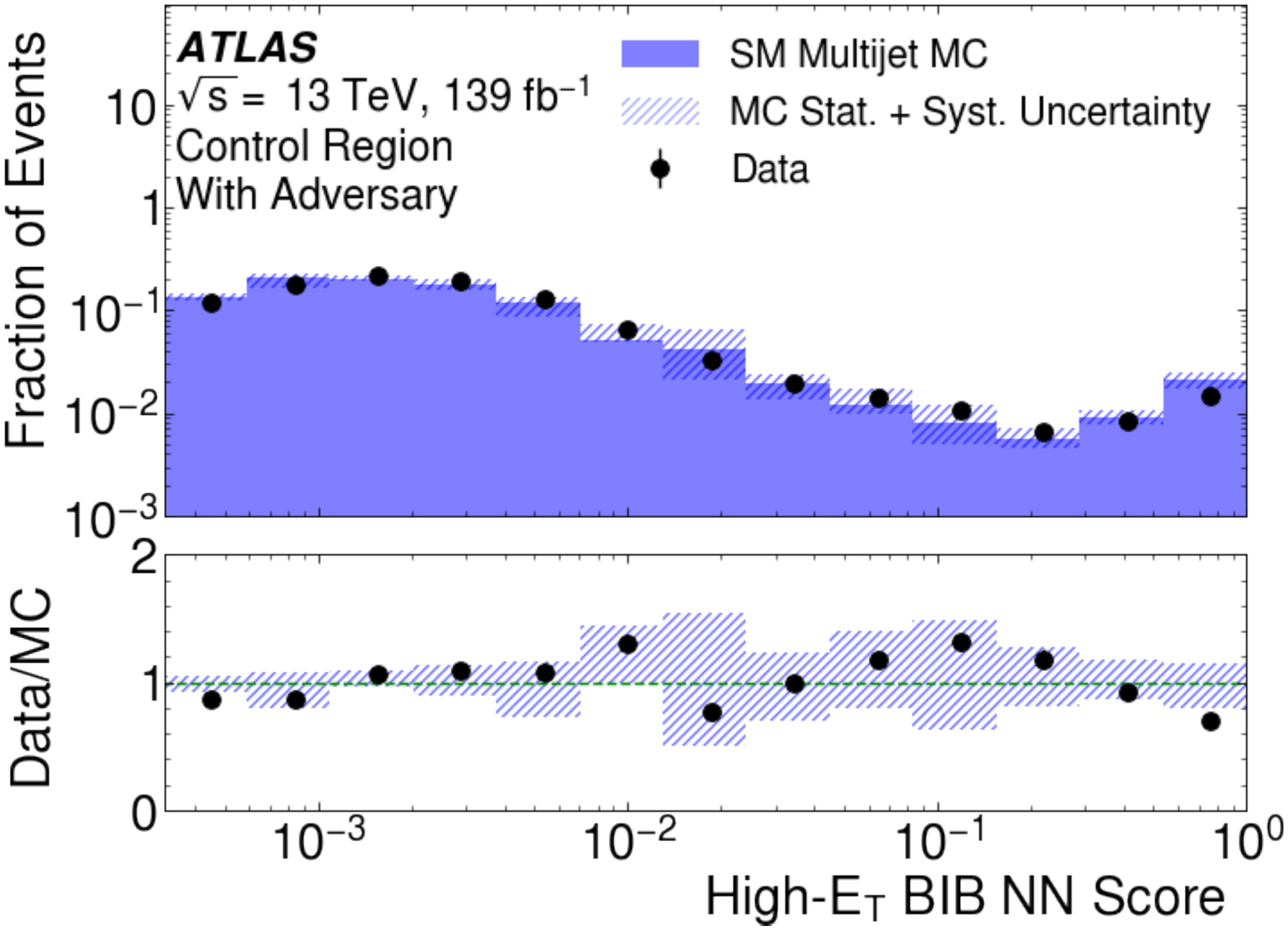}}
\caption{The NN output scores in the dijet control region for (a) the low-\ET training with no adversary network and (b) with an adversary network included, (c) for the high-\ET training with no adversary network and (d) with an adversary network included. Statistical uncertainties are shown in all plots. In cases where training with adversary networks is considered, the systematic uncertainty related to modelling discrepancies is included as well.}
\label{fig:NNcontrol_main_bib}
\end{figure}

The NN learns the differences between simulation and real data and feeds this information back to the jet-tagging neural network such that the latter network avoids taking advantage of the MC mismodelling. The whole network tries to minimise the loss function. This is achieved by simultaneously minimising the principal network's loss function (learning to discriminate Signal from BIB and SM multijets) and maximising the adversary's loss function (learning not to discriminate between simulation and data). During training the cross-entropy function $H(\vec{x})$ is used, combining the cross-entropy of the main network where $\vec{x}$ is the signal, multijet and BIB jets and the cross-entropy of the adversarial network where $\vec{x}$ is the input variables of the control region MC simulation and data jets. The adversary part is modulated by a scale factor $\lambda$, with a value of $\lambda=10$ chosen to minimize the impact of mismodelling in the final NN output while maintaining good separation. The adversarial cross-entropy term is subtracted from the main cross-entropy term to give the total loss function.
The adversary greatly improves the level of agreement between data and SM multijet simulation in the jet-tagging NN outputs, reducing the impact of any potential mismodeling of the NN input variables along the analysis chain.
Figure~\ref{fig:NNcontrol_main_bib} shows the BIB NN score in the dijet control region in data and in simulation, before and after including the adversarial network in the training.
The remaining small discrepancies in the modelling of the NN output variables are covered by a dedicated uncertainty, which is described in Section~\ref{sec:systematics} and applied to the signal efficiency only, as the background estimate in this analysis is data-driven.

Over-fitting during training is monitored by checking the loss in the training and testing sets periodically. No over-training is seen, which would be characterised as a large deviation between training and testing loss.

\begin{figure}[h!]
\centering
\subfloat[]{\includegraphics[width=0.48\linewidth]{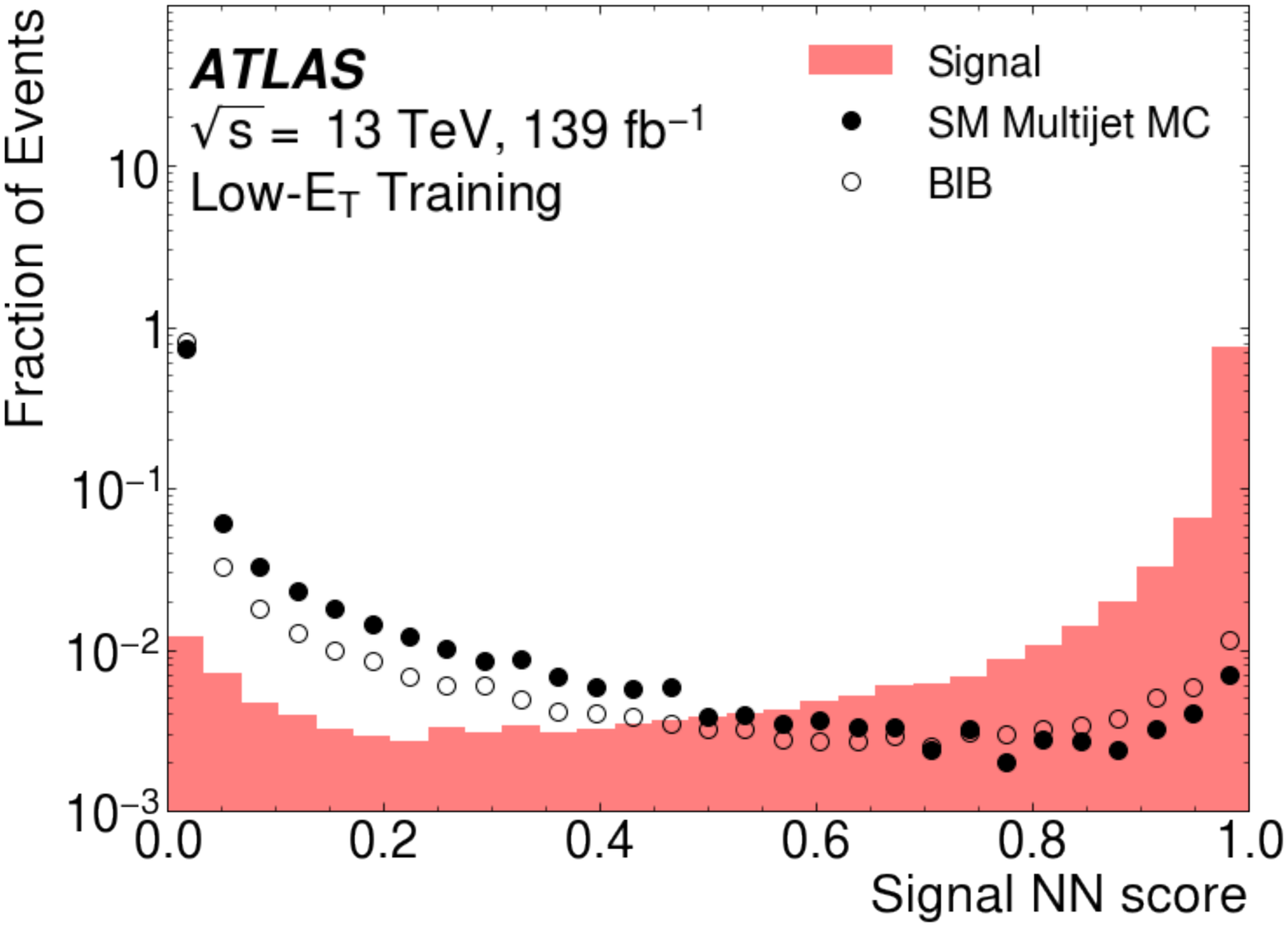}}
\subfloat[]{\includegraphics[width=0.48\linewidth]{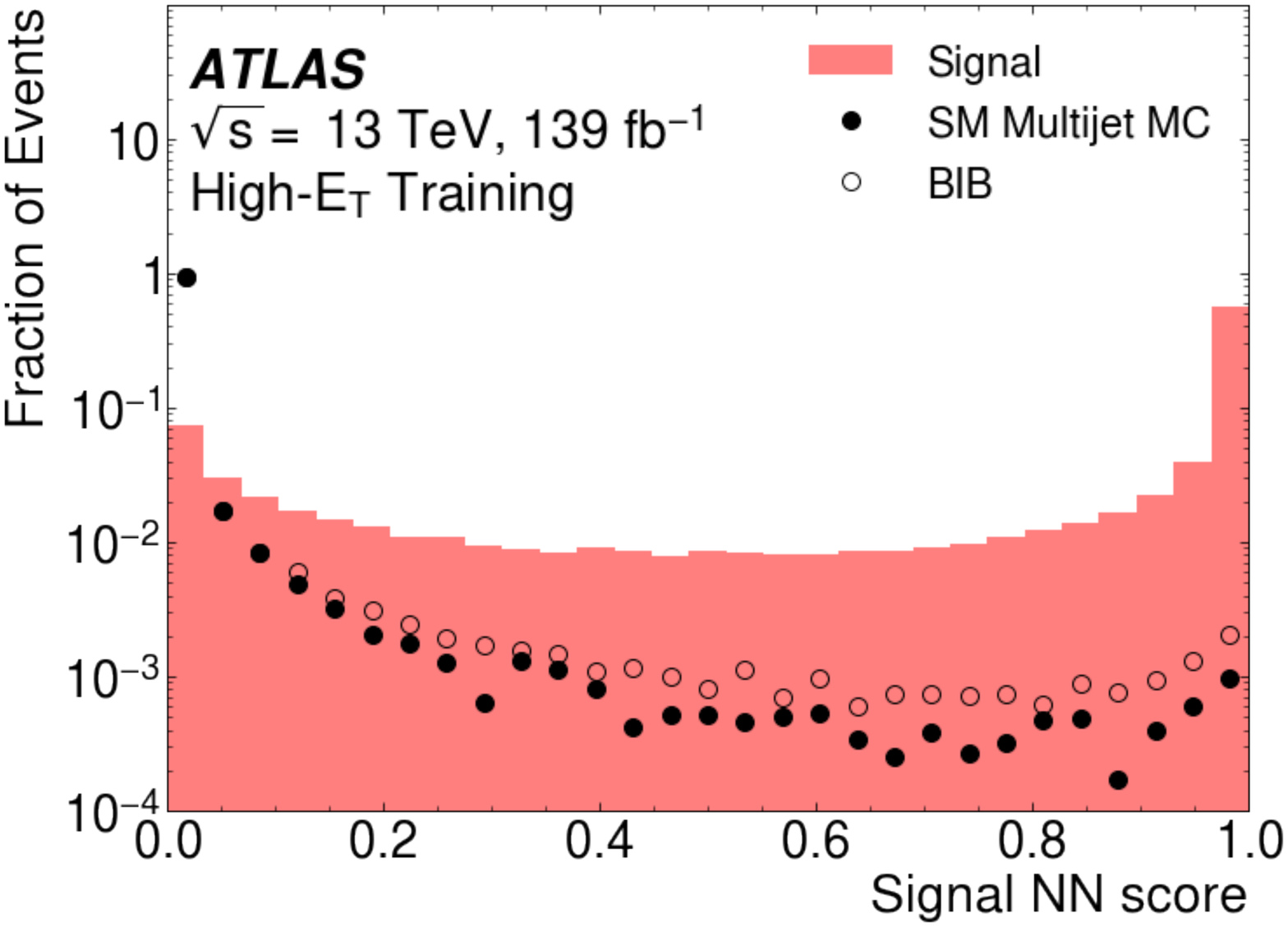}}\\
\subfloat[]{\includegraphics[width=0.48\linewidth]{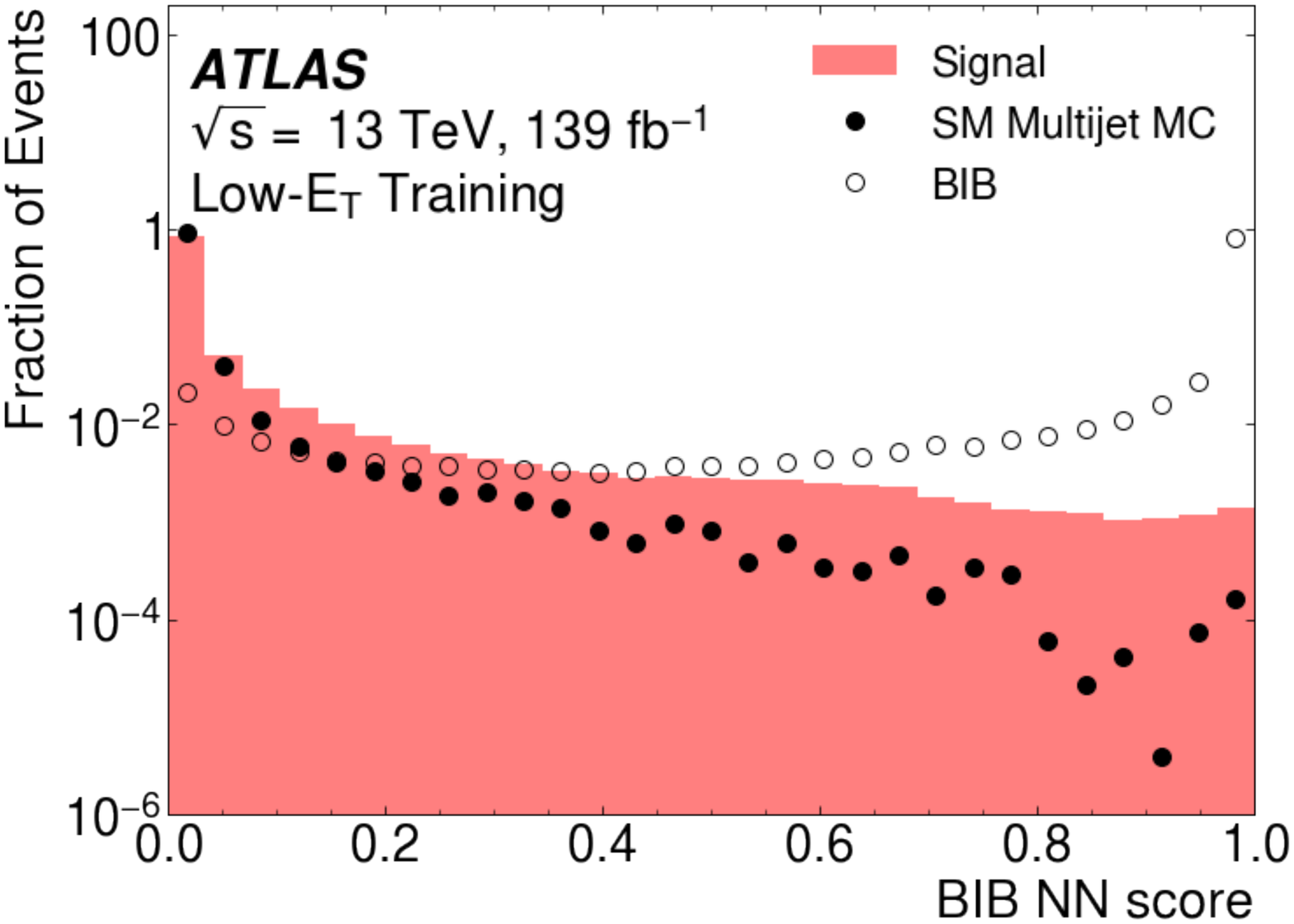}}
\subfloat[]{\includegraphics[width=0.48\linewidth]{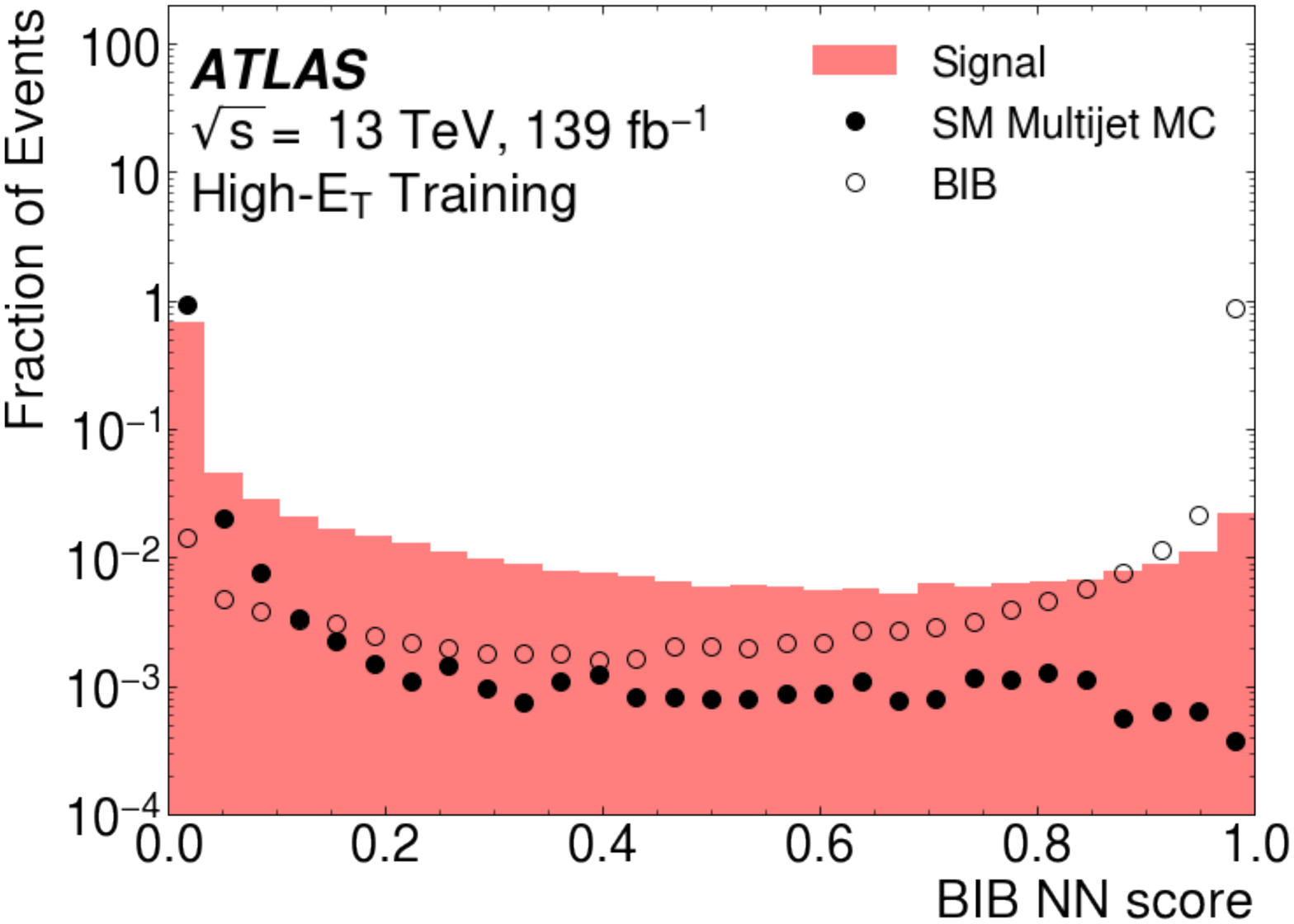}}\\
\subfloat[]{\includegraphics[width=0.48\linewidth]{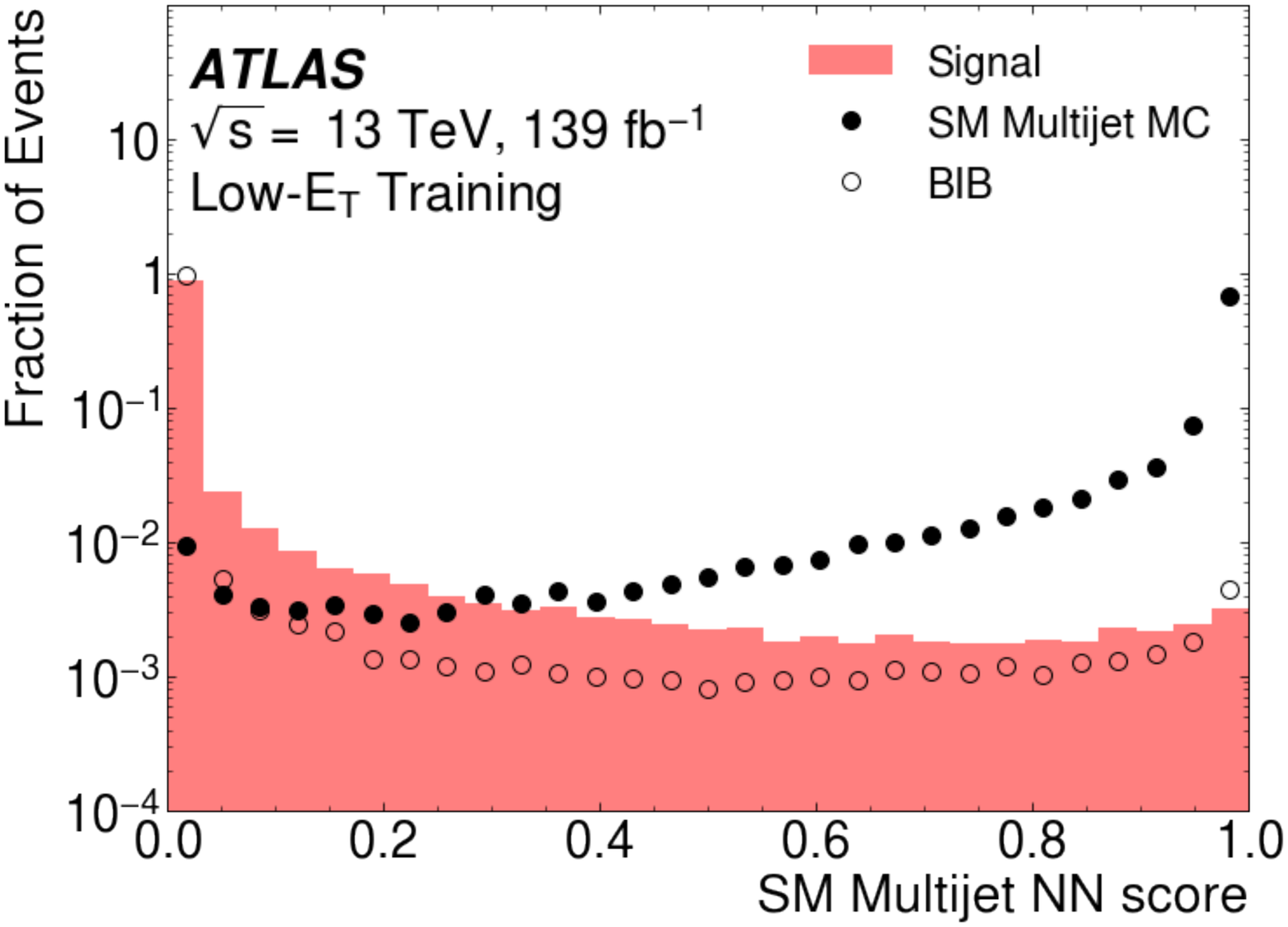}}
\subfloat[]{\includegraphics[width=0.48\linewidth]{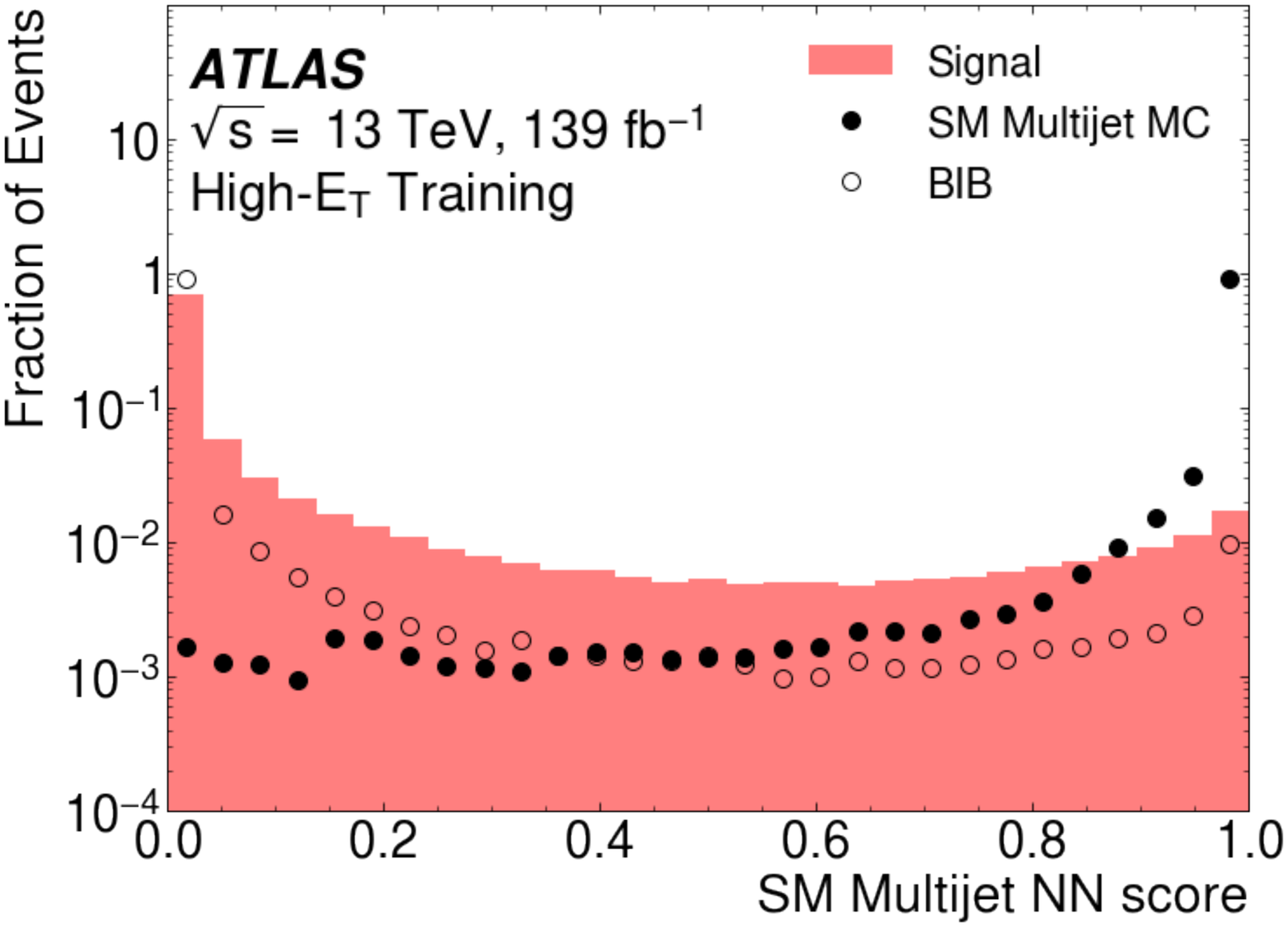}}
\caption{The NN output scores for (a,b) signal, (c,d) BIB and (e,f) SM multijets shown for low-\ET and high-\ET training. The jets used here are part of the validation set, and follow the selection used when choosing jets to train the NN in signal, multijets and BIB.}
\label{fig:nn_scores}
\end{figure}

The performance of the final trained algorithm can be seen in Figure~\ref{fig:nn_scores}. 
The training took between 50 and 100 run-throughs of the input data to converge, depending on the training.
In the event selection, both NNs (low- and high-\ET) are evaluated in every clean jet of every signal sample. This provides a better selection than using only the low-\ET (high-\ET) NN in the low-mass (high-mass) signal samples because of the residual presence of high-\pt jets in the low-mass samples and low-\pt jets in the high-mass samples.

\FloatBarrier
\subsection{Event selection}
 
A per-event boosted decision tree (BDT) is designed with the foremost objective of discriminating BIB events from signal events where at least two jets have a considerable displacement.
As explained at the beginning of this section, the CalRatio triggers contain an algorithm to reject BIB. However, this algorithm is not $100\%$ effective, and hence some BIB contamination is expected in the main dataset.
The BIB data sample contains SM multijet events on top of the BIB event that caused them to be selected by the trigger.
Consequently, even if no multijet sample is explicitly used in the training, the per-event BDT is able to discriminate signal from BIB as well as from multijet background.
A combination of signal samples in the HS model is used to train the per-event BDT against BIB data. As in the NN training, only part of each signal sample is used for training, leaving the other part for interpretation.

Four CalRatio jet candidates are defined as the two clean jets with the highest low-\ET signal-NN-score and the two clean jets with the highest high-\ET signal-NN-score in the event (\FCRljet, \SCRljet, \FCRhjet, \SCRhjet).
Analogously, four BIB jet candidates are defined as the two clean jets with the highest low-\ET BIB-NN-score and the two clean jets with the highest high-\ET BIB-NN-score in the event (\FBIBljet, \SBIBljet, \FBIBhjet, \SBIBhjet).
The NN outputs of these jets are used as input variables for the per-event BDTs.
Other event-level variables are used in the training: examples include the distance $\Delta R$ between the two high-\ET CalRatio jet candidates (\FCRhjet, \SCRhjet) and \mhtoht , where $H_{\mathrm T}$ is the scalar sum of the transverse momenta of the jets.
 
\begin{figure}[ht]
\centering
\subfloat[]{\includegraphics[width=0.497\linewidth]{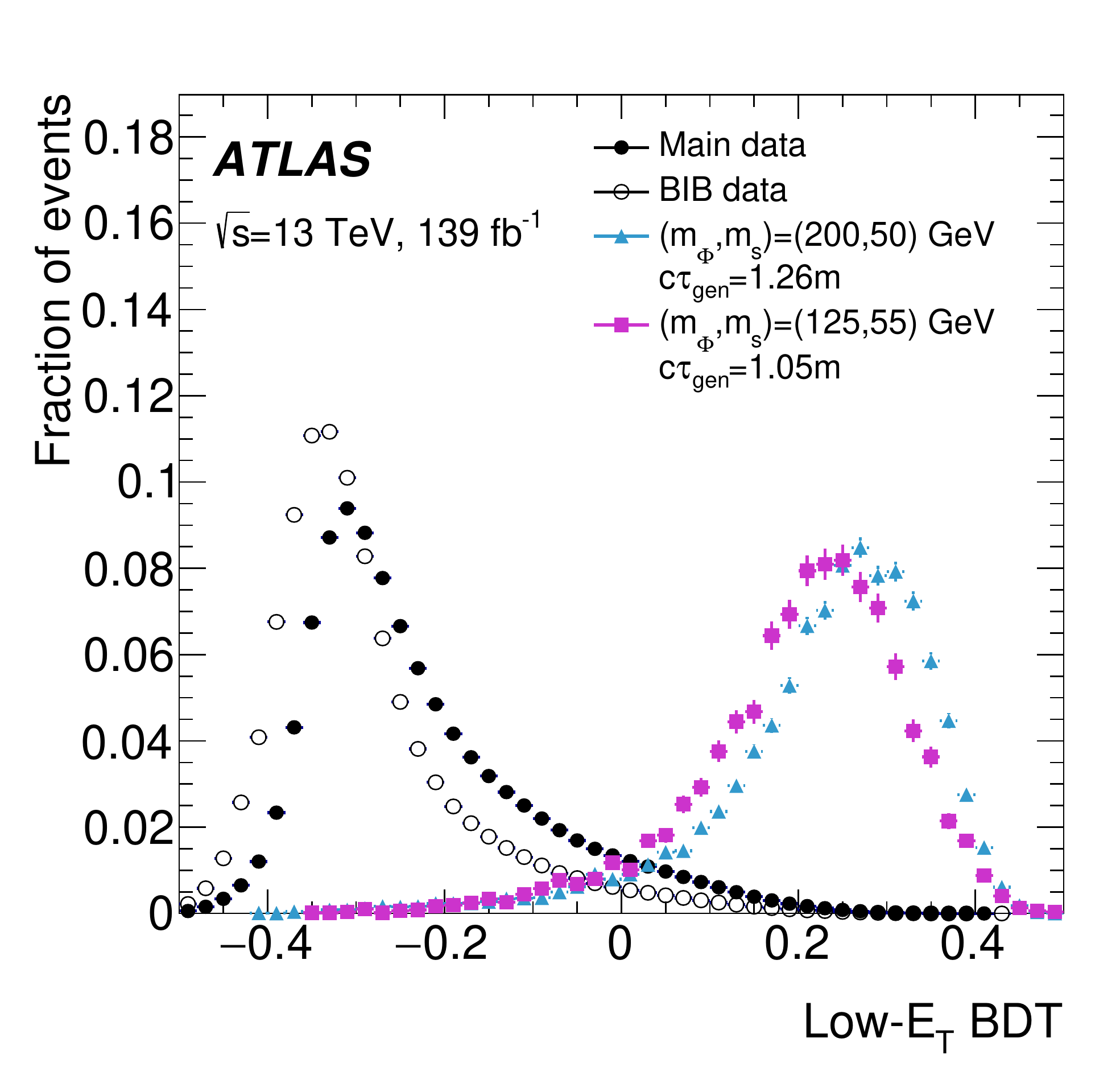}}
\subfloat[]{\includegraphics[width=0.497\linewidth]{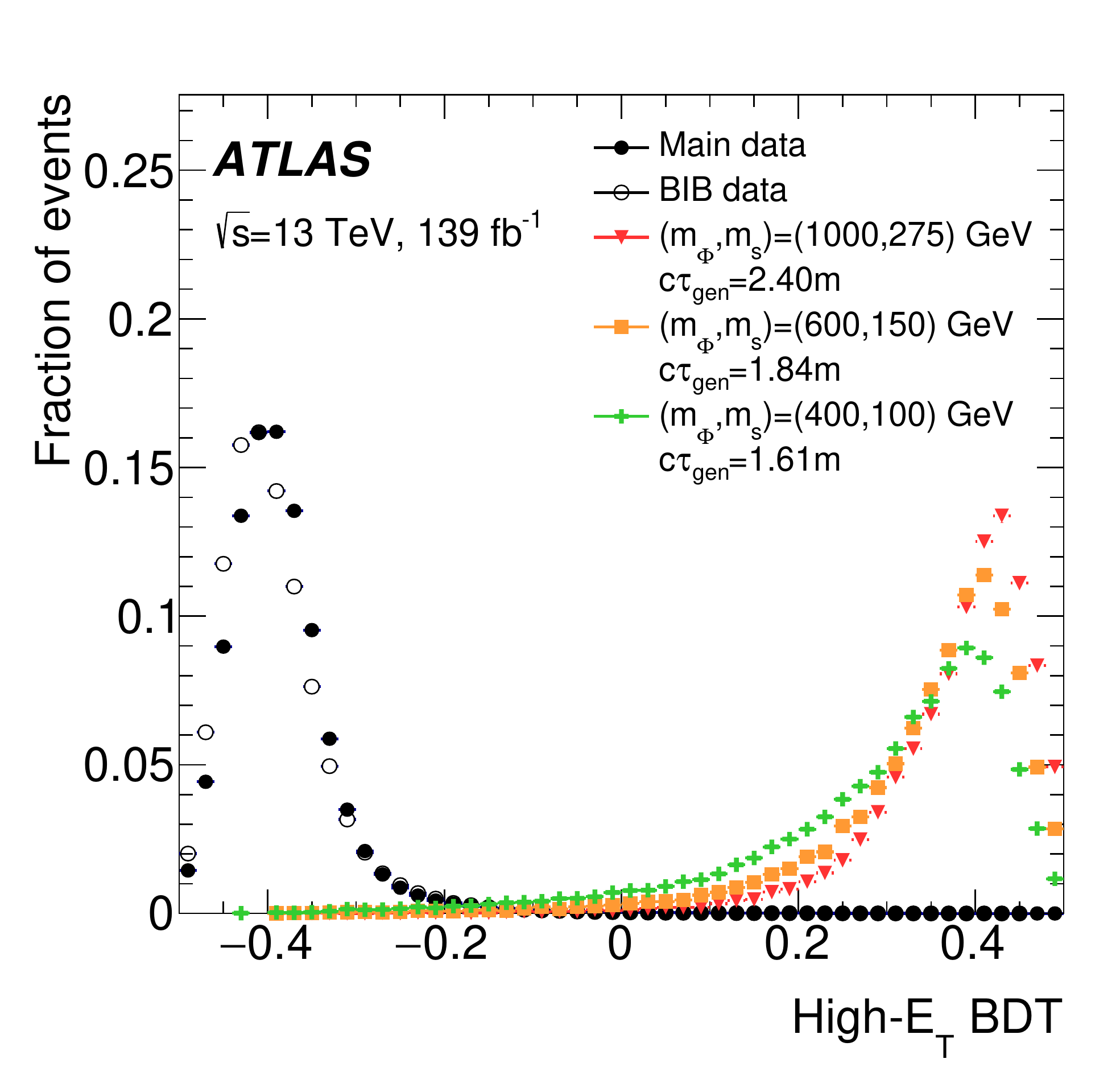}}
\caption{Distribution of the (a) low-\et per-event BDT and (b) high-\et per-event BDT outputs in main data, BIB data and some of the benchmark signal samples after preselection.
Only statistical uncertainties are shown.
}
\label{fig:eventBDT}
\end{figure}

Following the same strategy as used for the per-jet NN and to obtain optimal signal-to-background discrimination at any jet \pt, two versions of the per-event BDT are trained: the high-\et BDT is trained for the analysis of the high-mass ($\mPhi = 400$ to 1000~\GeV) signal samples, with a high jet-\pt distribution, and the low-\et BDT is trained for low-mass ($\mPhi = 125$ to 200~\GeV) signal samples, with a softer jet-\pt distribution.
The two versions use the same set of input variables and the same BIB background sample. They differ only in the signal samples used during training.
Only signal events where at least one of the LLPs decays after passing through the inner detector are considered in the training.
Figure~\ref{fig:eventBDT} shows the distribution of the per-event BDT outputs from five signal samples, as well as from the main data and BIB data.
The BDT output in the BIB dataset peaks at slightly lower values than in the main data.
Using the jet's time, measured as the energy-weighted average of the timing for each calorimeter cell related to the jet and corrected by the corresponding time-of-flight from the interaction point, and $z$-coordinate measurements of the jets, it was checked that events with low BDT values (\evbdtLITopo$<-0.2$ or \evbdtLITau$<-0.35$) have the typical characteristics of BIB, while events with intermediate values ($-0.2<$~\evbdtLITopo$<0.05$ or $-0.35<$~\evbdtLITau$<0.0$) are most like multijet events.
 
Taking this separation between BIB and multijets into account, the per-event BDT serves two purposes in the analysis:
first, it is used as part of the event cleaning described below in this section to reject BIB; second, it is used in the data-driven background estimation in the search region (see Section~\ref{sec:background}).
 
The simulated distributions of the per-jet NN and other inputs to the per-event BDTs show mostly good agreement with data.
A dedicated procedure to evaluate the impact of residual mismodelling of NN and BDT input variables on the signal efficiency is described in Section~\ref{sec:systematics}.

Two selections are defined, referred to as the high-\et selection and the low-\et selection, which are optimised to give maximum sensitivity for high-\ET models and low-\ET models, respectively.
 
Event-cleaning selections are applied to remove as much BIB and detector-related background as possible: the per-event BDT output must satisfy \evbdtLITopo$>0.05$ and \evbdtLITau$>0.0$ in the low-\et and high-\et selections, respectively; trigger matching, where at least one of the CalRatio jet candidates must be matched to the jet that fired the trigger and fulfil the requirements applied at trigger level; a timing window of $-3 < t < 15$~\ns, where $t$ is the jet time for any of the CalRatio jet candidates or BIB jet candidates (this selection helps to remove remaining BIB jets and jet candidates produced by detector noise while retaining signal jets originating from relatively slowly moving LLPs with timing up to 15~ns);
a veto on events where one of the CalRatio jet candidates falls in the transition region between the ECal barrel and endcaps ($1.45 < |\eta| < 1.55$), where poor coverage by the electromagnetic calorimeter produces jets with artificially low fractions of energy in the ECal; and a veto on events where one of the CalRatio jet candidates or BIB jet candidates has $\lograt < -1.5$.
These requirements ensure that in the final selection, the only remaining source of background is multijet events, as described in Section~\ref{sec:background}.
 
The final selections are optimised to maximise the signal-to-background ratio in each search region.
Variables with good signal-to-background discrimination at event level are used, such as \mhtoht and the product of the NN signal scores of the two relevant CalRatio jet candidates.
The quantity \mhtoht has high values for BIB events but it has a softer distribution for signal. Hence it is a good discriminator, especially in the high-\pt regime.
Making a selection based on the product of the NN signal scores of the two relevant CalRatio jet candidates helps to further reduce the background while keeping most signal events where one of the LLPs decays in the ATLAS calorimeters and the other one decays between the IP and the outer edge of the calorimeters. The low-\ET NN product is defined as the product of the two highest low-\ET NN signal-scores for clean jets in a given event. The high-\ET NN product is defined analogously.
The requirements shown in Table~\ref{tab:sel_cuts} are applied for the low-\et and high-\et selections.
 
\begin{table}[h]
\centering
\begin{tabular}{|c|c|}
\hline
Low-\ET selection    & High-\ET selection  \\ \hline
$\mhtoht <0.6$ & $\mhtoht <0.6$ \\
$(\sumLogRl)>2$ & $(\sumLogRh)>1$ \\
$\FCRljetpt>80\GeV$ & $\FCRhjetpt>70\GeV$ \\
$\SCRljetpt>80\GeV$ & $\SCRhjetpt>80\GeV$ \\
low-\et NN product $>0.7$ & high-\et NN product $>0.5$\\
\hline
\end{tabular}
\vspace{0.25cm}
\caption{Final selection requirements in the low-\ET and high-\ET selections.}
\label{tab:sel_cuts}
\end{table}
 
The signal efficiency for signal events to pass the low- and high-\ET selections depends on the momenta, decay positions and decay products of the two LLPs in the event.
For the benchmark HS models which are analysed using the low-\ET selection, the efficiencies range between 0.5\% and 0.005\%.
Benchmark models analysed using the high-\ET selection have signal efficiencies varying between nearly 9.3\% and 1.3\%.
The efficiencies typically increase with the mean LLP \pT in the sample.
Further, the efficiencies are highest for samples where most decays take place in the calorimeter. Finally, the efficiency depends on what particles the LLPs decay to.
LLP decays to bottom quarks are typically dominant in the samples considered by this analysis. Hence, the efficiency for events where the LLPs decay to $b$-quarks lead to similar efficiencies to the ranges mentioned above. Events where LLPs decay to $c$-quarks also have a similar efficiency. LLP decays to pairs of taus typically have an efficiency around half of the nominal efficiency. Finally samples where the LLPs decay to top quarks lead to a reduction in efficiency by about an order of magnitude.
The signal efficiencies parameterized as a function of kinematic quantities and the decay mode are available in HEPData~\cite{hepdata.115578}.

\section{Background estimation}
\label{sec:background}
 
A data-driven ABCD method is used to estimate the contribution from the dominant background (SM multijet events) to the final selection.
The ABCD method relies on the assumption that the distribution of background events can be factorised in the plane of two relatively statistically independent variables.
In this plane, the method uses three control regions (B, C and D) to estimate the contribution of background events in the search region (A).
In the case of no signal contamination in regions B, C and D, the number of background events in region A can be predicted using $N_\mathrm{A}=(N_\mathrm{B}\cdot N_\mathrm{C})/N_\mathrm{D}$, where $N_{X}$ is the number of background events in region $X$.
In reality, there is non-zero signal contamination in the control regions. This is accounted for by using a modified ABCD method, which involves fitting to background and signal models simultaneously.
The background component of the yield in each of the regions A, B, C and D is constrained to obey the standard ABCD relation, within the bounds of the ABCD method uncertainty (described below), while a signal strength parameter uniformly scales the signal yield in each region.

\begin{figure}[ht]
\centering
\subfloat[]{\includegraphics[width=0.33\linewidth]{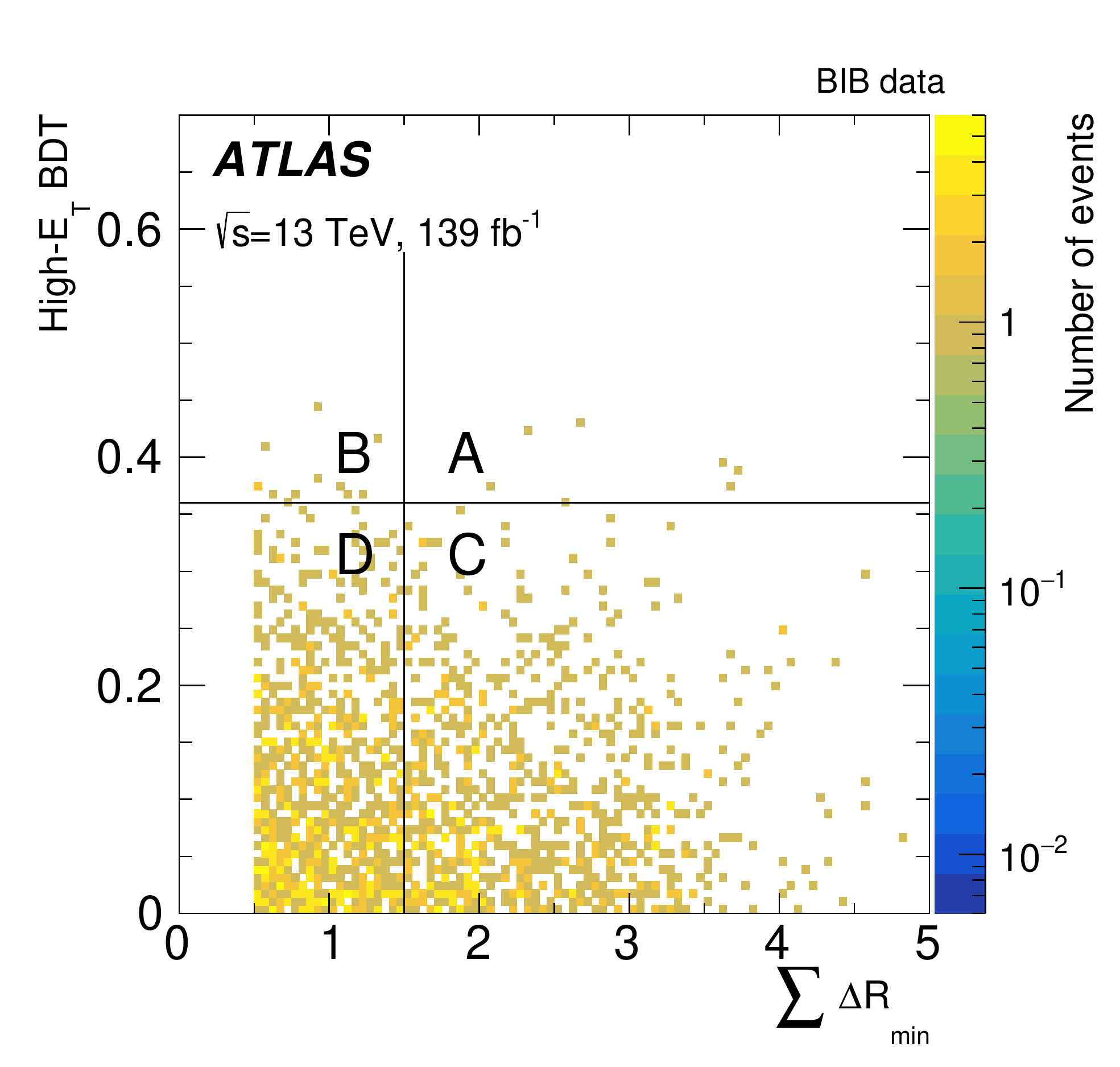}}
\subfloat[]{\includegraphics[width=0.33\linewidth]{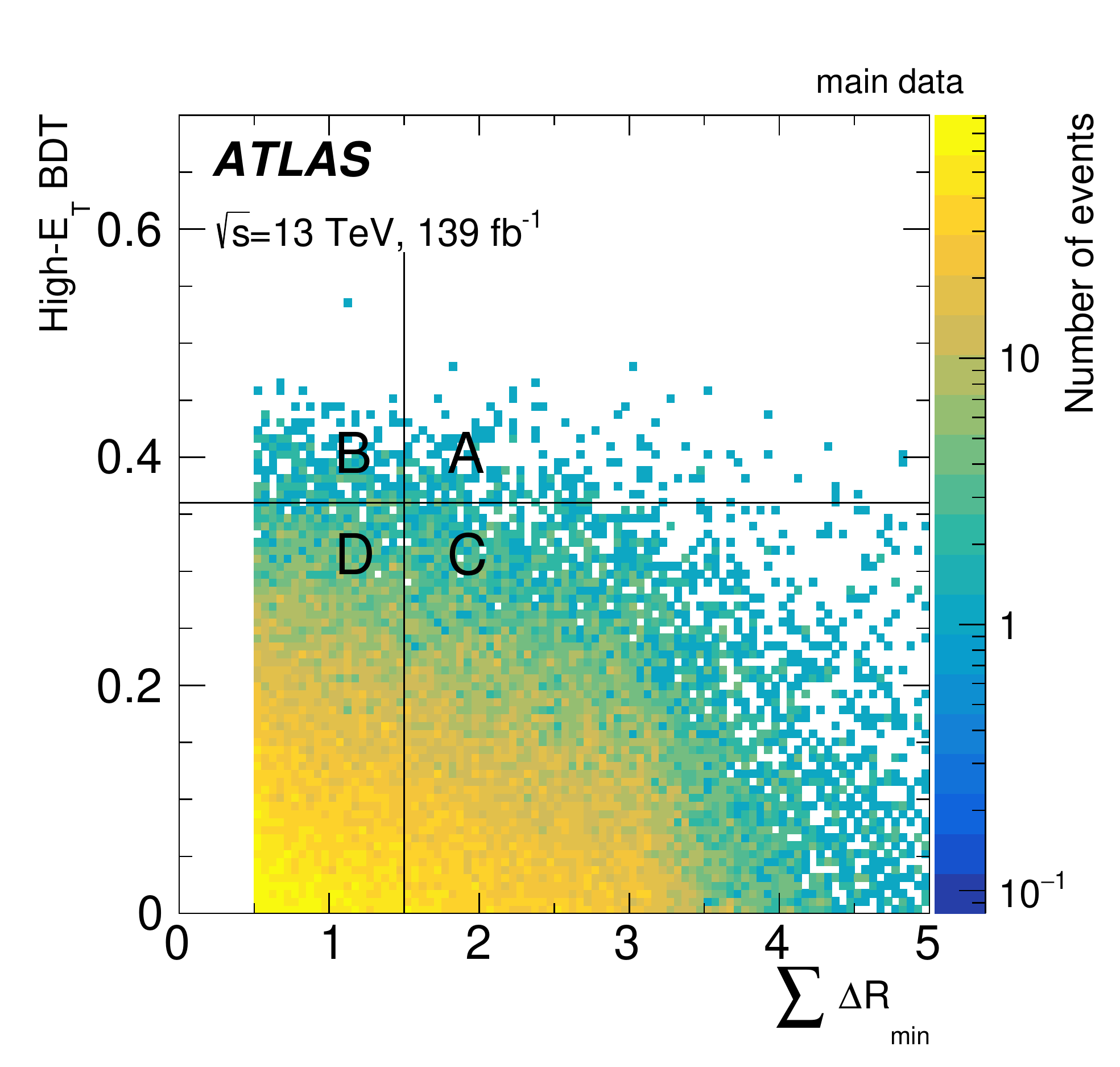}}
\subfloat[]{\includegraphics[width=0.33\linewidth]{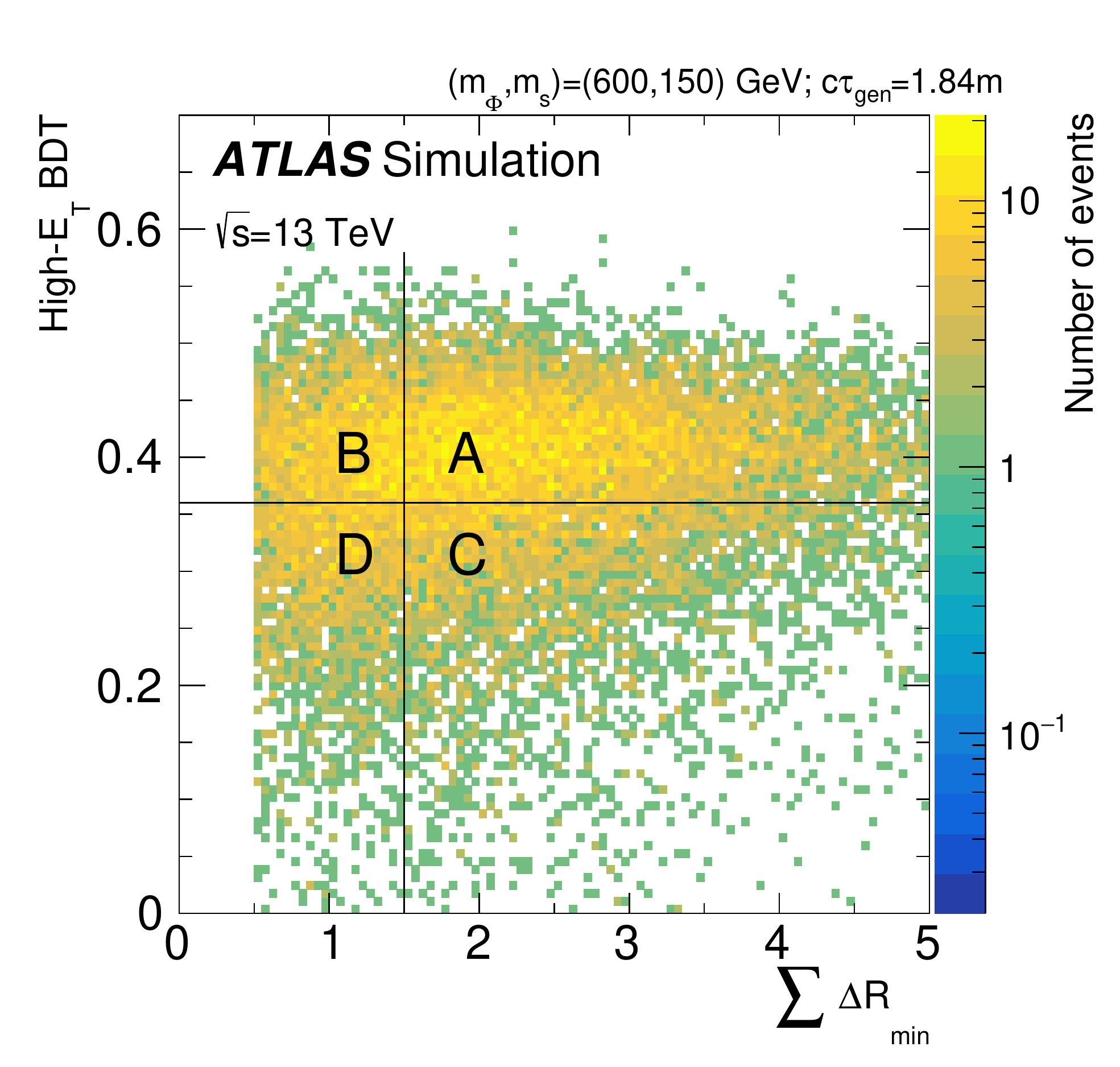}}\\
 
\subfloat[]{\includegraphics[width=0.33\linewidth]{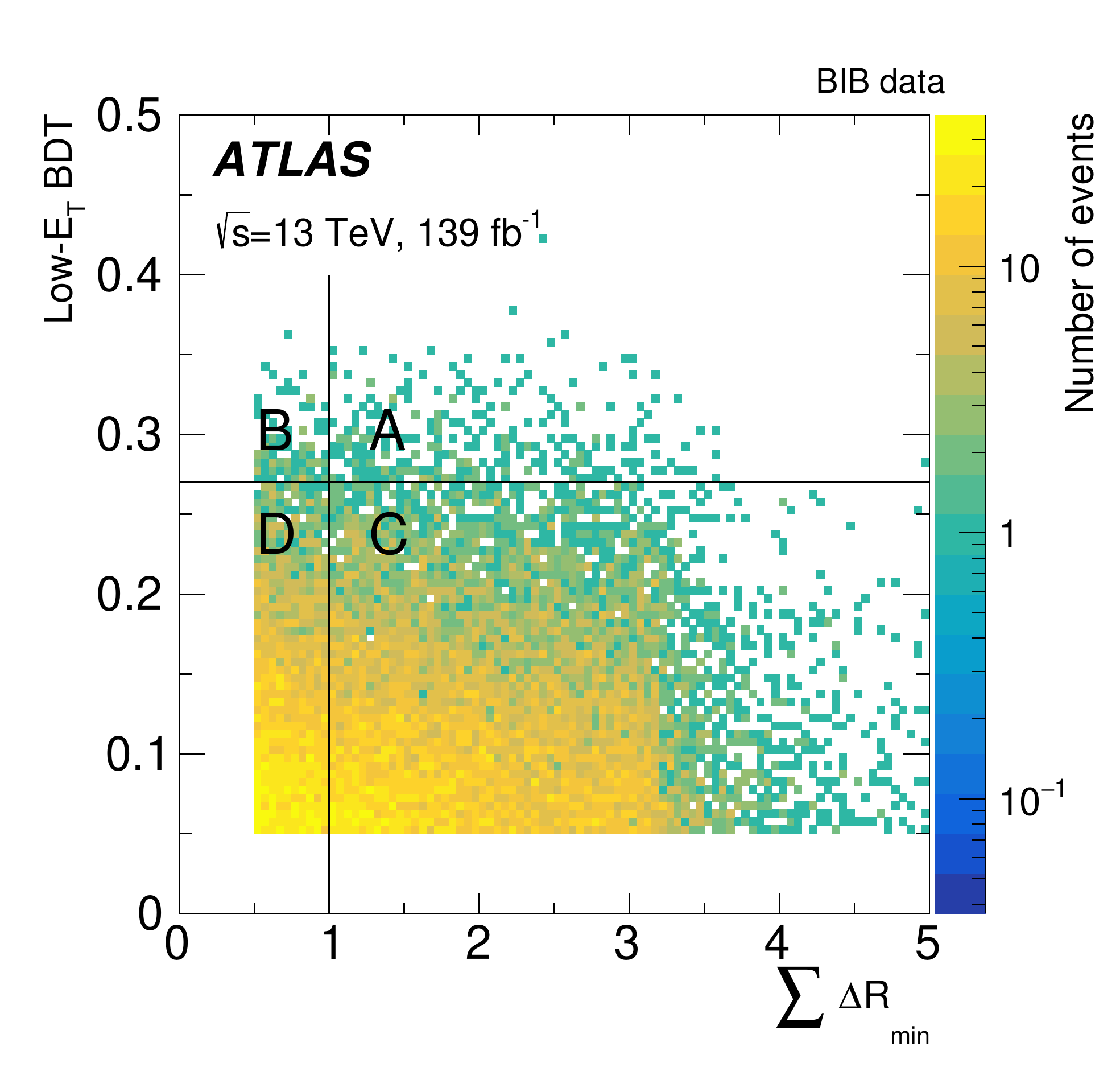}}
\subfloat[]{\includegraphics[width=0.33\linewidth]{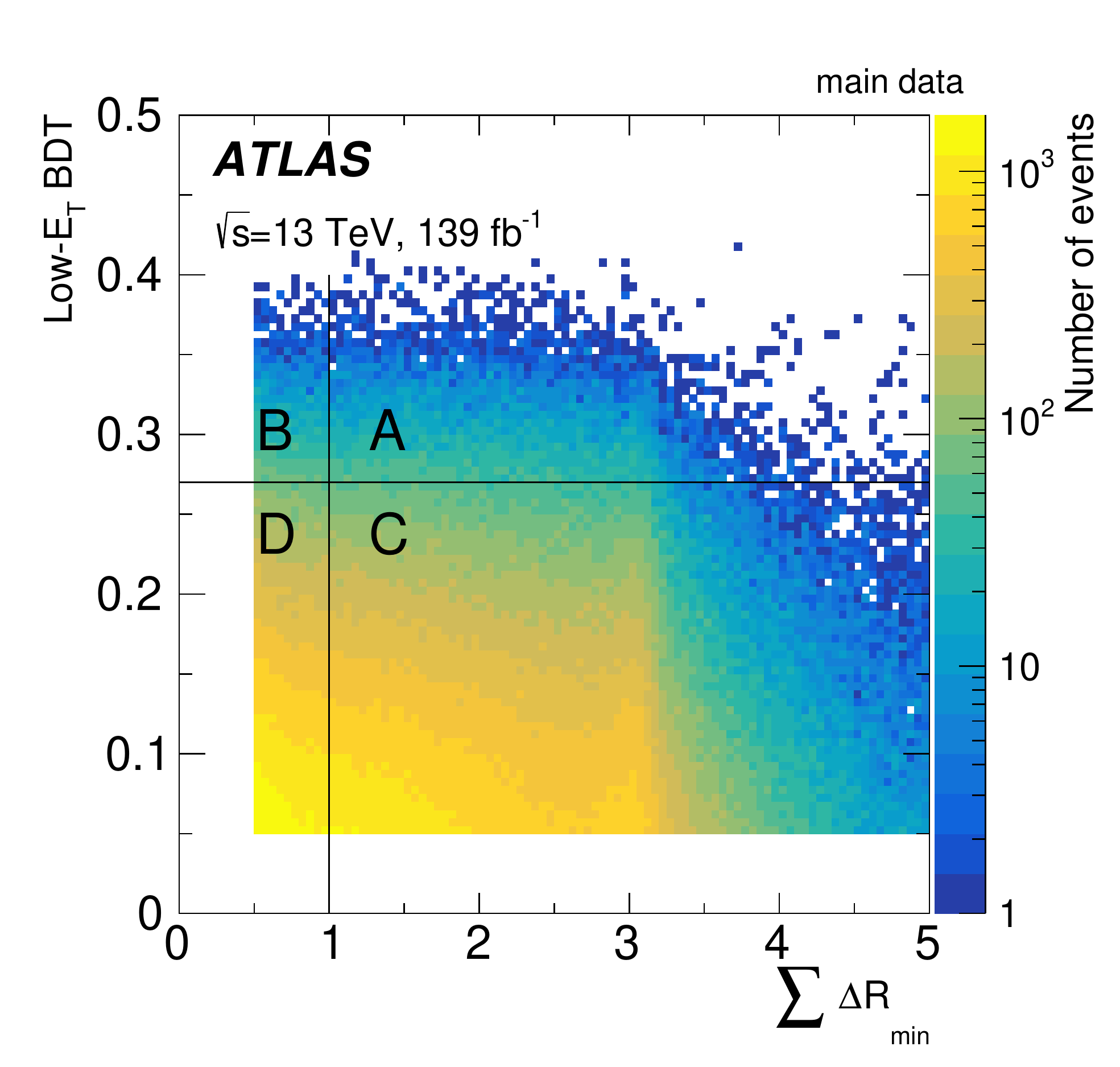}}
\subfloat[]{\includegraphics[width=0.33\linewidth]{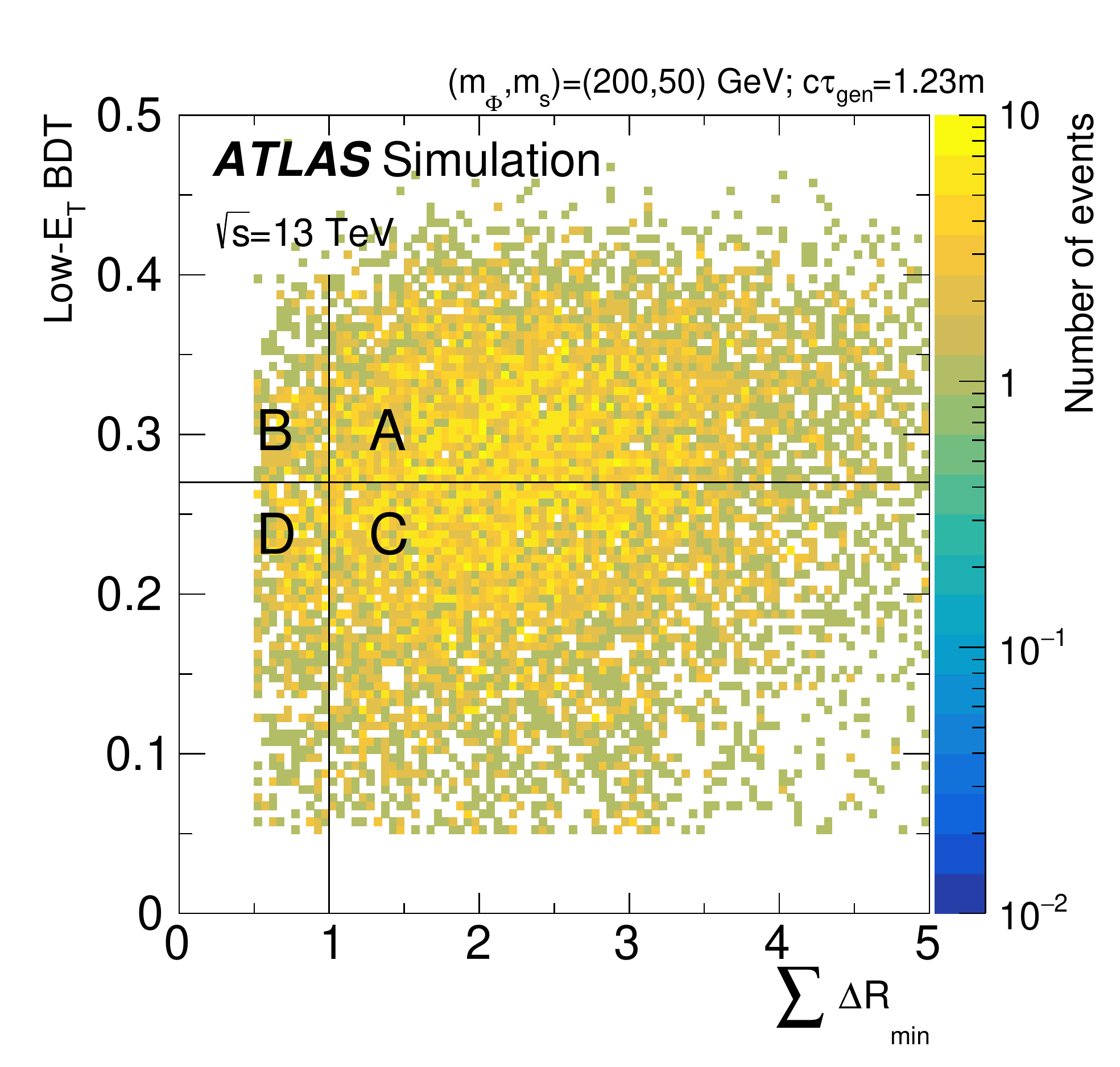}}
\caption{The distributions of $\sumMinDR$(jet, tracks) vs \evbdtLITau for (a) BIB events, (b) main data and (c) a signal sample after event cleaning for the high-\ET selection.
Panels (d,e,f) show the equivalent distributions for the low-\ET selection.
The signal sample with $\mPhi = 600 \GeV$ and $\mS = 150 \GeV$ is shown for the high-$\ET$ selection, while the sample with $\mPhi = 200 \GeV$ and $\mS = 50 \GeV$ is shown for the low-$\ET$ selection.
}
\label{fig:ABCDvariables2DEvClean}
\end{figure}

Events passing the high-\ET and low-\ET selections defined in the previous section are divided into four subregions according to two variables: $\sumMinDR$(jet, tracks) and \evbdtLITau or \evbdtLITopo, depending on the selection.
The variables are uncorrelated (Pearson correlation coefficient $|r| < 0.03$ in the main dataset after the event cleaning, with additional tests for correlations described below) and have good separation between signal and multijet background, as shown in Figure~\ref{fig:ABCDvariables2DEvClean}.
 
Region A is defined by $\evbdtLITau \geq 0.36$ and $\sumMinDR \geq1.5$ for the high-$\ET$ analysis and by
$\evbdtLITopo \geq 0.27$ and $\sumMinDR \geq 1.0$ for the low-$\ET$ analysis.
Regions B, C, and D are obtained by reversing one or both of these selections.

In cases where more than one background population is included in the final selection, the condition that the two variables defining the plane are statistically independent is only guaranteed if their contributions have the same shape in the ABCD plane.
In this search there are three major sources of background (BIB, cosmics and SM multijets) with different distributions in the plane, which results in the two variables defining it having some correlation.
It is therefore necessary to make sure that the selection above has a high rejection power for two of the three sources.
Specifically, the contribution from BIB and cosmics in the ABCD plane must be negligible after the object and event selection requirements, leaving only the contribution of SM multijet events to be estimated by the ABCD method.
 
Two checks were performed to confirm that the contribution of background events from non-collision background is negligible after the selection.
 
First, the number of events satisfying each stage of the selection for the main dataset and the BIB dataset is shown in Table~\ref{tab:Selection_cutFlow} for the high-\et and low-\et selections, along with the fraction of signal events passing each cut for several benchmark samples.
For both the high-\et and low-\et selections, the number of BIB events satisfying all selection criteria is well within the uncertainty in the number of events passing all selections in the main dataset.
Considering that the efficiency of BIB identification in this dataset is approximately $70\%$, the BIB contamination in the main dataset can be considered negligible.
Furthermore, the events from the BIB dataset that pass the selection were checked and found to display properties of multijet events.
In particular, their $\phi$ and $z$ vs time distributions do not show the typical shape of BIB.
The events from the main dataset that pass the event cleaning were also checked and were found not to display the properties of BIB.
 
The second check is to confirm that the contamination from cosmic rays in the ABCD plane is negligible.
This is done using the cosmics dataset.
The number of events in this dataset passing the full selections is checked after weighting by the following two factors.
The first factor takes into account the difference between the number of bunch crossings where protons are present in both beams while the CalRatio triggers were enabled, and the number of empty bunch crossings, during which the cosmics dataset was collected. The protons-to-empty live-time ratio depends on the beam conditions, and its value lies in the range 2 to 3.5.
The second factor takes into account the fact that events in the cosmics dataset will have no related collision activity, and therefore will be largely trackless: events in the cosmics dataset will be far more likely than events from the main dataset to pass the requirements on jet \minDR at the HLT and in the analysis preselection. The second factor is calculated as the ratio of the number of events entering the ABCD plane in the main dataset to the number passing the selection if all tracks in the event are ignored. This factor is ${<}0.1$ in all data-taking periods.
The final estimated number of events in the ABCD plane is $2.5 \pm 0.8$ (none of which are in region A) in the high-$\ET$ selection and $1.8 \pm 0.5$ ($0.3 \pm 0.2$ in region A) in the low-$\ET$ selection. These yields are negligible in comparison with the expected number of data events in the ABCD plane in the main dataset, which is over $300$ in both selections.

\begin{table}[ht]
\resizebox{\textwidth}{!}{
\begin{tabular}{|ll|r|r|r|r|r|r|}
\hline
& High-$\ET$ selection:                                     &   Main data     &   BIB        &  $\mPhi=1000\GeV$     &  $\mPhi=600\GeV$       &  $\mPhi=400\GeV$     \\ 
&                                                           &                 &              &  $\mS=275\GeV$        &  $\mS=150\GeV$         &  $\mS=100\GeV$      \\ 
&                                                           &                 &              &  ($c\tau_{gen}=2.40$~m)  & ($c\tau_{gen}=1.84$~m)        &   ($c\tau_{gen}=1.61$~m)     \\ 
\hline
\textbf{Preselection:}         & trigger, 2 clean jets                                     & 40\,743\,867   &  2\,200\,854&  \numR{23.96 }\%~~~~~  & \numR{19.48 }\%~~~~~  &  \numR{15.09}\%~~~~~ \\
&  $\sumMinDR >0.5$                                         & 28\,248\,024   &     1\,399\,351&  \numR{23.26 }\%~~~~~  &  \numR{19.05 }\%~~~~~ &  \numR{14.8}\%~~~~~  \\
\textbf{Event cleaning:}       & $\evbdtLITau>0.0$                                         & 75\,224       &     3141    & \numR{23.11 }\%~~~~~  &  \numR{18.65 }\%~~~~~ &  \numR{14.04}\%~~~~~  \\
& Trigger matching                                          &  58\,190       &     2026     &  \numR{ 20.50 }\%~~~~~  &  \numR{ 17.46 }\%~~~~~ &  \numR{13.57}\%~~~~~ \\
& $-3 < t < 15\ns$              &  54\,108     &     1837     &  \numR{ 19.91 }\%~~~~~ &  \numR{ 16.94 }\%~~~~~ &  \numR{13.19}\%~~~~~  \\
& $\lograt > -1.5$ for \FCRjet, \SCRjet, \FBIBjet, \SBIBjet &  50\,516  &    1733     &  \numR{ 18.81 }\%~~~~~ &  \numR{ 16.08 }\%~~~~~ &  \numR{12.52}\%~~~~~ \\
& $|\eta| \notin [1.45, 1.55]$ for \FCRjet, \SCRjet         &  47\,037     &     1627     &  \numR{ 17.66 }\%~~~~~ &  \numR{ 15.01 }\%~~~~~ &  \numR{ 11.70}\%~~~~~ \\
\textbf{High-$\ET$ selection:} & $\mhtoht<0.6$                                             &  40\,464   &    1295     &  \numR{ 14.33 }\%~~~~~  &  \numR{ 13.21 }\%~~~~~ &  \numR{ 10.89}\%~~~~~ \\
& $\FCRhjetpt>70\GeV $                                     &  39\,266       &     1270     & \numR{ 14.29 }\%~~~~~  &  \numR{ 13.13 }\%~~~~~ &  \numR{ 10.78}\%~~~~~  \\
& $\SCRhjetpt>80\GeV $                                     &  21\,787       &     531      &  \numR{ 12.79 }\%~~~~~  &  \numR{ 11.21 }\%~~~~~ &  \numR{  8.47}\%~~    \\
& $\sumLogRh>1$                                             &  13\,183       &     341      &  \numR{ 11.52 }\%~~~~~  &  \numR{  9.6 }\%~~    &  \numR{  6.8}\%~~   \\
& High-\ET NN product $> 0.5$                               &  393          &     15       &  \numR{  9.25 }\%~~     &  \numR{  6.87 }\%~~    &  \numR{  4.24}\%~~  \\
\textbf{Region A :}            &                                                           &  22          &     1        &  \numR{  7.40 }\%~~     &  \numR{  4.83 }\%~~    &  \numR{  2.61}\%~~  \\
\textbf{Region B :}            &                                                           &  7            &     0        &  \numR{  0.63 }\%     &  \numR{  0.54 }\%    &  \numR{  0.30}\%       \\
\textbf{Region C :}            &                                                           &  233          &     7        &  \numR{  1.12 }\%~~     &  \numR{  1.37 }\%~~    &  \numR{  1.18}\%~~    \\
\textbf{Region D :}            &                                                           &  131          &     7        &  \numR{  0.10 }\%      &  \numR{  0.14 }\%    &  \numR{ 0.15}\%       \\
\hline
\end{tabular}
}
 
\resizebox{\textwidth}{!}{
\begin{tabular}{|ll|r|r|r|r|r|r|}
\hline
& Low-$\ET$ selection:                                      &   Main data     &   BIB        & $\mPhi=200\GeV$  &  $\mPhi=125\GeV$  & $\mPhi=60\GeV$    \\ 
&                                                           &                 &              & $\mS=50\GeV$    &  $\mS=55\GeV$ )    & $\mS=5\GeV$       \\ 
&                                                           &                 &              & ($c\tau_{gen}=1.25$~m)    &  ($c\tau_{gen}=1.05$~m)    & ($c\tau_{gen}=0.22$~m)    \\ 
\hline
\textbf{Preselection: }       & trigger, 2 clean jets                                     &   40\,743\,867  &  2\,200\,854 & \numR{7.06}\%~~    & \numR{1.1}\%~~~~     & \numR{0.79}\%~~     \\
& $\sumMinDR >0.5$                                          &   28\,248\,024  &  1\,399\,351 & \numR{6.87}\%~~    & \numR{1.05}\%~~~~     & \numR{0.73}\%~~     \\
\textbf{Event cleaning:}       & $\evbdtLITopo>0.05$                                       &   1\,288\,596   &  44\,035    & \numR{6.32}\%~~    & \numR{0.93}\%~~     & \numR{0.51}\%~~     \\
& Trigger matching                                          &   1\,138\,961   &  36\,266    & \numR{6.01}\%~~    & \numR{0.86}\%~~     & \numR{0.47}\%~~     \\
&  $-3 < t < 15\ns$             &   1\,123\,239   &  35\,245    & \numR{5.87}\%~~    & \numR{0.84}\%~~     & \numR{0.45}\%~~     \\
&  $\lograt > -1.5$ for \FCRjet, \SCRjet, \FBIBjet, \SBIBjet&   1\,038\,019   &  33\,100    & \numR{5.52}\%~~    & \numR{0.79}\%~~     & \numR{0.43}\%~~     \\
&  $|\eta| \notin [1.45, 1.55]$ for \FCRjet, \SCRjet        &   976\,805     &  31\,292    & \numR{5.12}\%~~    & \numR{0.73}\%~~     & \numR{0.39}\%~~     \\
\textbf{Low-$\ET$ selection:}  & $\mhtoht<0.6$                                             &   965\,748     &  30\,712    & \numR{4.99}\%~~    & \numR{0.71}\%~~     & \numR{0.38}\%~~     \\
& $\FCRljetpt>80\GeV $                                     &   315\,530     &  10\,048    & \numR{4.03}\%~~    & \numR{0.54}\%~~     & \numR{0.23}\%~~     \\
& $\SCRljetpt>80\GeV $                                     &   73\,484      &  2810     & \numR{1.75}\%~~    & \numR{0.25}\%~~     & \numR{0.1}\%~~     \\
& $\sumLogRl>2$                                             &   3375       &  93        & \numR{0.59}\%    & \numR{0.06}\%     & \num{0.02}\%~~      \\
& Low-\ET NN product $> 0.7$                                 &   307         &  10        & \numR{0.53}\%    & \num{0.04}\%~~      & \num{0.006}\%     \\
\textbf{Region A :}            &                                                           &   23          &  0         & \numR{0.46}\%    & \num{0.03}\%~~      & \num{0.003}\%     \\
\textbf{Region B :}            &                                                           &   3           &  0         & \num{0.01}\%     & \num{0}\%~~~~~~~         & \num{0}\%~~~~~~~         \\
\textbf{Region C :}            &                                                           &   220         &  7         & \num{0.06}\%     & \num{0.01}\%~~      & \num{0.002}\%     \\
\textbf{Region D :}            &                                                           &   61          &  3         & 0\%~~~~~     &  0\%~~~~~~~              & 0\%~~~~~~~               \\
\hline
\end{tabular}
}
\vspace{0.25cm}
\caption{Sequential impact of each requirement on the number of events passing the selection for the high-$\ET$ (top table) and low-$\ET$ (bottom table) selections. The signal columns represent the cumulative fraction of events passing the selection than the number of events. }
\label{tab:Selection_cutFlow}
\end{table}

The validity of the ABCD method is tested by applying it to a number of validation regions (VRs), which are orthogonal to region A.
A first set of validation regions is defined using the nominal event selections but looking only into part of regions B, C and D.
Restricting the VR ABCD plane to intermediate values of the BDT output (using part of nominal regions C and D) allows a test of the background estimation method in the whole \sumMinDR range in the absence of signal contamination. Likewise, restricting the VR ABCD plane to low values of \sumMinDR and allowing any BDT output value (using part of nominal regions B and D) permits a test of the method at large values of the BDT output.
As an example, the nominal high-$\ET$ event selection is validated in region $\textrm{VRCD}_\textrm{high-ET}$, defined using the nominal high-$\ET$ selection but restricted to the range $0.0 < \evbdtLITau < 0.3$ and $0.5 < \sumMinDR < 5.0$.
First, the ABCD method is tested by dividing this region into four subregions defined by the boundaries of $\evbdtLITau = 0.15$ and $\sumMinDR = x$, where the value of $x$ is varied from 1 to 4.
Then the ABCD method is tested again in this same region by setting the $\sumMinDR$(jet, tracks) boundary at 1.5 and allowing the \evbdtLITau boundary to take values from 0.05 to 0.25.
 
\begin{figure}[ht]
\centering
\subfloat[]{\includegraphics[width=0.3\linewidth]{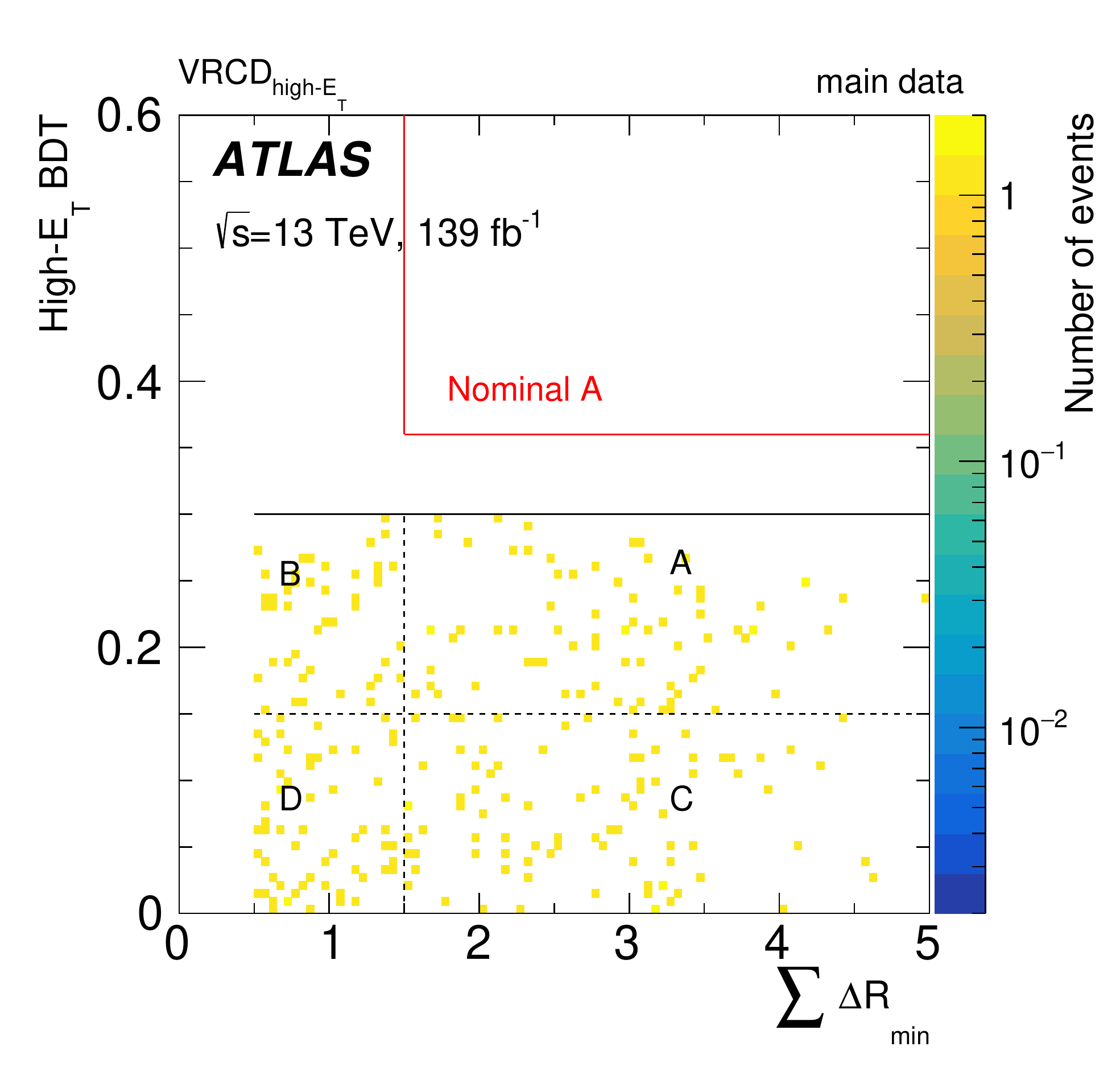}}
\subfloat[]{\includegraphics[width=0.3\linewidth]{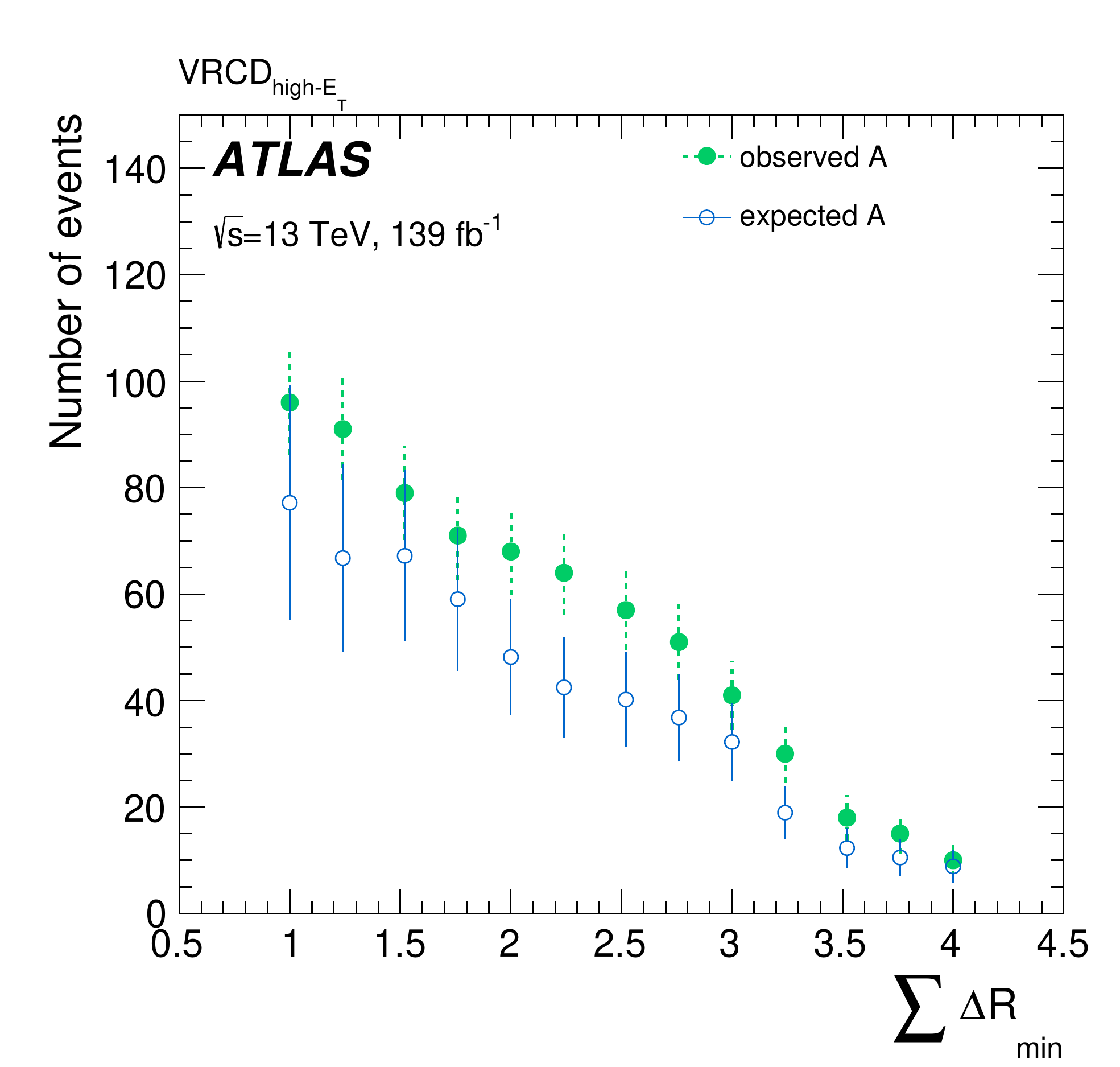}}
\subfloat[]{\includegraphics[width=0.3\linewidth]{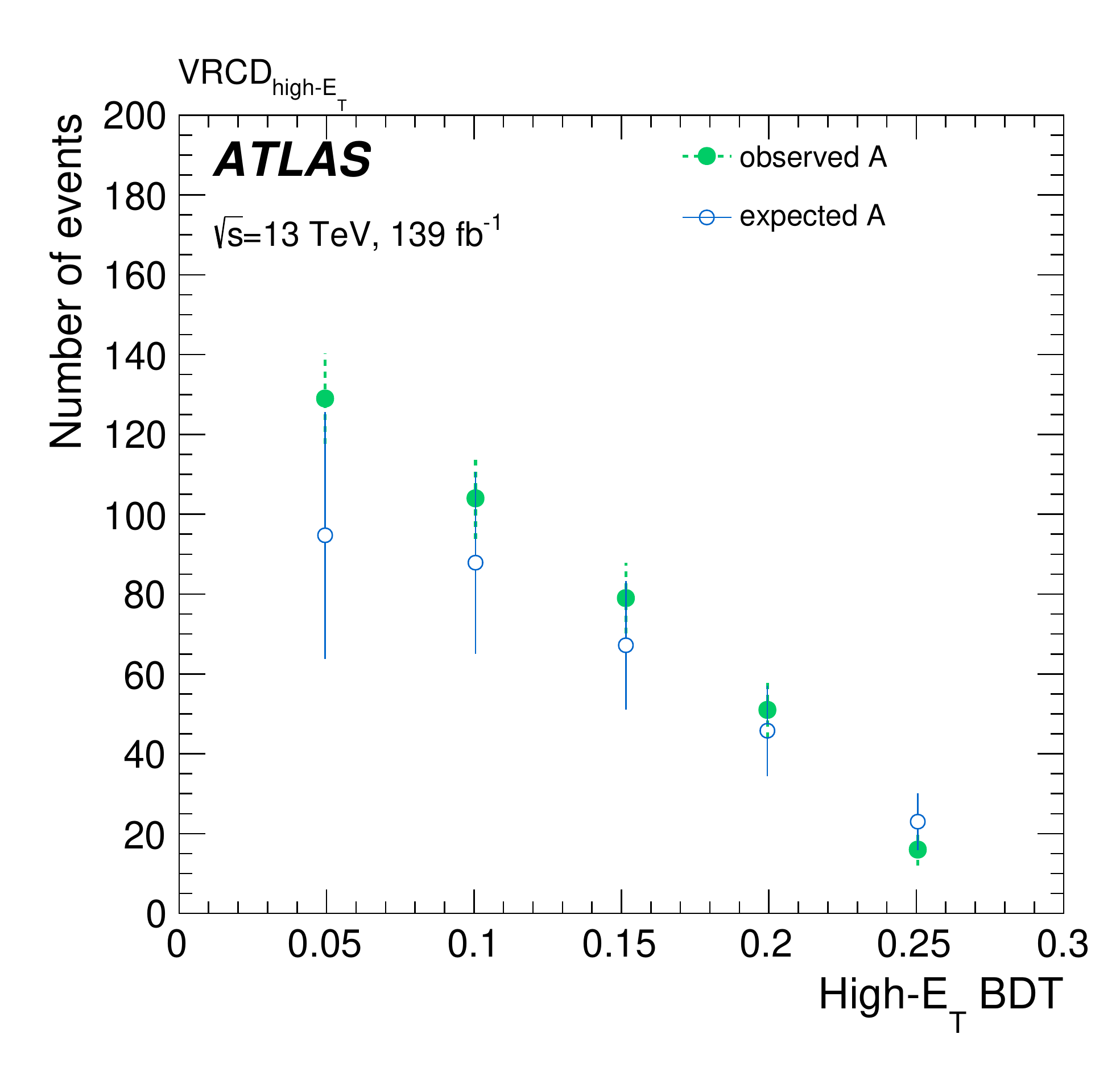}} \\
\subfloat[]{\includegraphics[width=0.3\linewidth]{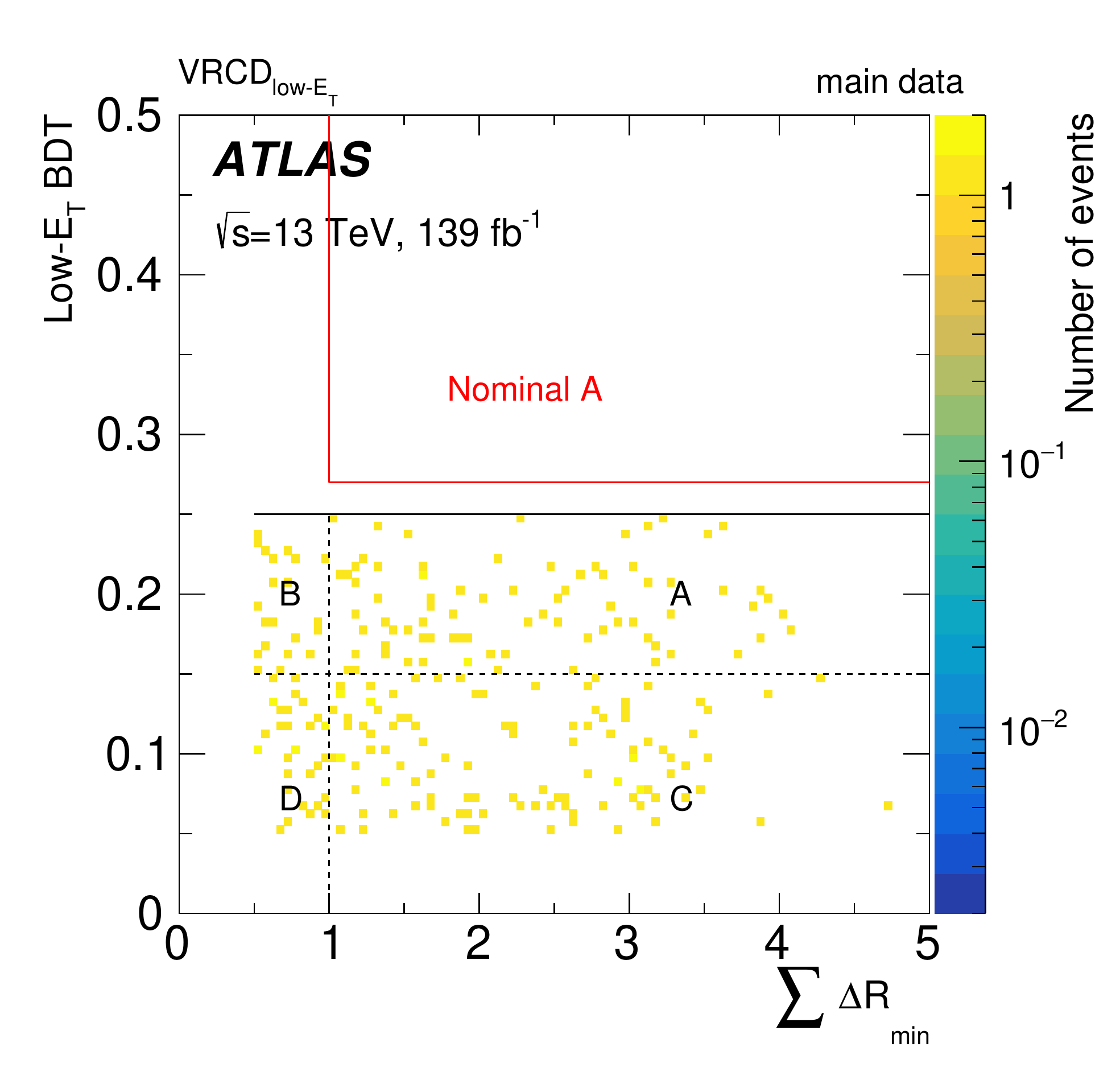}}
\subfloat[]{\includegraphics[width=0.3\linewidth]{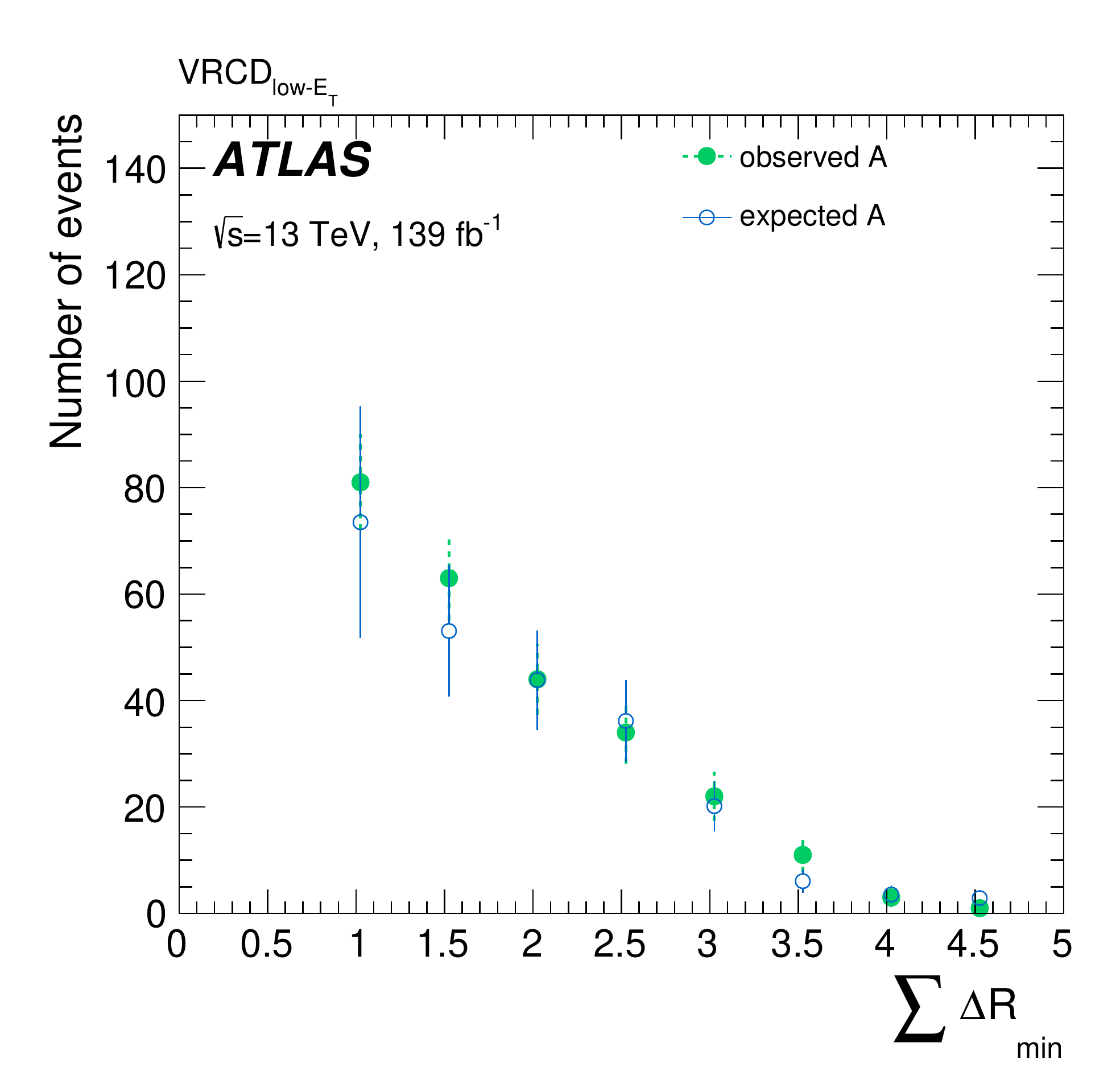}}
\subfloat[]{\includegraphics[width=0.3\linewidth]{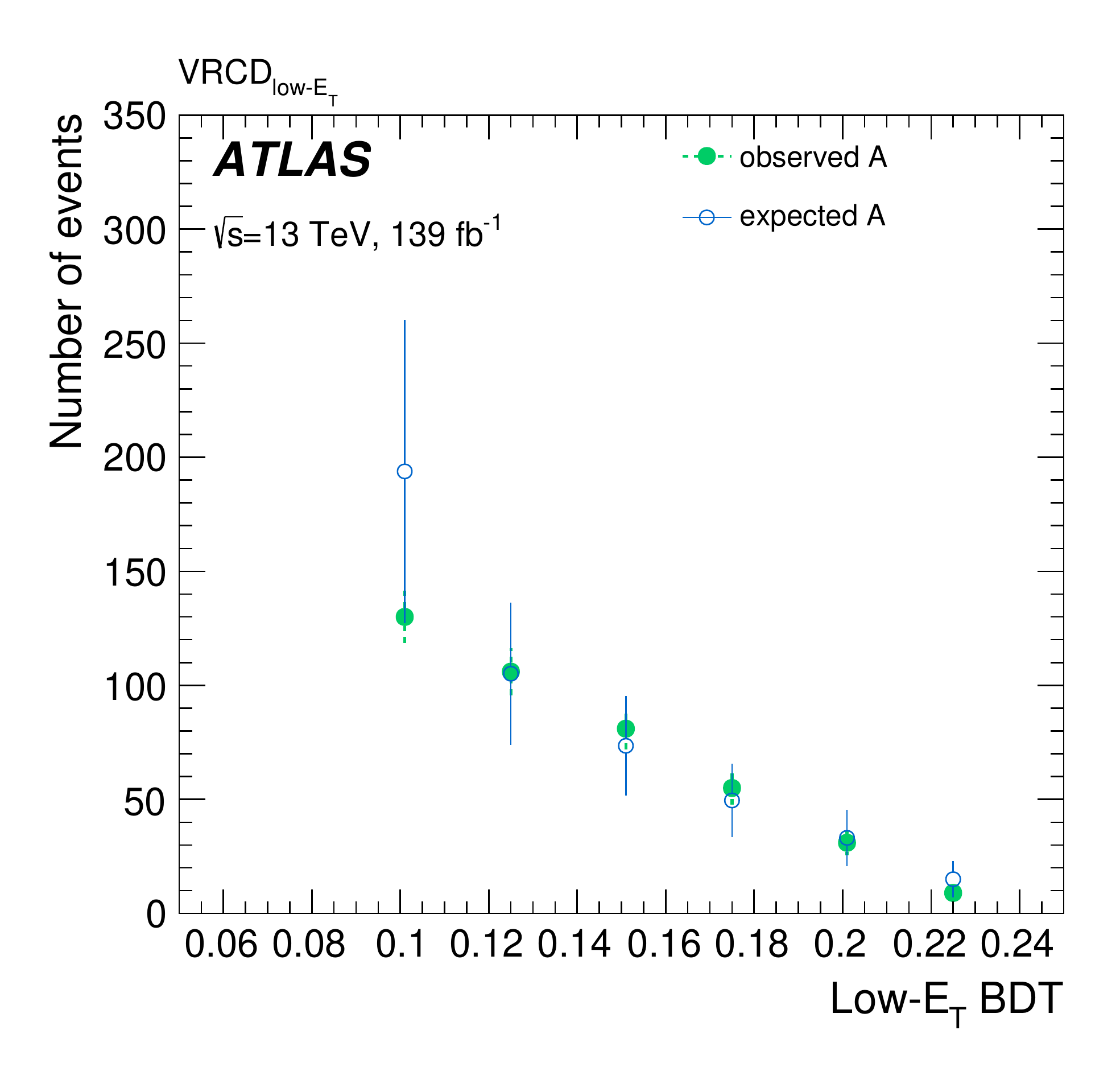}}
\caption{(a) The per-event BDT vs $\sumMinDR$(jet, tracks) distribution for the main dataset in the $\textrm{VRCD}_\textrm{high-\ET}$ validation region. Region A in the nominal high-\ET selection is shown for reference. Also shown is a comparison of the observed and expected events in region A of this VR. The $x$-axis shows the different (b) $\sumMinDR$(jet, tracks) and (c) BDT boundaries used to test the ABCD method. Statistical uncertainties are shown in (b) and (c).
Panels (d,e,f) show the equivalent distributions in the $\textrm{VRCD}_\textrm{low-\ET}$ validation region.
}
\label{fig:ABCDplaneHVRCD}
\end{figure}

The ABCD plane defined in this VR can be seen in Figure~\ref{fig:ABCDplaneHVRCD} for $\textrm{VRCD}_\textrm{high-ET}$ and the low-\ET selection, $\textrm{VRCD}_\textrm{low-ET}$, along with the level of agreement between the expected and observed numbers of events in region A in all the tested boundary selections.
It should be noted that the tests in each VR are statistically correlated. Therefore, statistical fluctuations can affect several tests.

A second set of VRs is defined by inverting the selection on the product of the signal NN scores of the two CalRatio jets in the analysis selections.
 
Given that signal contamination is low in all the validation regions, the plain ABCD method assuming no signal is applied.
The statistical precision of this closure test is better than the final statistical uncertainty of the number of events observed in region A. Good closure, within one standard deviation, is observed in all the validation regions defined in this section.

\section{Systematic uncertainties}
\label{sec:systematics}
 
The uncertainty in the data-driven ABCD method for the background estimate is studied in the dijet control region.
In this control region, an alternative ABCD plane is defined using the same variables as in the analysis, but adjusting the boundaries in regions A, B, C and D to reduce the effect of statistical fluctuations in the estimation of the number of dijet events in region A by this method.
The observed number of events in region A is compared with the estimate given by the ABCD method for both the high-\ET and the low-\ET ABCD planes.
In all the tests performed, the observed and expected numbers of events agree within the statistical uncertainties. This confirms that the two variables forming the plane are statistically independent and therefore no systematic uncertainty is added for the background estimation method.
 
The uncertainty in the combined 2015--2018 integrated luminosity is 1.7\% \cite{ATLAS-CONF-2019-021}, obtained using the LUCID-2 detector \cite{LUCID2} for the primary luminosity measurements.
 
Events in MC simulation are reweighted to obtain the correct pile-up distribution.
The uncertainty in the pile-up reweighting of the reconstructed events in the MC simulation is estimated by comparing the distribution of the number of primary vertices in the MC simulation with the distribution in data as a function of the instantaneous luminosity.
Differences between these distributions are reduced by scaling the mean number of $pp$ interactions per bunch crossing in the MC simulation and the $\pm 1 \sigma$ uncertainties are assigned to these scaling factors~\cite{ATLAS-CONF-2019-021,LUCID2}.
The effect on the signal event yields varies between
2\% and 8\% depending on the signal model.
 
The jet energy scale and jet energy resolution introduce uncertainties in the signal yield of ${<}1\%$ to 5\% and 1\% to 6\%, respectively, depending on the signal model. These uncertainties are calculated using the procedure detailed in Ref.~\cite{Aaboud:2017jcu}.
Since the jets used in this analysis are required to have a low fraction of calorimeter energy in the ECal, the jet energy uncertainties are rederived as a function of ECal energy fraction as well as $\eta$.
The additional jet energy uncertainties are found to have an effect of 1\% to 6\% on the signal yield, and are conservatively taken in quadrature with the regular jet-energy uncertainties.
The low-\ET models are more sensitive to all jet energy uncertainties than the high-\ET models.
 
The uncertainty in the signal trigger efficiency is estimated by studying how well modelled the three most important trigger variables (jet \et, \logR, and \pt of tracks within the jet) are between HLT-reconstructed quantities and offline-reconstructed quantities in data and MC simulation. A tag-and-probe technique is applied to a pure sample of multijet events obtained using standard jet triggers in both data and MC simulation.
Scale factors that represent the degree of mismodelling in each variable are derived and then applied in an emulation of the CalRatio triggers.
The change in yield relative to the nominal (unscaled) trigger emulation after the full analysis selection is taken as the size of the systematic uncertainty, which is between 1\% and 7\% depending on the signal model.
 
A systematic uncertainty is included to account for potential mismodelling of input variables used in the machine-learning techniques applied in the analysis.
Using the same control sample of dijet events defined for the evaluation of the systematic uncertainty in the data-driven background estimate, the distributions of the inputs and outputs of the per-jet NNs and the per-event BDTs are studied.
They are found to agree fairly well between data and MC simulation.
The residual differences are translated into a systematic uncertainty in the signal efficiency, using the following procedure.
For each mis-modeled variable, the residual differences are quantified through a transfer factor between simulation and data.
In an ensemble of pseudo-experiments, the NN and BDT input variables for each signal event are varied by this transfer factor; in each pseudo-experiment the transfer factor is modulated by a random Gaussian, with a mean of zero and a width determined by measurements in the control region.
The final per-event BDTs are then re-evaluated, and the overall signal efficiency of the sample is evaluated for each pseudo-experiment.
This sequence of steps allows for the statistical determination of the systematic uncertainty in the final signal efficiency.
The value of the resulting uncertainty can thus be obtained from the distributions of efficiencies for the ensemble, and can be as large as 6\% depending on the signal model.
 
Finally, an uncertainty due to the NLO-reweighting of the signal samples is obtained by comparing the NLO \MADGRAPH predictions for a 125~\GeV Higgs boson mediator with predictions at next-to-next-to-leading-order accuracy in
QCD from \POWHEGBOX[v2]~\cite{Hamilton:2013fea,Hamilton:2015nsa,Alioli:2010xd,Nason:2004rx,Frixione:2007vw}.
This results in an additional signal efficiency uncertainty of 1\%--7\% for most samples.

\section{Results and statistical interpretation}
\label{sec:interpretation}
 
A data-driven background estimation and signal hypothesis test is performed simultaneously in all regions.
The procedure for the simultaneous fit is explained in detail in the previous iteration of this analysis~\cite{EXOT-2017-25} and is summarised in the following.
A profile likelihood function is constructed from the product of the probabilities of observing a given number of events in each region of the ABCD plane, given the expected number of events in that region. This expected number of events is given by the sum of the predicted signal yield in that region (scaled by a parameter of interest called the signal strength) and the expected number of background events. The background component of the expected yield in each region is constrained to satisfy the ABCD relation introduced in Section~\ref{sec:background}. The introduction of a signal component will therefore dynamically modify the remaining allowed background prediction in this set-up.
Additionally, a nuisance parameter which represents the total uncertainty in the signal efficiency, and consequently the signal yield, is introduced.
 
The background estimates before (\emph{a priori}) and after (\emph{a posteriori}) unblinding region A are shown in Table~\ref{table:ABCD_table}. After unblinding, the modified ABCD method takes account of the observed number of events in region A when making the final fit.
As a consequence, the background estimate changes relative to the \emph{a priori case}, since the fit now uses all available information to determine the expected background in each region.
As shown in the table, there is a slight excess in the observed number of events in region A over the \emph{a priori} expected background in both the high- and low-\ET selections.
In the \emph{a priori} fit, these excesses correspond to background-only hypothesis $p$-values of 0.083 and 0.076 in the high-\ET and low-\ET selections, respectively.
The significance is reduced to slightly under one standard deviation in the \emph{a posteriori} (background-only fit) estimate.
 
The \cls method~\cite{Read:2002hq}, using the ``alternative test statistic'' $\tilde{q}$~\cite{Cowan:2010js} is used to set upper limits on the production cross-section times BR of the LLP signals considered.
The \pyhf~\cite{pyhf,pyhf_joss} framework is used to implement the likelihood function and extract the upper limits.
An asymptotic approximation~\cite{Cowan:2010js} is used for these results.
This approximation was tested against the full frequentist pseudo-experiment-based method for a variety of signal samples and was found to give consistent limits.
Since each signal sample was generated for a given lifetime assumption, it is necessary to extrapolate the limits across lifetimes.
This extrapolation is performed using a reweighting method, which is described in detail in the previous iteration of this analysis~\cite{EXOT-2017-25} and summarised briefly here.
LLPs follow an exponential decay distribution, where the decay constant is given by the mean lifetime $\tau_{\textrm{gen}}$ in a given sample. To extrapolate to a different mean proper lifetime $\tau_{\textrm{new}}$, a weight $w$ is calculated as a function of the given particle's proper decay time $t$:
 
\begin{equation*}
w(t) = \frac{\tau_\textrm{gen}}{\exp(-t/\tau_\textrm{gen})} \cdot \frac{\exp(-t/\tau_\textrm{new})}{\tau_\textrm{new}}.
\end{equation*}
The per-event weight is the product of the weights obtained for each LLP in a given event.

\begin{table}[h!t]
\begin{center}
\setlength{\tabcolsep}{0.2pc}{\small
\def\arraystretch{1.2}
\begin{tabular*}{\textwidth}{@{\extracolsep{\fill}}|l|r|r|r|r|}
\hline
High-$E_{\mathrm{T}}$ selection             &  A                  &  B               &  C             & D            \\
\hline
Observed data                                & 22                  &  7               &  233           &  131          \\
\emph{a priori} & ~ & ~ & ~ & ~\\
Estimated background                        & $12.4 \pm 4.7$      &  $7    \pm 2.6$  &  $233 \pm 15$~  &  $131\pm11$   \\
\emph{a posteriori (background-only fit)} & ~ & ~ & ~ & ~\\
Fitted background                           & $18.8 \pm 3.5 $     &  $10.2 \pm 3.2$  & $236  \pm 15$~  & $128\pm 11$   \\
\emph{a posteriori (signal-plus-background fit)} & ~ & ~ & ~ & ~ \\
Fitted background                           & $10.0 \pm 6.0$      &  $5.7  \pm 2.4$  & $230  \pm 15$~  & $131 \pm11$   \\
Fitted signal ($(\mH, \mS) = (600, 150)\GeV$) & $12.2 \pm 8.7$      &  $1.4  \pm 1.0$  & $3.4  \pm 2.5$ & $<1$          \\
\hline    \hline
Low-$E_{\mathrm{T}}$ selection              &  A                  &  B               &  C             & D             \\
\hline
Observed data                                & 23                  &  3               &  220           &  61          \\
\emph{a priori} & ~ & ~ & ~ & ~ \\
Estimated background                        & $10.8 \pm 6.6$      &  $3    \pm 1.7$  &  $220  \pm 15$~ &  $61\pm7.8$   \\
\emph{a posteriori (background-only fit)} & ~ & ~ & ~ & ~ \\
Fitted background                           & $20.6 \pm 4.0 $     &  $5.4  \pm 2.3 $  & $222   \pm 15$~ & $59\pm 7.7$    \\
\emph{a posteriori (signal-plus-background fit)} & ~ & ~ & ~ & ~\\
Fitted background                           & $8.4 \pm 7.7$      &  $2.4  \pm 1.5$  & $217   \pm 15$~  & $61 \pm 7.8$ \\
Fitted signal ($(\mH, \mS) = (125, 55)\GeV$)  & $14.6 \pm 9.9$     &  $<1$             & $3.2  \pm 2.2$  & $<1$         \\
\hline
\end{tabular*}
}
\end{center}
\caption{Application of the modified ABCD method to the final high-\et and low-\et selections.
The \emph{a priori} estimate refers to the `pre-unblinding' case, where the data in region A are ignored by removing the Poisson constraint in that region and the signal strength is fixed to zero.
This matches the simple $N^\mathrm{bkg}_\mathrm{A}=(N^\mathrm{bkg}_\mathrm{B}\cdot N^\mathrm{bkg}_\mathrm{C})/N^\mathrm{bkg}_\mathrm{D}$ relation.
The \emph{a posteriori} estimate refers to the `post-unblinding' case, including the observed data in region A in the background-only global fit, obtained by fixing the signal strength to 0 (background-only fit) or allowing it to float (signal-plus-background fit).
The table also shows one set of representative signal yields in each selection for the signal-plus-background fit.
Only statistical uncertainties are included in the quoted error of the background, while the uncertainties in the signal include those from both statistical and experimental sources.
}
\label{table:ABCD_table}
\end{table}
 
\begin{figure}[ht]
\centering
\subfloat[]{\includegraphics[width=0.48\linewidth]{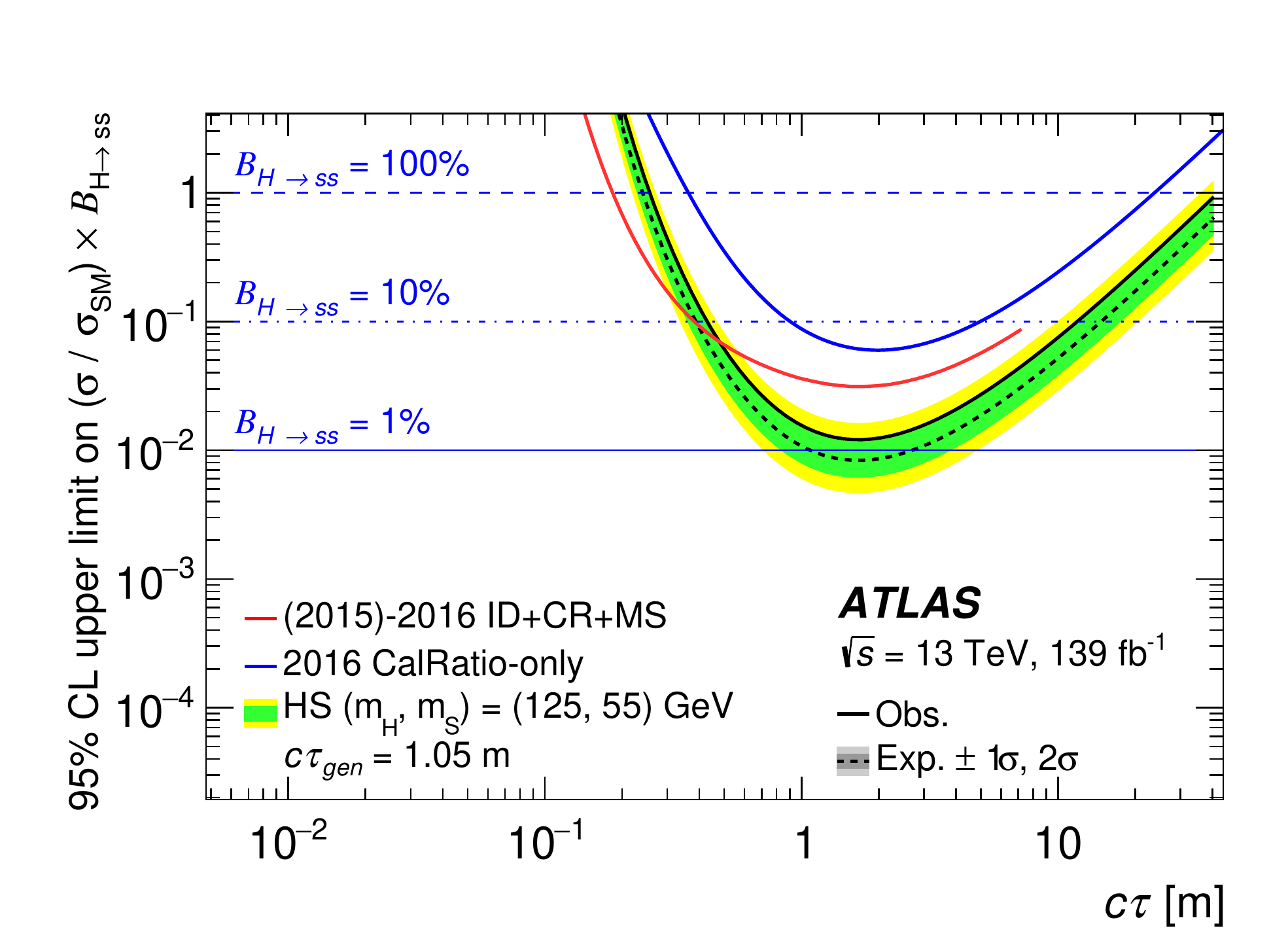}}
\subfloat[]{\includegraphics[width=0.48\linewidth]{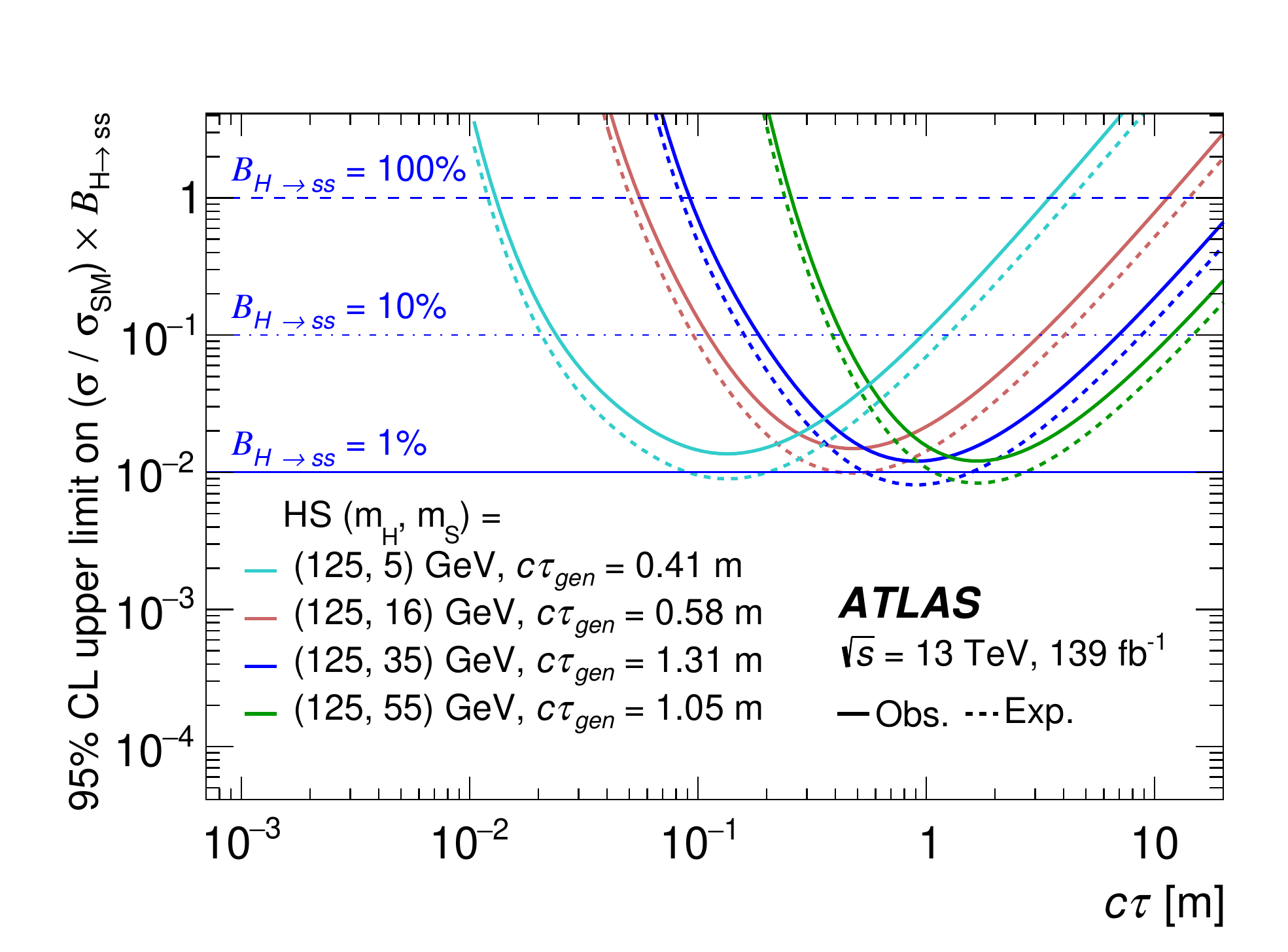}}\\
\caption{
(a) 95\% CL expected and observed limits on the BR of SM Higgs bosons to pairs of neutral LLPs ($B_{H\rightarrow ss}$), showing the $\pm 1 \sigma$ (green) and $\pm 2 \sigma$ (yellow) expected limit bands, as well as a comparison with the results from previous ATLAS searches~\cite{EXOT-2018-61, EXOT-2017-25}, and (b) summary of 95\% CL expected and observed limits on the BR of a SM Higgs boson mediator to pairs of neutral LLPs considered in this analysis. 
The cross-section for SM Higgs boson gluon--gluon fusion production is assumed to be $48.6$~pb.
}
\label{fig:limit-summary-higgs}
\end{figure}

\begin{figure}[ht]
\centering
\subfloat[]{\includegraphics[width=0.48\linewidth]{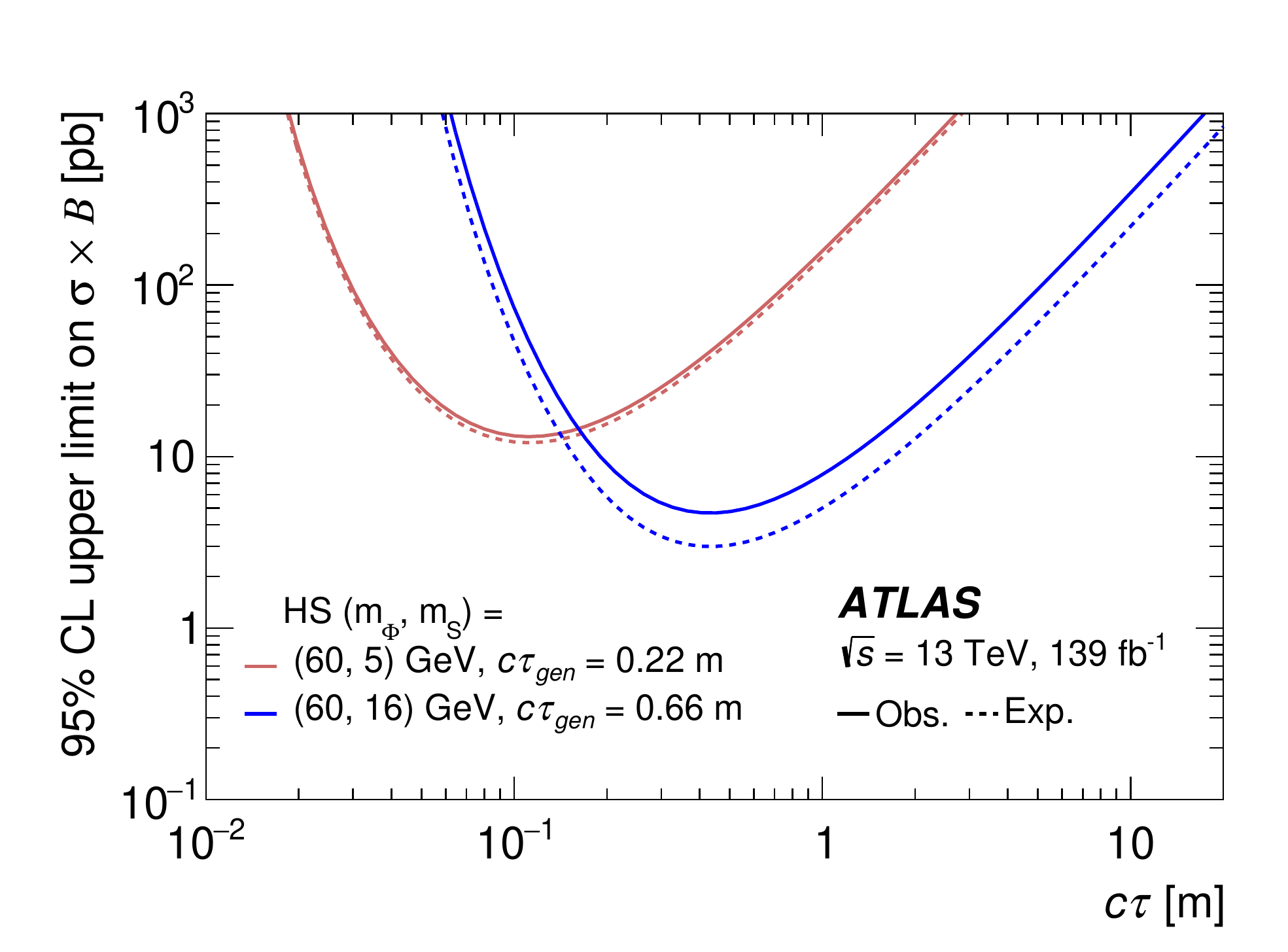}}
\subfloat[]{\includegraphics[width=0.48\linewidth]{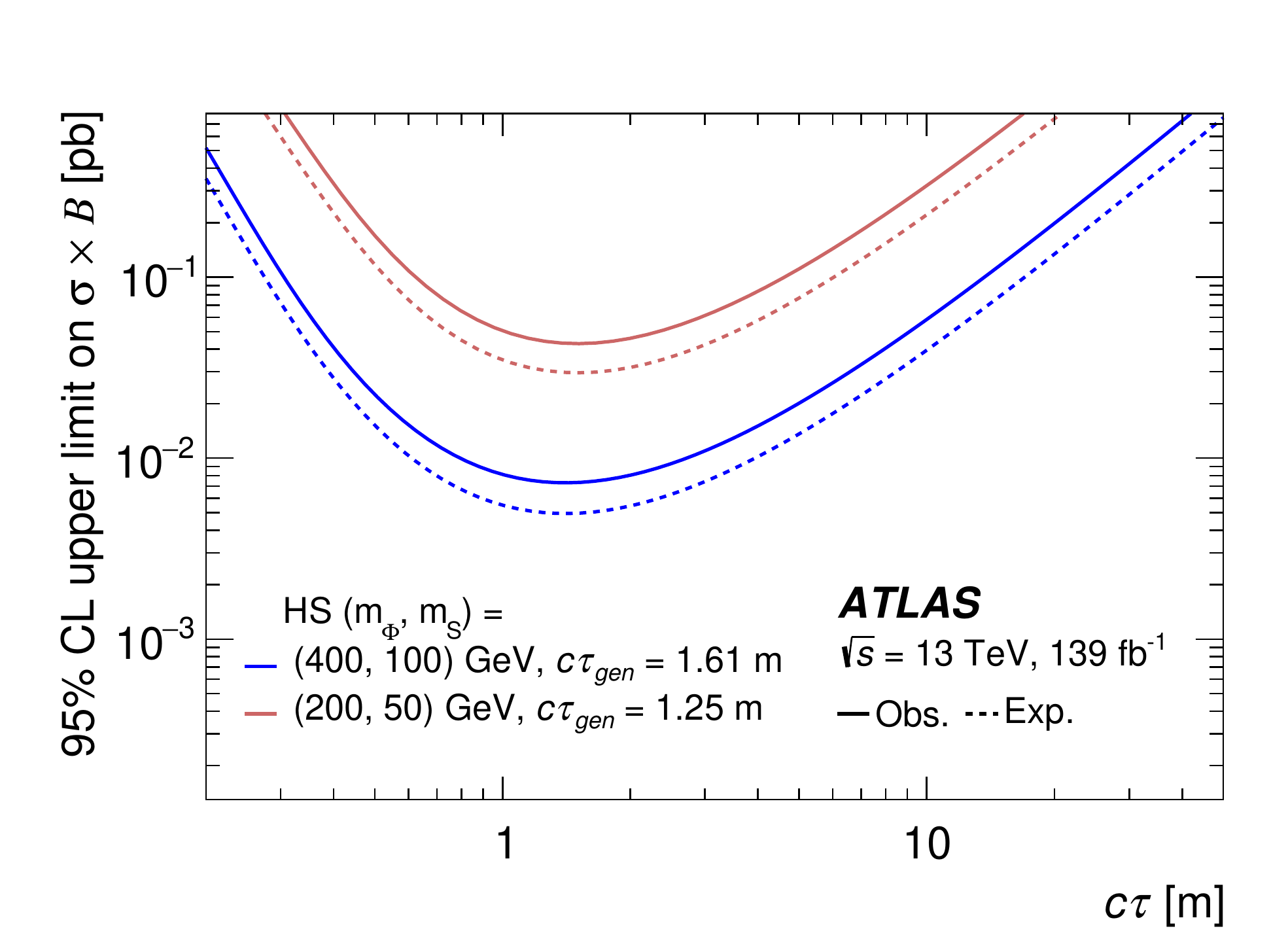}}\\
\subfloat[]{\includegraphics[width=0.48\linewidth]{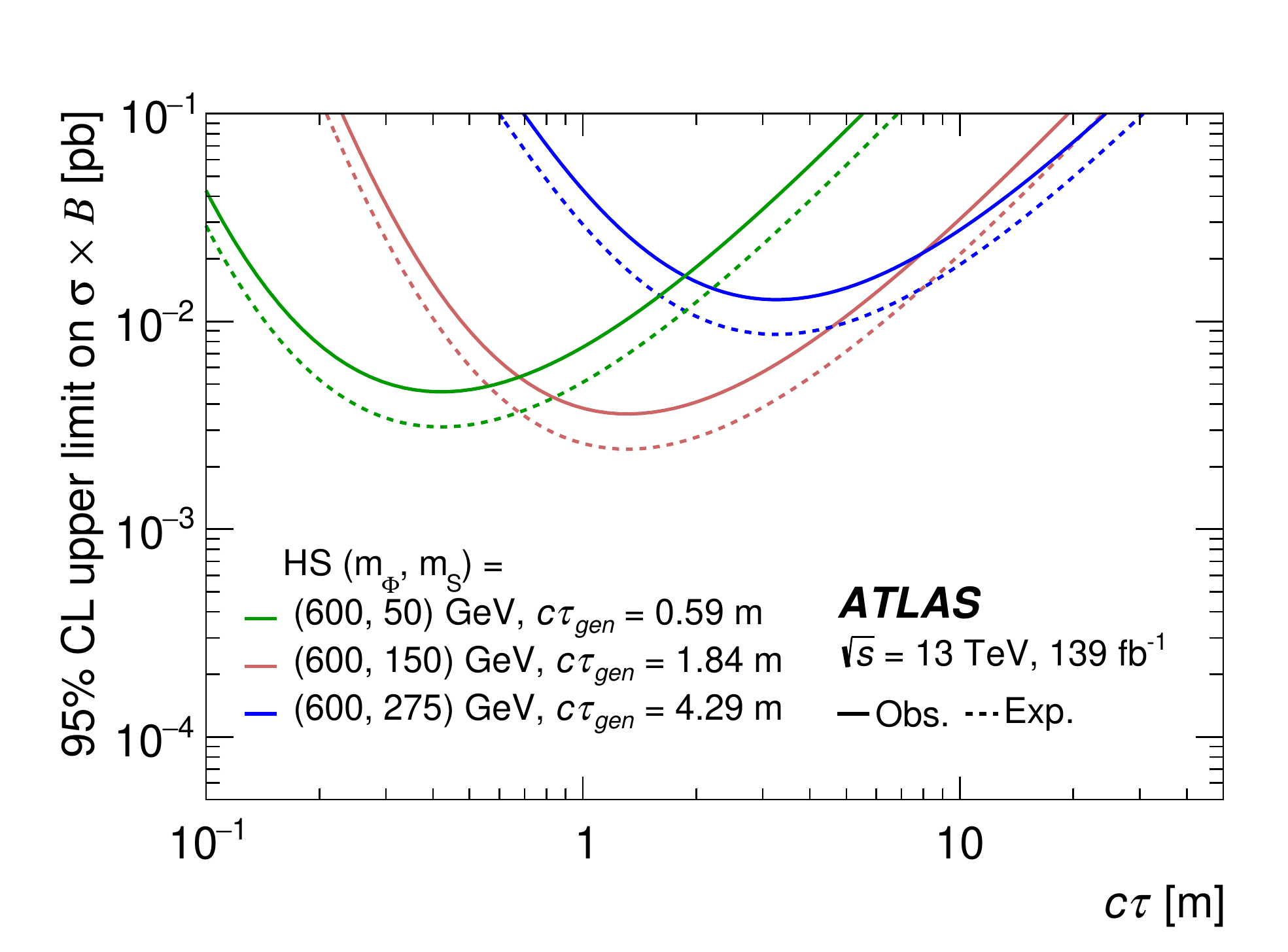}}
\subfloat[]{\includegraphics[width=0.48\linewidth]{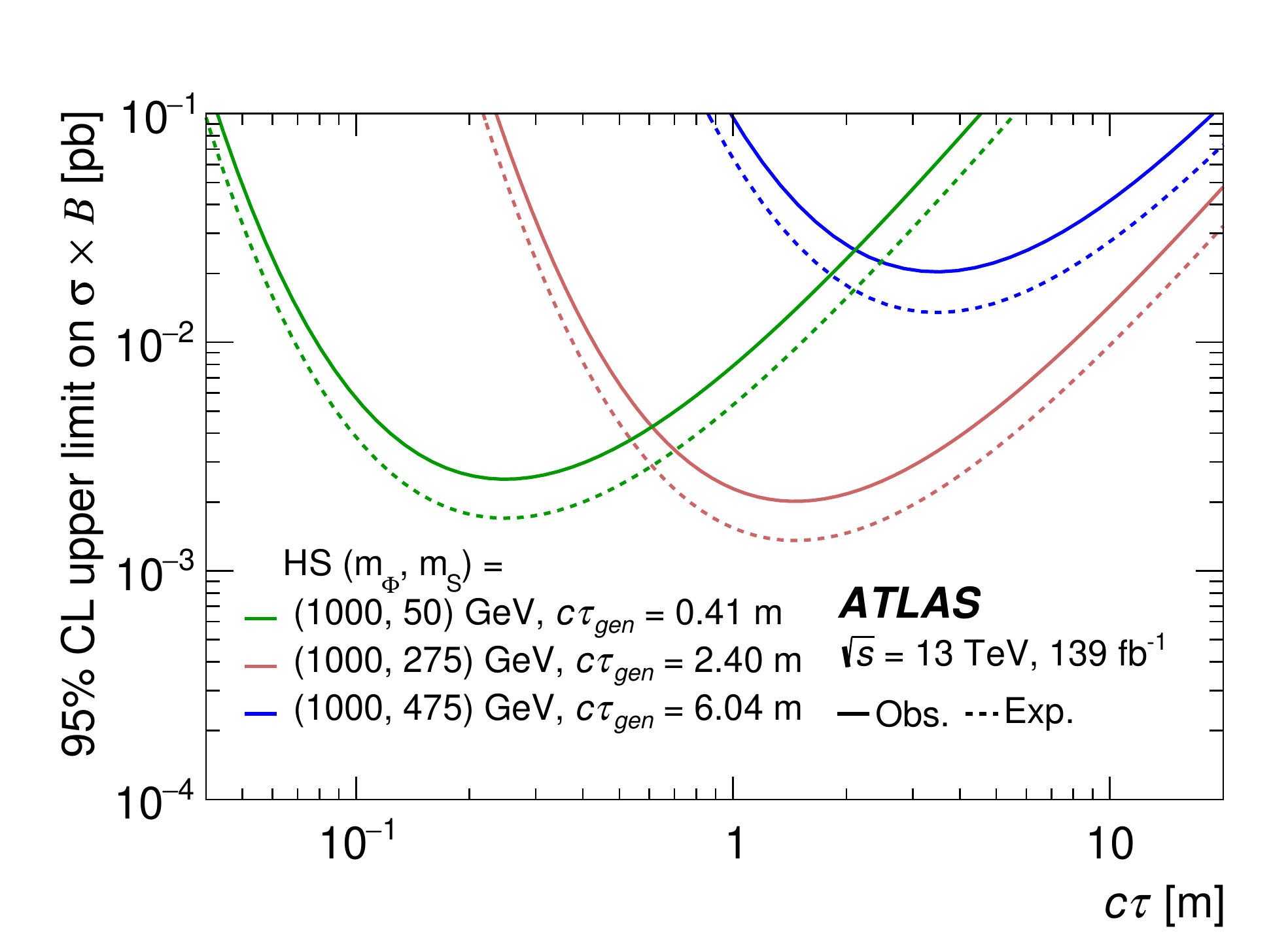}}
\caption{Summary of 95\% CL expected and observed limits for the hidden sector (HS) models considered in this analysis, where the mediator is not the SM Higgs boson: (a) shows the case where $\mH=60\gev$; (b) shows intermediate masses of $\mH=200\gev$ and $\mH=400\gev$; (c) shows the case where $\mH=600\gev$; (d) shows the highest \mH hypothesis with $\mH=1\tev$.
}
\label{fig:limit-summary}
\end{figure}

Figure~\ref{fig:limit-summary-higgs} shows a summary of the observed limits for the samples where the mediator is assumed to be the SM Higgs boson, with mass $125\gev$.
SM Higgs boson BRs to long-lived neutral scalars above 10\% are excluded for $c\tau$ values between a few centimetres and about 20 metres, depending on the mass of the LLP.
Figure~\ref{fig:limit-summary} shows a summary of observed limits on the cross-section times BR for other models, where the mediator is heavier or lighter than the SM Higgs boson.
The excluded ranges for each model are summarised in Table~\ref{table:summaryLimits}.
For mediators of mass 60~\GeV, cross-sections above 100~pb are excluded between about 30~mm and 5~m, depending on the LLP mass. For mediators heavier than the SM Higgs boson, cross-section times BR values above 0.1~pb are typically excluded between a few tens of millimetres and up to 20 metres. For high-\ET models where the scalar mass is much smaller than the mediator mass, cross-section times BR values above 0.01~pb can be excluded for $c\tau$ values of a few hundreds of millimetres to a few metres, depending on the model.
 
The analysis improves on the limits from the previous iteration of the analysis for many models. The improvements are typically between a factor of two and a factor of five. The low-\ET models are the ones for which the limits are most improved.
The previous analysis used only a fraction of the 13~\TeV dataset: 11~fb$^{-1}$ and 33~fb$^{-1}$ for the analysis of low-\ET and high-\ET models, respectively. Part of the gain is therefore explained by the increase in the integrated luminosity of the data sample. The rest of the gain, particularly for the low-\ET selection, comes from improvements in the low-\ET trigger and in the displaced-jet identification.

\begin{table}[h!t]
\begin{center}{\small
 
\begin{tabular}{|l|c|c|c|c|}
\hline
Model &  \multicolumn{4}{c|}{Excluded $\ctau$ range [m]} \\ \hline
& $B_{H\rightarrow ss} = 1$ & $B_{H\rightarrow ss} = 0.5$  & $B_{H\rightarrow ss} = 0.1 $  & $B_{H\rightarrow ss} = 0.05$ \\
& $\sigma=48.6 $~pb
& $\sigma=48.6$~pb
& $\sigma=48.6$~pb
& $\sigma=48.6$~pb \\  \hline
$ \mPhi =125 \GeV, \mS =5  \GeV$ & 0.01--3.47~~ & 0.02--2.40~~ & 0.02--0.96 & 0.03--0.62 \\
$ \mPhi =125 \GeV, \mS =16  \GeV$ & 0.06--11.4 & 0.07--7.86~~ & 0.11--3.15 & 0.15--2.01 \\
$ \mPhi =125 \GeV, \mS =35  \GeV$ & 0.09--24.8 & 0.11--17.2 & 0.19--6.94 & 0.25--4.48 \\
$ \mPhi =125 \GeV, \mS =55  \GeV$ & 0.26--40.5 & 0.29--29.1 & ~~0.43--11.9 & 0.54--7.73 \\
~ & ~ & ~ & ~ & ~  \\
\hline
~  & $\sigma \times B=500$~pb
& $\sigma \times B=100$~pb
& $\sigma \times B=50$~pb
& $\sigma \times B=10$~pb \\ \hline
$ \mPhi =60 \GeV, \mS =5  \GeV$ & 0.02--1.88~~ & 0.03--0.76 & 0.04--0.49 & -- \\
$\mPhi =60 \GeV, \mS =16  \GeV$ & 0.07--12.1 & 0.09--5.15 & 0.11--3.49 & 0.19--1.23 \\
~ & ~ & ~ & ~ & ~  \\
\hline
& $\sigma \times B=0.1 $~pb & $\sigma \times B=0.05$~pb & $\sigma \times B= 0.01$~pb & $\sigma \times B= 0.005$~pb \\ \hline
$ \mPhi =1000 \GeV, \mS =50  \GeV$ & 0.04--4.54~~ & 0.05--3.11~~ & 0.08--1.18 & 0.11--0.70 \\
$ \mPhi =1000 \GeV, \mS =275  \GeV$ & 0.24--29.8 & 0.27--20.5 & 0.43--7.95 & 0.56--4.91 \\
$ \mPhi =1000 \GeV, \mS =475  \GeV$ & 0.99--18.8 & 1.32--11.6 &  --  &  --  \\
$ \mPhi =600 \GeV, \mS =50  \GeV$ & 0.08--5.52~~ & 0.10--3.73~~ & 0.17--1.28 & 0.31--0.59 \\
$ \mPhi =600 \GeV, \mS =150  \GeV$ & 0.23--19.4 & 0.28--13.2 & 0.48--4.78 & 0.72--2.61 \\
$ \mPhi =600 \GeV, \mS =275  \GeV$ & 0.70--24.4 & 0.93--15.6 & -- & -- \\
$ \mPhi =400 \GeV, \mS =100  \GeV$ & 0.31--13.7 & 0.38--9.11~~ & 0.80--2.72 & -- \\
$ \mPhi =200 \GeV, \mS =50  \GeV$ & 0.63--4.62~~ & 1.02--2.33~~ & -- & -- \\ \hline
\end{tabular}
}
\end{center}
\caption{Summary of the $c\tau$ values which are excluded at 95\% CL for the models considered in this analysis.
}
\label{table:summaryLimits}
\end{table}

\FloatBarrier

\section{Conclusion}
\label{sec:conclusion}
 
In this paper, an analysis targeting pair-produced neutral long-lived particles decaying in the ATLAS calorimeters is presented.
The analysis used the full LHC Run~2 dataset of 13~\TeV proton--proton collisions, the size of which is 139~\ifb.
The analysis is structured in a very similar way to the one described in a previous paper 
using only the 2016 dataset but has several improvements, including a new method to select the most signal-like displaced jets with the help of a deep neural network trained using an adversary network.
No significant excess of events in the signal region is observed relative to the data-driven background prediction.
The CL$_\mathrm{s}$ method is used to set 95\% CL limits on the cross-section times branching ratio as a function of $c$ times the long-lived particle mean proper lifetime.
Limits are set on hidden-sector benchmark models with mediator masses ranging from 60~\GeV to 1~\TeV and long-lived scalar masses ranging from 5~\GeV to 475~\GeV.
The improvements to the analysis and additional data lead to an improvement on the limits for mediator masses above and below 200~\GeV by a factor of around 1.5--2 and 3--5, respectively, compared with the previous version of the analysis.
For models with a SM Higgs boson mediator, branching ratios to neutral scalars above 10\% are excluded for long-lived particle mean proper lifetimes times $c$ between approximately 20~mm and 10~m, depending on the model.

\section*{Acknowledgements}
 
% The next lines are included from the .//acknowledgements/Acknowledgements.tex input file

We thank CERN for the very successful operation of the LHC, as well as the
support staff from our institutions without whom ATLAS could not be
operated efficiently.
 
We acknowledge the support of
ANPCyT, Argentina;
YerPhI, Armenia;
ARC, Australia;
BMWFW and FWF, Austria;
ANAS, Azerbaijan;
SSTC, Belarus;
CNPq and FAPESP, Brazil;
NSERC, NRC and CFI, Canada;
CERN;
ANID, Chile;
CAS, MOST and NSFC, China;
Minciencias, Colombia;
MEYS CR, Czech Republic;
DNRF and DNSRC, Denmark;
IN2P3-CNRS and CEA-DRF/IRFU, France;
SRNSFG, Georgia;
BMBF, HGF and MPG, Germany;
GSRI, Greece;
RGC and Hong Kong SAR, China;
ISF and Benoziyo Center, Israel;
INFN, Italy;
MEXT and JSPS, Japan;
CNRST, Morocco;
NWO, Netherlands;
RCN, Norway;
MEiN, Poland;
FCT, Portugal;
MNE/IFA, Romania;
JINR;
MES of Russia and NRC KI, Russian Federation;
MESTD, Serbia;
MSSR, Slovakia;
ARRS and MIZ\v{S}, Slovenia;
DSI/NRF, South Africa;
MICINN, Spain;
SRC and Wallenberg Foundation, Sweden;
SERI, SNSF and Cantons of Bern and Geneva, Switzerland;
MOST, Taiwan;
TAEK, Turkey;
STFC, United Kingdom;
DOE and NSF, United States of America.
In addition, individual groups and members have received support from
BCKDF, CANARIE, Compute Canada and CRC, Canada;
COST, ERC, ERDF, Horizon 2020 and Marie Sk{\l}odowska-Curie Actions, European Union;
Investissements d'Avenir Labex, Investissements d'Avenir Idex and ANR, France;
DFG and AvH Foundation, Germany;
Herakleitos, Thales and Aristeia programmes co-financed by EU-ESF and the Greek NSRF, Greece;
BSF-NSF and GIF, Israel;
Norwegian Financial Mechanism 2014-2021, Norway;
NCN and NAWA, Poland;
La Caixa Banking Foundation, CERCA Programme Generalitat de Catalunya and PROMETEO and GenT Programmes Generalitat Valenciana, Spain;
G\"{o}ran Gustafssons Stiftelse, Sweden;
The Royal Society and Leverhulme Trust, United Kingdom.
 
The crucial computing support from all WLCG partners is acknowledged gratefully, in particular from CERN, the ATLAS Tier-1 facilities at TRIUMF (Canada), NDGF (Denmark, Norway, Sweden), CC-IN2P3 (France), KIT/GridKA (Germany), INFN-CNAF (Italy), NL-T1 (Netherlands), PIC (Spain), ASGC (Taiwan), RAL (UK) and BNL (USA), the Tier-2 facilities worldwide and large non-WLCG resource providers. Major contributors of computing resources are listed in Ref.~\cite{ATL-SOFT-PUB-2021-003}.
 
% End of text imported from the .//acknowledgements/Acknowledgements.tex input file

\clearpage
 
\printbibliography

@misc{hepdata.115578,
  author = "",
  title = "{Search for long-lived neutral particles in  pp collisions at sqrt{s} = 13 TeV that decay into displaced hadronic jets in the ATLAS calorimeter}",
  howpublished = "{HEPData (collection)}",
  year = 2022,
  note = "\url{https://doi.org/10.17182/hepdata.115578}"
}

@article{Cepeda:2019klc,
  author = "{Working Group 2 on the Physics of the HL-LHC}",
  editor = "Dainese, Andrea and Mangano, Michelangelo and Meyer, Andreas B. and Nisati, Aleandro and Salam, Gavin and Vesterinen, Mika Anton",
  title = "{Report from Working Group 2}: {Higgs Physics at the HL-LHC and HE-LHC}",
  eprint = "1902.00134",
  archivePrefix = "arXiv",
  primaryClass = "hep-ph",
  reportNumber = "CERN-LPCC-2018-04",
  doi = "10.23731/CYRM-2019-007.221",
  journal = "CERN Yellow Rep. Monogr.",
  volume = "7",
  pages = "221--584",
  year = "2019"
}

@misc{tensorflow,
title={ {TensorFlow}: Large-Scale Machine Learning on Heterogeneous Systems},
howpublished = {\url{https://www.tensorflow.org/}},
note={Software available from tensorflow.org},
author={
    Martín~Abadi and
    Ashish~Agarwal and
    Paul~Barham and
    Eugene~Brevdo and
    Zhifeng~Chen and
    Craig~Citro and
    Greg~S.~Corrado and
    Andy~Davis and
    Jeffrey~Dean and
    Matthieu~Devin and
    Sanjay~Ghemawat and
    Ian~Goodfellow and
    Andrew~Harp and
    Geoffrey~Irving and
    Michael~Isard and
    Yangqing Jia and
    Rafal~Jozefowicz and
    Lukasz~Kaiser and
    Manjunath~Kudlur and
    Josh~Levenberg and
    Dandelion~Mané and
    Rajat~Monga and
    Sherry~Moore and
    Derek~Murray and
    Chris~Olah and
    Mike~Schuster and
    Jonathon~Shlens and
    Benoit~Steiner and
    Ilya~Sutskever and
    Kunal~Talwar and
    Paul~Tucker and
    Vincent~Vanhoucke and
    Vijay~Vasudevan and
    Fernanda~Viégas and
    Oriol~Vinyals and
    Pete~Warden and
    Martin~Wattenberg and
    Martin~Wicke and
    Yuan~Yu and
    Xiaoqiang~Zheng},
  year={2015},
}

@misc{keras,
  title={Keras},
  author={Chollet, Fran\c{c}ois and others},
  year={2015},
  howpublished={\url{https://keras.io}},
}

@inproceedings{lstm, author = {Hochreiter, Sepp and Schmidhuber, J\"{u}rgen}, title = {LSTM Can Solve Hard Long Time Lag Problems}, year = {1996}, publisher = {MIT Press}, address = {Cambridge, MA, USA}, booktitle = {Proceedings of the 9th International Conference on Neural Information Processing Systems}, pages = {473–479}, numpages = {7}, location = {Denver, Colorado}, series = {NIPS'96} }

@article{Arvanitaki:2012ps,
  author         = "Arvanitaki, Asimina and Craig, Nathaniel and Dimopoulos,
  Savas and Villadoro, Giovanni",
  title          = "{Mini-Split}",
  journal        = "JHEP",
  volume         = "02",
  year           = "2013",
  pages          = "126",
  doi            = "10.1007/JHEP02(2013)126",
  eprint         = "1210.0555",
  archivePrefix  = "arXiv",
  primaryClass   = "hep-ph",
  SLACcitation   = "%%CITATION = ARXIV:1210.0555;%%"
}

@article{ArkaniHamed:2012gw,
  author         = "Arkani-Hamed, Nima and Gupta, Arpit and Kaplan, David E.
    and Weiner, Neal and Zorawski, Tom",
  title          = "{Simply Unnatural Supersymmetry}",
  year           = "2012",
  eprint         = "1212.6971",
  archivePrefix  = "arXiv",
  primaryClass   = "hep-ph",
  SLACcitation   = "%%CITATION = ARXIV:1212.6971;%%"
}

@article{Giudice:1998bp,
  author         = "Giudice, G. F. and Rattazzi, R.",
  title          = "{Theories with gauge-mediated supersymmetry breaking}",
  journal        = "Phys. Rept.",
  volume         = "322",
  year           = "1999",
  pages          = "419-499",
  doi            = "10.1016/S0370-1573(99)00042-3",
  eprint         = "hep-ph/9801271",
  archivePrefix  = "arXiv",
  primaryClass   = "hep-ph",
  reportNumber   = "CERN-TH-97-380",
  SLACcitation   = "%%CITATION = HEP-PH/9801271;%%"
}

@article{Barbier:2004ez,
  author         = "Barbier, R. and others",
  title          = "{R-Parity-violating supersymmetry}",
  journal        = "Phys. Rept.",
  volume         = "420",
  year           = "2005",
  pages          = "1-202",
  doi            = "10.1016/j.physrep.2005.08.006",
  eprint         = "hep-ph/0406039",
  archivePrefix  = "arXiv",
  primaryClass   = "hep-ph",
  SLACcitation   = "%%CITATION = HEP-PH/0406039;%%"
}

@article{Csaki:2015fea,
  author         = "Csaki, Csaba and Kuflik, Eric and Slone, Oren and
    Volansky, Tomer",
  title          = "{Models of dynamical R-parity violation}",
  journal        = "JHEP",
  volume         = "06",
  year           = "2015",
  pages          = "045",
  doi            = "10.1007/JHEP06(2015)045",
  eprint         = "1502.03096",
  archivePrefix  = "arXiv",
  primaryClass   = "hep-ph",
  SLACcitation   = "%%CITATION = ARXIV:1502.03096;%%"
}

@article{Fan:2011yu,
  author         = "Fan, JiJi and Reece, Matthew and Ruderman, Joshua T.",
  title          = "{Stealth supersymmetry}",
  journal        = "JHEP",
  volume         = "11",
  year           = "2011",
  pages          = "012",
  doi            = "10.1007/JHEP11(2011)012",
  eprint         = "1105.5135",
  archivePrefix  = "arXiv",
  primaryClass   = "hep-ph",
  SLACcitation   = "%%CITATION = ARXIV:1105.5135;%%"
}

@article{Fan:2012jf,
  author         = "Fan, JiJi and Reece, Matthew and Ruderman, Joshua T.",
  title          = "{A stealth supersymmetry sampler}",
  journal        = "JHEP",
  volume         = "07",
  year           = "2012",
  pages          = "196",
  doi            = "10.1007/JHEP07(2012)196",
  eprint         = "1201.4875",
  archivePrefix  = "arXiv",
  primaryClass   = "hep-ph",
  SLACcitation   = "%%CITATION = ARXIV:1201.4875;%%"
}

@article{Chacko:2015fbc,
  author         = "Chacko, Zackaria and Curtin, David and Verhaaren,
  Christopher B.",
  title          = "{A quirky probe of neutral naturalness}",
  journal        = "Phys. Rev. D",
  volume         = "94",
  year           = "2016",
  number         = "1",
  pages          = "011504",
  doi            = "10.1103/PhysRevD.94.011504",
  eprint         = "1512.05782",
  archivePrefix  = "arXiv",
  primaryClass   = "hep-ph",
  SLACcitation   = "%%CITATION = ARXIV:1512.05782;%%"
}

@article{Burdman:2006tz,
  author         = "Burdman, Gustavo and Chacko, Z. and Goh, Hock-Seng and
    Harnik, Roni",
  title          = "{Folded supersymmetry and the LEP paradox}",
  journal        = "JHEP",
  volume         = "02",
  year           = "2007",
  pages          = "009",
  doi            = "10.1088/1126-6708/2007/02/009",
  eprint         = "hep-ph/0609152",
  archivePrefix  = "arXiv",
  primaryClass   = "hep-ph",
  reportNumber   = "SLAC-PUB-12115",
  SLACcitation   = "%%CITATION = HEP-PH/0609152;%%"
}

@article{Cai:2008au,
  author         = "Cai, Haiying and Cheng, Hsin-Chia and Terning, John",
  title          = "{A quirky little Higgs model}",
  journal        = "JHEP",
  volume         = "05",
  year           = "2009",
  pages          = "045",
  doi            = "10.1088/1126-6708/2009/05/045",
  eprint         = "0812.0843",
  archivePrefix  = "arXiv",
  primaryClass   = "hep-ph",
  SLACcitation   = "%%CITATION = ARXIV:0812.0843;%%"
}

@article{Chacko:2005pe,
  author         = "Chacko, Z. and Goh, Hock-Seng and Harnik, Roni",
  title          = "{Natural Electroweak Breaking from a Mirror
    Symmetry}",
journal        = "Phys. Rev. Lett.",
volume         = "96",
year           = "2006",
pages          = "231802",
doi            = "10.1103/PhysRevLett.96.231802",
eprint         = "hep-ph/0506256",
archivePrefix  = "arXiv",
primaryClass   = "hep-ph",
SLACcitation   = "%%CITATION = HEP-PH/0506256;%%"
}

@article{Strassler:2006im,
  author         = "Strassler, Matthew J. and Zurek, Kathryn M.",
  title          = "{Echoes of a hidden valley at hadron colliders}",
  journal        = "Phys. Lett. B",
  volume         = "651",
  year           = "2007",
  pages          = "374-379",
  doi            = "10.1016/j.physletb.2007.06.055",
  eprint         = "hep-ph/0604261",
  archivePrefix  = "arXiv",
  primaryClass   = "hep-ph",
  SLACcitation   = "%%CITATION = HEP-PH/0604261;%%"
}

@Article{Strassler:951513,
  author         = "{M. J. Strassler, K. M. Zurek}",
  title          = "{Discovering the Higgs through highly-displaced vertices}",
  journal        = "Phys. Lett. B",
  volume         = "661",
  year           = "2006",
  pages          = "263-267",
  doi            = "10.1016/j.physletb.2008.02.008",
  eprint         = "hep-ph/0605193",
  archivePrefix  = "arXiv",
  primaryClass   = "hep-ph",
}

@article{Baumgart:2009tn,
  author         = "Baumgart, Matthew and Cheung, Clifford and Ruderman,
  Joshua T. and Wang, Lian-Tao and Yavin, Itay",
  title          = "{Non-abelian dark sectors and their collider signatures}",
  journal        = "JHEP",
  volume         = "04",
  year           = "2009",
  pages          = "014",
  doi            = "10.1088/1126-6708/2009/04/014",
  eprint         = "0901.0283",
  archivePrefix  = "arXiv",
  primaryClass   = "hep-ph",
  SLACcitation   = "%%CITATION = ARXIV:0901.0283;%%"
}

@article{Kaplan:2009ag,
  author         = "Kaplan, David E. and Luty, Markus A. and Zurek, Kathryn
    M.",
  title          = "{Asymmetric dark matter}",
  journal        = "Phys. Rev. D",
  volume         = "79",
  year           = "2009",
  pages          = "115016",
  doi            = "10.1103/PhysRevD.79.115016",
  eprint         = "0901.4117",
  archivePrefix  = "arXiv",
  primaryClass   = "hep-ph",
  reportNumber   = "FERMILAB-PUB-09-345-A-T",
  SLACcitation   = "%%CITATION = ARXIV:0901.4117;%%"
}

@article{Chan:2011aa,
  author         = "Chan, Yuk Fung and Low, Matthew and Morrissey, David E.
    and Spray, Andrew P.",
  title          = "{LHC signatures of a minimal supersymmetric hidden
    valley}",
journal        = "JHEP",
volume         = "05",
year           = "2012",
pages          = "155",
doi            = "10.1007/JHEP05(2012)155",
eprint         = "1112.2705",
archivePrefix  = "arXiv",
primaryClass   = "hep-ph",
SLACcitation   = "%%CITATION = ARXIV:1112.2705;%%"
}

@article{Dienes:2011ja,
  author         = "Dienes, Keith R. and Thomas, Brooks",
  title          = "{Dynamical dark matter: I. Theoretical overview}",
  journal        = "Phys. Rev. D",
  volume         = "85",
  year           = "2012",
  pages          = "083523",
  doi            = "10.1103/PhysRevD.85.083523",
  eprint         = "1106.4546",
  archivePrefix  = "arXiv",
  primaryClass   = "hep-ph",
  SLACcitation   = "%%CITATION = ARXIV:1106.4546;%%"
}

@article{Dienes:2012yz,
  author         = "Dienes, Keith R. and Su, Shufang and Thomas, Brooks",
  title          = "{Distinguishing dynamical dark matter at the LHC}",
  journal        = "Phys. Rev. D",
  volume         = "86",
  year           = "2012",
  pages          = "054008",
  doi            = "10.1103/PhysRevD.86.054008",
  eprint         = "1204.4183",
  archivePrefix  = "arXiv",
  primaryClass   = "hep-ph",
  reportNumber   = "UH511-1191-12",
  SLACcitation   = "%%CITATION = ARXIV:1204.4183;%%"
}

@article{Cui:2014twa,
  author         = "Cui, Yanou and Shuve, Brian",
  title          = "{Probing baryogenesis with displaced vertices at the
    LHC}",
journal        = "JHEP",
volume         = "02",
year           = "2015",
pages          = "049",
doi            = "10.1007/JHEP02(2015)049",
eprint         = "1409.6729",
archivePrefix  = "arXiv",
primaryClass   = "hep-ph",
reportNumber   = "UMD-PP-014-014",
SLACcitation   = "%%CITATION = ARXIV:1409.6729;%%"
}

@article{Helo:2013esa,
  author         = "Helo, Juan C. and Kovalenko, Sergey and Hirsch, Martin",
  title          = "{Heavy neutrino searches at the LHC with displaced
    vertices}",
journal        = "Phys. Rev. D",
volume         = "89",
year           = "2014",
pages          = "073005",
doi            = "10.1103/PhysRevD.89.073005",
eprint         = "1312.2900",
archivePrefix  = "arXiv",
primaryClass   = "hep-ph",
SLACcitation   = "%%CITATION = ARXIV:1312.2900;%%",
relatedstring  = "Erratum:"
}

@article{Batell:2016zod,
  author         = "Batell, Brian and Pospelov, Maxim and Shuve, Brian",
  title          = "{Shedding light on neutrino masses with dark forces}",
  journal        = "JHEP",
  volume         = "08",
  year           = "2016",
  pages          = "052",
  doi            = "10.1007/JHEP08(2016)052",
  eprint         = "1604.06099",
  archivePrefix  = "arXiv",
  primaryClass   = "hep-ph",
  SLACcitation   = "%%CITATION = ARXIV:1604.06099;%%"
}

@Article{Chung2006,
  author         = "{S. Chang,  P.J. Fox, N. Weiner}",
  title          = "{Naturalness and Higgs decays in the MSSM with a singlet}",
  journal        = "JHEP",
  volume         = "08",
  year           = "2006",
  pages          = "068",
  doi            = "10.1088/1126-6708/2006/08/068",
  eprint         = "hep-ph/0511250",
  archivePrefix  = "arXiv",
  primaryClass   = "hep-ph",
}

@Article{Chang2008,
  author         = "{S. Chang,  R. Derm\'{i}u{s}ek, J.F. Gunion, N. Weiner}",
  title          = "{Nonstandard Higgs Boson Decays}",
  journal        = "Ann. Rev. Nucl. Part. Sci.",
  volume         = "58",
  year           = "2008",
  pages          = "75-98",
  doi            = "10.1146/annurev.nucl.58.110707.171200",
  eprint         = "0801.4554",
  archivePrefix  = "arXiv",
  primaryClass   = "hep-ph",
}

@article{Jedamzik:2006xz,
  author         = "{ATLAS Collaboration}",
  author         = "Jedamzik, Karsten",
  title          = "{Big bang nucleosynthesis constraints on hadronically and
    electromagnetically decaying relic neutral particles}",
journal        = "Phys. Rev. D",
volume         = "74",
year           = "2006",
pages          = "103509",
doi            = "10.1103/PhysRevD.74.103509",
eprint         = "hep-ph/0604251",
archivePrefix  = "arXiv",
primaryClass   = "hep-ph",
SLACcitation   = "%%CITATION = HEP-PH/0604251;%%"
}

@Article{HVD0,
  author         = "{D0 Collaboration}",
  title          = "{Search for resonant pair production of neutral long-lived particles decaying to $b\bar{b}$ in $p\bar{p}$ collisions at $\sqrt(s)=1.96~\TeV$}",
  journal        = "Phys. Rev. Lett.",
  volume         = "103",
  year           = "2009",
  pages          = "071801",
  doi            = " 10.1103/PhysRevLett.103.071801",
  eprint         = "0906.1787",
  archivePrefix  = "arXiv",
  primaryClass   = "hep-ex",
}

@Article{HVCDF,
  author         = "{CDF Collaboration}",
  title          = "{Search for heavy metastable particles decaying to jet pairs in $p\bar{p}$ collisions at $\sqrt{s} = 1.96~\TeV$}",
  journal        = "Phys. Rev. D",
  volume         = "85",
  year           = "2012",
  pages          = "012007",
  doi            = "10.1103/PhysRevD.85.012007",
  eprint         = "1109.3136",
  archivePrefix  = "arXiv",
  primaryClass   = "hep-ex",
}

@article{LHCb:2017xxn,
  author = "{LHCb Collaboration}",
  title = "{Updated search for long-lived particles decaying to jet pairs}",
  eprint = "1705.07332",
  archivePrefix = "arXiv",
  primaryClass = "hep-ex",
  reportNumber = "LHCB-PAPER-2016-065, CERN-EP-2017-083",
  doi = "10.1140/epjc/s10052-017-5178-x",
  journal = "Eur. Phys. J. C",
  volume = "77",
  number = "12",
  pages = "812",
  year = "2017"
}

@article{CERN-EP-2016-029,
  author         = "{ATLAS Collaboration}",
  title          = "{Beam-induced and cosmic-ray backgrounds observed in the
    ATLAS detector during the LHC 2012 proton-proton running
    period}",
collaboration  = "ATLAS",
journal        = "JINST",
volume         = "11",
year           = "2016",
number         = "05",
pages          = "P05013",
doi            = "10.1088/1748-0221/11/05/P05013",
eprint         = "1603.09202",
archivePrefix  = "arXiv",
primaryClass   = "hep-ex",
reportNumber   = "CERN-EP-2016-029",
SLACcitation   = "%%CITATION = ARXIV:1603.09202;%%"
}

@Article{MadGraphNLO,
  author         = "{J. Alwall et al}",
  title          = "{The automated computation of tree-level and next-to-leading order differential cross
    sections, and their matching to parton shower simulations}",
journal        = "JHEP",
volume         = "07",
year           = "2014",
pages          = "079",
doi            = " 10.1007/JHEP07(2014)079",
eprint         = "1405.0301",
archivePrefix  = "arXiv",
primaryClass   = "hep-ex",
}

@article{Ball:2012cx,
  author         = "Ball, Richard D. and others",
  title          = "{Parton distributions with LHC data}",
  journal        = "Nucl. Phys. B",
  volume         = "867",
  year           = "2013",
  pages          = "244-289",
  doi            = "10.1016/j.nuclphysb.2012.10.003",
  eprint         = "1207.1303",
  archivePrefix  = "arXiv",
  primaryClass   = "hep-ph",
  reportNumber   = "EDINBURGH-2012-08, IFUM-FT-997, FR-PHENO-2012-014,
  RWTH-TTK-12-25, CERN-PH-TH-2012-037, SFB-CPP-12-47",
  SLACcitation   = "%%CITATION = ARXIV:1207.1303;%%"
}

@article{Sjostrand:2014zea,
  author         = "Sjöstrand, Torbjörn and Ask, Stefan and Christiansen,
  Jesper R. and Corke, Richard and Desai, Nishita and Ilten,
  Philip and Mrenna, Stephen and Prestel, Stefan and
    Rasmussen, Christine O. and Skands, Peter",
  title          = "{An introduction to PYTHIA 8.2}",
  journal        = "Comput. Phys. Commun.",
  volume         = "191",
  year           = "2015",
  pages          = "159-177",
  doi            = "10.1016/j.cpc.2015.01.024",
  eprint         = "1410.3012",
  archivePrefix  = "arXiv",
  primaryClass   = "hep-ph",
  reportNumber   = "LU-TP-14-36, MCNET-14-22, CERN-PH-TH-2014-190,
  FERMILAB-PUB-14-316-CD, DESY-14-178, SLAC-PUB-16122",
  SLACcitation   = "%%CITATION = ARXIV:1410.3012;%%"
}

@ARTICLE{Sjostrand:2007gs,
  author = {Sj\"{o}strand, Torbjorn and Mrenna, Stephen and Skands, Peter},
  title = {{A brief introduction to PYTHIA 8.1}},
  journal = {Comput. Phys. Commun.},
  year = {2008},
  volume = {178},
  pages = {852-867},
  archiveprefix = {arXiv},
  doi = {10.1016/j.cpc.2008.01.036},
  eprint = {0710.3820},
  primaryclass = {hep-ph},
  reportnumber = {CERN-LCGAPP-2007-04, LU-TP-07-28, FERMILAB-PUB-07-512-CD-T},
  slaccitation = {%%CITATION = ARXIV:0710.3820;%%}
}

@ARTICLE{Geant4,
  author = {Agostinelli, S. and others},
  title = {{GEANT4 - a simulation toolkit}},
  journal = {Nucl. Instrum. Meth.},
  year = {2003},
  volume = {A506},
  pages = {250-303},
  doi = {10.1016/S0168-9002(03)01368-8},
  reportnumber = {SLAC-PUB-9350, FERMILAB-PUB-03-339},
  slaccitation = {%%CITATION = NUIMA,A506,250;%%}
}

@article{LUCID2,
  author={G. Avoni and others},
  title={The new LUCID-2 detector for luminosity measurement and monitoring in ATLAS},
  journal={JINST},
  volume={13},
  number={07},
  pages={P07017},
  doi="10.1088/1748-0221/13/07/P07017",
  year={2018},
}

@article{Aaboud:2017jcu,
  author         = "{ATLAS Collaboration}",
  title          = "{Jet energy scale measurements and their systematic
    uncertainties in proton-proton collisions at $\sqrt{s} =
    13$ TeV with the ATLAS detector}",
collaboration  = "ATLAS",
journal        = "Phys. Rev. D",
volume         = "96",
year           = "2017",
number         = "7",
pages          = "072002",
doi            = "10.1103/PhysRevD.96.072002",
eprint         = "1703.09665",
archivePrefix  = "arXiv",
primaryClass   = "hep-ex",
reportNumber   = "CERN-EP-2017-038",
SLACcitation   = "%%CITATION = ARXIV:1703.09665;%%"
}

@software{pyhf,
  author = {Lukas Heinrich and Matthew Feickert and Giordon Stark},
  title = "{pyhf: v0.6.3}",
  version = {0.6.3},
  doi = {10.5281/zenodo.1169739},
  url = {https://doi.org/10.5281/zenodo.1169739},
  note = {https://github.com/scikit-hep/pyhf/releases/tag/v0.6.3}
}

@article{pyhf_joss,
  doi = {10.21105/joss.02823},
  url = {https://doi.org/10.21105/joss.02823},
  year = {2021},
  publisher = {The Open Journal},
  volume = {6},
  number = {58},
  pages = {2823},
  author = {Lukas Heinrich and Matthew Feickert and Giordon Stark and Kyle Cranmer},
  title = {pyhf: pure-Python implementation of HistFactory statistical models},
  journal = {Journal of Open Source Software}
}

@inproceedings{adversary_ref,
  author    = {Gilles Louppe and
    Michael Kagan and
      Kyle Cranmer},
  editor    = {Isabelle Guyon and
    Ulrike von Luxburg and
      Samy Bengio and
      Hanna M. Wallach and
      Rob Fergus and
      S. V. N. Vishwanathan and
      Roman Garnett},
  title     = {Learning to Pivot with Adversarial Networks},
  booktitle = {Advances in Neural Information Processing Systems 30: Annual Conference
    on Neural Information Processing Systems 2017, December 4-9, 2017,
    Long Beach, CA, {USA}},
  pages     = {981--990},
  year      = {2017},
  url       = {https://proceedings.neurips.cc/paper/2017/hash/48ab2f9b45957ab574cf005eb8a76760-Abstract.html},
  bibsource = {dblp computer science bibliography, https://dblp.org}
}

@Article{Cacciari:2008gp,
     author    = "Cacciari, Matteo and Salam, Gavin P. and Soyez, Gregory",
     title     = "{The anti-\(k_{t}\) jet clustering algorithm}",
     journal   = "JHEP",
     volume    = "04",
     year      = "2008",
     pages     = "063",
     eprint    = "0802.1189",
     archivePrefix = "arXiv",
     primaryClass  =  "hep-ph",
     doi       = "10.1088/1126-6708/2008/04/063",
     SLACcitation  = "%%CITATION = 0802.1189;%%"
}

@Article{Fastjet,
      author         = "Cacciari, Matteo and Salam, Gavin P. and Soyez, Gregory",
      title          = "{FastJet user manual}",
      journal        = "Eur. Phys. J. C",
      volume         = "72",
      year           = "2012",
      pages          = "1896",
      doi            = "10.1140/epjc/s10052-012-1896-2",
      eprint         = "1111.6097",
      archivePrefix  = "arXiv",
      primaryClass   = "hep-ph",
      reportNumber   = "CERN-PH-TH-2011-297",
      SLACcitation   = "%%CITATION = ARXIV:1111.6097;%%"
}

@Article{Nason:2004rx,
      author         = "Nason, Paolo",
      title          = "{A new method for combining NLO QCD with shower Monte Carlo algorithms}",
      journal        = "JHEP",
      volume         = "11",
      pages          = "040",
      doi            = "10.1088/1126-6708/2004/11/040",
      year           = "2004",
      eprint         = "hep-ph/0409146",
      archivePrefix  = "arXiv",
}

@Article{Frixione:2007vw,
      author         = "Frixione, Stefano and Nason, Paolo and Oleari, Carlo",
      title          = "{Matching NLO QCD computations with parton shower
                        simulations: the POWHEG method}",
      journal        = "JHEP",
      volume         = "11",
      pages          = "070",
      doi            = "10.1088/1126-6708/2007/11/070",
      year           = "2007",
      eprint         = "0709.2092",
      archivePrefix  = "arXiv",
      primaryClass   = "hep-ph",
}

@Article{Alioli:2010xd,
      author         = "Alioli, Simone and Nason, Paolo and Oleari, Carlo and Re,
                        Emanuele",
      title          = "{A general framework for implementing NLO calculations in
                        shower Monte Carlo programs: the POWHEG BOX}",
      journal        = "JHEP",
      volume         = "06",
      pages          = "043",
      doi            = "10.1007/JHEP06(2010)043",
      year           = "2010",
      eprint         = "1002.2581",
      archivePrefix  = "arXiv",
      primaryClass   = "hep-ph",
}

@article{Hamilton:2013fea,
      author         = "Hamilton, Keith and Nason, Paolo and Re, Emanuele and
                        Zanderighi, Giulia",
      title          = "{NNLOPS simulation of Higgs boson production}",
      journal        = "JHEP",
      volume         = "10",
      year           = "2013",
      pages          = "222",
      doi            = "10.1007/JHEP10(2013)222",
      eprint         = "1309.0017",
      archivePrefix  = "arXiv",
      primaryClass   = "hep-ph",
      reportNumber   = "MCNET-13-11, CERN-PH-TH-2013-205, OUTP-13-18P",
      SLACcitation   = "%%CITATION = ARXIV:1309.0017;%%"
}

@article{Hamilton:2015nsa,
      author         = "Hamilton, Keith and Nason, Paolo and Zanderighi, Giulia",
      title          = "{Finite quark-mass effects in the NNLOPS POWHEG+MiNLO
                        Higgs generator}",
      journal        = "JHEP",
      volume         = "05",
      year           = "2015",
      pages          = "140",
      doi            = "10.1007/JHEP05(2015)140",
      eprint         = "1501.04637",
      archivePrefix  = "arXiv",
      primaryClass   = "hep-ph",
      reportNumber   = "CERN-PH-TH-2015-006",
      SLACcitation   = "%%CITATION = ARXIV:1501.04637;%%"
}

@Article{Read:2002hq,
    Author = {Read, Alexander L.},
    Journal = {J. Phys. G},
    Pages = {2693},
    doi = "10.1088/0954-3899/28/10/313",
    Title = "{Presentation of search results: the \(CL_S\) technique}",
    Volume = {28},
    Year = {2002}
}

@Article{Cowan:2010js,
   Author = {Cowan, Glen and Cranmer, Kyle and Gross, Eilam and Vitells, Ofer},
   Title = {{Asymptotic formulae for likelihood-based tests of new physics}},
   Journal = {Eur. Phys. J. C},
   Volume = {71},
   Year = {2011},
   Pages = {1554},
   doi = "10.1140/epjc/s10052-011-1554-0",
   Eprint = "1007.1727",
   Archiveprefix = {arXiv},
   Primaryclass = {physics.data-an},
   related = "Cowan:2010js-err",
   relatedstring  = "Erratum:",
}

@Article{PERF-2007-01,
    author         = "{ATLAS Collaboration}",
    title          = "{The ATLAS Experiment at the CERN Large Hadron Collider}",
    journal        = "JINST",
    volume         = "3",
    year           = "2008",
    pages          = "S08003",
    doi            = "10.1088/1748-0221/3/08/S08003",
    primaryClass   = "hep-ex",
}

@Article{SOFT-2010-01,
    author         = "{ATLAS Collaboration}",
    title          = "{The ATLAS Simulation Infrastructure}",
    journal        = "Eur. Phys. J. C",
    volume         = "70",
    year           = "2010",
    pages          = "823",
    doi            = "10.1140/epjc/s10052-010-1429-9",
    eprint         = "1005.4568",
    archivePrefix  = "arXiv",
    primaryClass   = "physics.ins-det",
}

@Article{DAPR-2012-01,
    author         = "{ATLAS Collaboration}",
    title          = "{Characterisation and mitigation of beam-induced backgrounds observed in the ATLAS detector during the 2011 proton--proton run}",
    journal        = "JINST",
    volume         = "8",
    year           = "2013",
    pages          = "P07004",
    doi            = "10.1088/1748-0221/8/07/P07004",
    reportNumber   = "CERN-PH-EP-2012-313",
    eprint         = "1303.0223",
    archivePrefix  = "arXiv",
    primaryClass   = "hep-ex",
}

@Article{TRIG-2012-02,
    author         = "{ATLAS Collaboration}",
    title          = "{Triggers for displaced decays of long-lived neutral particles in the ATLAS detector}",
    journal        = "JINST",
    volume         = "8",
    year           = "2013",
    pages          = "P07015",
    doi            = "10.1088/1748-0221/8/07/P07015",
    reportNumber   = "CERN-PH-EP-2013-047",
    eprint         = "1305.2284",
    archivePrefix  = "arXiv",
    primaryClass   = "hep-ex",
}

@Article{PERF-2014-07,
    author         = "{ATLAS Collaboration}",
    title          = "{Topological cell clustering in the ATLAS calorimeters and its performance in LHC Run~1}",
    journal        = "Eur. Phys. J. C",
    volume         = "77",
    year           = "2017",
    pages          = "490",
    doi            = "10.1140/epjc/s10052-017-5004-5",
    reportNumber   = "CERN-PH-EP-2015-304",
    eprint         = "1603.02934",
    archivePrefix  = "arXiv",
    primaryClass   = "hep-ex",
}

@Article{EXOT-2017-05,
    author         = "{ATLAS Collaboration}",
    title          = "{Search for long-lived particles produced in \(pp\) collisions at \(\sqrt{s} = 13\,\text{TeV}\) that decay into displaced hadronic jets in the ATLAS muon spectrometer}",
    journal        = "Phys. Rev. D",
    volume         = "99",
    year           = "2019",
    pages          = "052005",
    doi            = "10.1103/PhysRevD.99.052005",
    reportNumber   = "CERN-EP-2018-241",
    eprint         = "1811.07370",
    archivePrefix  = "arXiv",
    primaryClass   = "hep-ex",
}

@Article{EXOT-2017-25,
    author         = "{ATLAS Collaboration}",
    title          = "{Search for long-lived neutral particles in \(pp\) collisions at \(\sqrt{s} = 13\,\text{TeV}\) that decay into displaced hadronic jets in the ATLAS calorimeter}",
    journal        = "Eur. Phys. J. C",
    volume         = "79",
    year           = "2019",
    pages          = "481",
    doi            = "10.1140/epjc/s10052-019-6962-6",
    reportNumber   = "CERN-EP-2018-351",
    eprint         = "1902.03094",
    archivePrefix  = "arXiv",
    primaryClass   = "hep-ex",
}

@Article{HIGG-2017-05,
    author         = "{ATLAS Collaboration}",
    title          = "{Search for the Higgs boson produced in association with a vector boson and decaying into two spin-zero particles in the \(H \rightarrow aa \rightarrow 4b\) channel in \(pp\) collisions at \(\sqrt{s} = 13\,\text{TeV}\) with the ATLAS detector}",
    journal        = "JHEP",
    volume         = "10",
    year           = "2018",
    pages          = "031",
    doi            = "10.1007/JHEP10(2018)031",
    reportNumber   = "CERN-EP-2018-128",
    eprint         = "1806.07355",
    archivePrefix  = "arXiv",
    primaryClass   = "hep-ex",
}

@Article{DAPR-2018-01,
    author         = "{ATLAS Collaboration}",
    title          = "{ATLAS data quality operations and performance for 2015--2018 data-taking}",
    journal        = "JINST",
    volume         = "15",
    year           = "2020",
    pages          = "P04003",
    doi            = "10.1088/1748-0221/15/04/P04003",
    reportNumber   = "CERN-EP-2019-207",
    eprint         = "1911.04632",
    archivePrefix  = "arXiv",
    primaryClass   = "physics.ins-det",
}

@Article{EXOT-2018-57,
    author         = "{ATLAS Collaboration}",
    title          = "{Search for exotic decays of the Higgs boson into long-lived particles in \(pp\) collisions at \(\sqrt{s} = 13\,\text{TeV}\) using displaced vertices in the ATLAS inner detector}",
    journal        = "JHEP",
    volume         = "11",
    year           = "2021",
    pages          = "229",
    doi            = "10.1007/JHEP11(2021)229",
    eprint         = "2107.06092",
    archivePrefix  = "arXiv",
    primaryClass   = "hep-ex",
}

@Article{EXOT-2018-61,
    author         = "{ATLAS Collaboration}",
    title          = "{Search for long-lived neutral particles produced in \(pp\) collisions at \(\sqrt{s} = 13\,\text{TeV}\) decaying into displaced hadronic jets in the ATLAS inner detector and muon spectrometer}",
    journal        = "Phys. Rev. D",
    volume         = "101",
    year           = "2020",
    pages          = "052013",
    doi            = "10.1103/PhysRevD.101.052013",
    reportNumber   = "CERN-EP-2019-240",
    eprint         = "1911.12575",
    archivePrefix  = "arXiv",
    primaryClass   = "hep-ex",
}

@Article{PIX-2018-001,
    author         = "Abbott, B. and others",
    title          = "{Production and integration of the ATLAS Insertable B-Layer}",
    journal        = "JINST",
    volume         = "13",
    year           = "2018",
    pages          = "T05008",
    doi            = "10.1088/1748-0221/13/05/T05008",
    eprint         = "1803.00844",
    archivePrefix  = "arXiv",
    primaryClass   = "physics.ins-det",
}

@Article{EXOT-2019-23,
    author         = "{ATLAS Collaboration}",
    title          = "{Search for neutral long-lived particles in \(pp\) collisions at  \(\sqrt{s} = 13\,\text{TeV}\) that decay into displaced hadronic jets in the ATLAS calorimeter}",
    journal        = "JHEP",
    volume         = "06",
    year           = "2022",
    pages          = "005",
    doi            = "10.1007/JHEP06(2022)005",
    reportNumber   = "CERN-EP-2022-002",
    eprint         = "2203.01009",
    archivePrefix  = "arXiv",
    primaryClass   = "hep-ex",
}

@Article{TRIG-2019-02,
    author         = "{ATLAS Collaboration}",
    title          = "{Performance of the ATLAS Level-1 topological trigger in Run~2}",
    journal        = "Eur. Phys. J. C",
    volume         = "82",
    year           = "2021",
    pages          = "7",
    doi            = "10.1140/epjc/s10052-021-09807-0",
    reportNumber   = "CERN-EP-2021-040",
    eprint         = "2105.01416",
    archivePrefix  = "arXiv",
    primaryClass   = "hep-ex",
}

@Article{TRIG-2019-04,
    author         = "{ATLAS Collaboration}",
    title          = "{Operation of the ATLAS trigger system in Run~2}",
    journal        = "JINST",
    volume         = "15",
    year           = "2020",
    pages          = "P10004",
    doi            = "10.1088/1748-0221/15/10/P10004",
    reportNumber   = "CERN-EP-2020-109",
    eprint         = "2007.12539",
    archivePrefix  = "arXiv",
    primaryClass   = "hep-ex",
}

@Booklet{ATL-SOFT-PUB-2021-001,
    author         = "{ATLAS Collaboration}",
    title          = "{The ATLAS Collaboration Software and Firmware}",
    howpublished   = "{ATL-SOFT-PUB-2021-001}",
    url            = "https://cds.cern.ch/record/2767187",
    year           = "2021",
}

@Booklet{ATL-SOFT-PUB-2021-003,
    author         = "{ATLAS Collaboration}",
    title          = "{ATLAS Computing Acknowledgements}",
    howpublished   = "{ATL-SOFT-PUB-2021-003}",
    url            = "https://cds.cern.ch/record/2776662",
}

@Report{ATLAS-TDR-2010-19,
    author         = "{ATLAS Collaboration}",
    title          = "{ATLAS Insertable B-Layer Technical Design Report}",
    type           = "ATLAS-TDR-19; CERN-LHCC-2010-013",
    year           = "2010",
    url            = "https://cds.cern.ch/record/1291633",
    related        = "ATLAS-TDR-2010-19-addm",
    relatedstring  = "Addendum:",
}

@Article{CMS-EXO-17-018,
    author         = "{CMS Collaboration}",
    title          = "{Search for long-lived particles with displaced vertices in multijet events in proton--proton collisions at \(\sqrt{s} = 13\,\text{TeV}\)}",
    journal        = "Phys. Rev. D",
    volume         = "98",
    year           = "2018",
    pages          = "092011",
    doi            = "10.1103/PhysRevD.98.092011",
    reportNumber   = "CERN-EP-2018-203",
    eprint         = "1808.03078",
    archivePrefix  = "arXiv",
    primaryClass   = "hep-ex",
}

@Article{CMS-EXO-19-001,
    author         = "{CMS Collaboration}",
    title          = "{Search for long-lived particles using nonprompt jets and missing transverse momentum with proton--proton collisions at \(\sqrt{s} =  13\,\text{TeV}\)}",
    journal        = "Phys. Lett. B",
    volume         = "797",
    year           = "2019",
    pages          = "134876",
    doi            = "10.1016/j.physletb.2019.134876",
    reportNumber   = "CERN-EP-2019-113",
    eprint         = "1906.06441",
    archivePrefix  = "arXiv",
    primaryClass   = "hep-ex",
}

@Article{CMS-EXO-19-005,
    author         = "{CMS Collaboration}",
    title          = "{Search for long-lived particles using delayed photons in proton--proton collisions at \(\sqrt{s} = 13\,\text{TeV}\)}",
    journal        = "Phys. Rev. D",
    volume         = "100",
    year           = "2019",
    pages          = "112003",
    doi            = "10.1103/PhysRevD.100.112003",
    reportNumber   = "CERN-EP-2019-185",
    eprint         = "1909.06166",
    archivePrefix  = "arXiv",
    primaryClass   = "hep-ex",
}

@Article{CMS-EXO-19-011,
    author         = "{CMS Collaboration}",
    title          = "{A deep neural network to search for new long-lived particles decaying to jets}",
    journal        = "Mach. Learn. Sci. Tech.",
    volume         = "1",
    year           = "2020",
    pages          = "035012",
    doi            = "10.1088/2632-2153/ab9023",
    reportNumber   = "CERN-EP-2019-281",
    eprint         = "1912.12238",
    archivePrefix  = "arXiv",
    primaryClass   = "hep-ex",
}

@Booklet{ATLAS-CONF-2015-029,
    author         = "{ATLAS Collaboration}",
    title          = "{Selection of jets produced in \(13~\text{TeV}\) proton--proton collisions with the ATLAS detector}",
    howpublished   = "{ATLAS-CONF-2015-029}",
    url            = "https://cds.cern.ch/record/2037702",
    year           = "2015",
}

@Booklet{ATLAS-CONF-2017-061,
    author         = "{ATLAS Collaboration}",
    title          = "{The ATLAS Tau Trigger in Run~2}",
    howpublished   = "{ATLAS-CONF-2017-061}",
    url            = "https://cds.cern.ch/record/2274201",
    year           = "2017",
}

@Booklet{ATLAS-CONF-2019-021,
    author         = "{ATLAS Collaboration}",
    title          = "{Luminosity determination in \(pp\) collisions at \(\sqrt{s} = 13\,\text{TeV}\) using the ATLAS detector at the LHC}",
    howpublished   = "{ATLAS-CONF-2019-021}",
    url            = "https://cds.cern.ch/record/2677054",
    year           = "2019",
}

@Booklet{ATL-PHYS-PUB-2014-021,
    author         = "{ATLAS Collaboration}",
    title          = "{ATLAS Pythia~8 tunes to \(7~\text{TeV}\) data}",
    howpublished   = "{ATL-PHYS-PUB-2014-021}",
    url            = "https://cds.cern.ch/record/1966419",
    year           = "2014",
}

@Booklet{ATL-PHYS-PUB-2016-017,
    author         = "{ATLAS Collaboration}",
    title          = "{The Pythia~8 A3 tune description of ATLAS minimum bias and inelastic measurements incorporating the Donnachie--Landshoff diffractive model}",
    howpublished   = "{ATL-PHYS-PUB-2016-017}",
    url            = "https://cds.cern.ch/record/2206965",
    year           = "2016",
}
 
\clearpage
% ATLAS Collaboration author list
% Reference date of EXOT-2019-23 is 2021-11-03
% Author list last updated on date 20-APR-22
% Data extracted on 20-Apr-2022 for paper reference EXOT-2019-23
% at 7:40pm
 
\begin{flushleft}
\hypersetup{urlcolor=black}
{\Large The ATLAS Collaboration}

\bigskip

\AtlasOrcid[0000-0002-6665-4934]{G.~Aad}$^\textrm{\scriptsize 100}$,    
\AtlasOrcid[0000-0002-5888-2734]{B.~Abbott}$^\textrm{\scriptsize 126}$,    
\AtlasOrcid[0000-0002-7248-3203]{D.C.~Abbott}$^\textrm{\scriptsize 101}$,    
\AtlasOrcid[0000-0002-2788-3822]{A.~Abed~Abud}$^\textrm{\scriptsize 36}$,    
\AtlasOrcid[0000-0002-1002-1652]{K.~Abeling}$^\textrm{\scriptsize 53}$,    
\AtlasOrcid[0000-0002-2987-4006]{D.K.~Abhayasinghe}$^\textrm{\scriptsize 93}$,    
\AtlasOrcid[0000-0002-8496-9294]{S.H.~Abidi}$^\textrm{\scriptsize 29}$,    
\AtlasOrcid[0000-0002-9987-2292]{A.~Aboulhorma}$^\textrm{\scriptsize 35e}$,    
\AtlasOrcid[0000-0001-5329-6640]{H.~Abramowicz}$^\textrm{\scriptsize 158}$,    
\AtlasOrcid[0000-0002-1599-2896]{H.~Abreu}$^\textrm{\scriptsize 157}$
% ...
\bigskip
\\

% ...
$^{29}$Physics Department, Brookhaven National Laboratory, Upton NY; United States of America.\\
% ...
$^{35}$$^{(a)}$Facult\'e des Sciences Ain Chock, R\'eseau Universitaire de Physique des Hautes Energies - Universit\'e Hassan II, Casablanca;$^{(b)}$Facult\'{e} des Sciences, Universit\'{e} Ibn-Tofail, K\'{e}nitra;$^{(c)}$Facult\'e des Sciences Semlalia, Universit\'e Cadi Ayyad, LPHEA-Marrakech;$^{(d)}$LPMR, Facult\'e des Sciences, Universit\'e Mohamed Premier, Oujda;$^{(e)}$Facult\'e des sciences, Universit\'e Mohammed V, Rabat;$^{(f)}$Mohammed VI Polytechnic University, Ben Guerir; Morocco.\\
$^{36}$CERN, Geneva; Switzerland.\\
% ...
$^{53}$II. Physikalisches Institut, Georg-August-Universit\"{a}t G\"ottingen, G\"ottingen; Germany.\\
% ...
$^{93}$Department of Physics, Royal Holloway University of London, Egham; United Kingdom.\\
% ...
$^{100}$CPPM, Aix-Marseille Universit\'e, CNRS/IN2P3, Marseille; France.\\
$^{101}$Department of Physics, University of Massachusetts, Amherst MA; United States of America.\\
% ...
$^{126}$Homer L. Dodge Department of Physics and Astronomy, University of Oklahoma, Norman OK; United States of America.\\
% ...
$^{157}$Department of Physics, Technion, Israel Institute of Technology, Haifa; Israel.\\
$^{158}$Raymond and Beverly Sackler School of Physics and Astronomy, Tel Aviv University, Tel Aviv; Israel.\\
% ...
$^{a}$ Also at Borough of Manhattan Community College, City University of New York, New York NY; United States of America.\\
$^{b}$ Also at Bruno Kessler Foundation, Trento; Italy.\\
$^{c}$ Also at Center for High Energy Physics, Peking University; China.\\
$^{d}$ Also at Centro Studi e Ricerche Enrico Fermi; Italy.\\
$^{e}$ Also at CERN, Geneva; Switzerland.\\
$^{f}$ Also at D\'epartement de Physique Nucl\'eaire et Corpusculaire, Universit\'e de Gen\`eve, Gen\`eve; Switzerland.\\
% ...

\end{flushleft}

% Created with Glance <Atlas.Glance@cern.ch>

\end{document}